\documentclass[aps,twocolumn, superscriptaddress, english, nofootinbib, 10pt]{revtex4-2}

\usepackage{float}
\usepackage{amssymb, amsmath, bm, dcolumn, epsf, graphicx, latexsym, slashed, simplewick}
\usepackage[utf8]{inputenc}
\usepackage{subfigure}
\usepackage{comment}
\usepackage{hyperref}
\usepackage{enumerate}
\usepackage{xspace}
\usepackage{fontawesome}

\usepackage[usenames,dvipsnames]{xcolor}
\hypersetup{
    colorlinks = true,
    citecolor = {MidnightBlue},
    linkcolor = {BrickRed},
    urlcolor = {BrickRed}
}

\newcommand{\ba}{\begin{eqnarray}}
\newcommand{\ea}{\end{eqnarray}}
\newcommand{\planck}{{\sl Planck}\xspace}

\newcommand{\nv}{\hat{\bf n}}
\newcommand{\hpx}{{\tt HEALPix}\xspace}

\newcommand{\flamingo}{\textsc{Flamingo}\xspace}

\begin{document}

\title{Joint measurements of thermal Sunyaev-Zel'dovich and cosmic infrared background cross-correlations with cosmic shear}

\author{Amy Wayland}
\email{amy.wayland@physics.ox.ac.uk}
\affiliation{Department of Physics, University of Oxford, Denys Wilkinson Building, Keble Road, Oxford OX1 3RH, United Kingdom}

\author{Adrien La Posta}
\affiliation{Department of Physics, University of Oxford, Denys Wilkinson Building, Keble Road, Oxford OX1 3RH, United Kingdom}

\author{David Alonso}
\affiliation{Department of Physics, University of Oxford, Denys Wilkinson Building, Keble Road, Oxford OX1 3RH, United Kingdom}

\author{Baptiste Jego}
\affiliation{Observatoire Astronomique de Strasbourg, UMR 7550, CNRS, Université de Strasbourg, F-67000 Strasbourg, France}

\author{Matthieu Béthermin}
\affiliation{Observatoire Astronomique de Strasbourg, UMR 7550, CNRS, Université de Strasbourg, F-67000 Strasbourg, France}

\begin{abstract}
  The cross-correlation between weak lensing (WL) and the thermal Sunyaev-Zel'dovich (tSZ) effect probes the connection between the matter distribution and the thermal state of baryons at low redshift ($z\lesssim1$). A key systematic in such measurements is contamination by extragalactic foregrounds, particularly the cosmic infrared background (CIB). Here we describe the CIB spectral energy distribution (SED) with a small number of physically motivated spectral parameters, calibrated against external measurements of the radiation field intensity in galaxies. Their uncertainty is then propagated into the recovered WL$\times$tSZ signal in component separation at the level of the multi-frequency cross-power spectra rather than the map level. Applying this method to cosmic shear from the Dark Energy Survey and frequency maps from \planck, we show that the resulting measurements of the WL$\times$tSZ power spectrum are insensitive to variations in the effective CIB spectral index, radio source contamination, and the scale and redshift dependence of the SED. The measurements are robust to realistic foreground uncertainties, supporting their use as probes of baryonic feedback and cosmology, and providing a framework for future high-precision surveys. We compare our measurements with the \flamingo suite of hydrodynamical simulations and find that they are compatible with the \flamingo predictions for a low amplitude of matter fluctuations, parametrised by $S_8$, strongly disfavouring the larger $S_8$ preferred by \planck. This result is independent of the impact of baryonic feedback as modelled in \flamingo.
\end{abstract}

\maketitle

\section{Introduction} \label{sec:introduction}
 The thermal Sunyaev-Zel'dovich (tSZ) effect, arising from the inverse-Compton scattering of cosmic microwave background (CMB) photons off free electrons in the hot intergalactic medium, provides a direct probe of the pressure of the baryonic gas that traces the large-scale matter distribution \cite{Sunyaev1972observations,astro-ph/9808050}. Because this gas is heated and redistributed by radiative cooling, star formation, and feedback from active galactic nuclei, tSZ observations are sensitive to astrophysical processes that remain difficult to model from first principles. Uncertainty in the resulting baryonic feedback is currently one of the leading systematics limiting the precision of weak lensing (WL) cosmology on small angular scales, and has been proposed as a possible contributor to the mild tension between low-redshift and CMB inferences of the amplitude of matter clustering \cite{Schneider2021,Amon2022,Chen2022,Arico2023,McCarthy2023}.

 The tSZ signal is quantified by the Compton-$y$ parameter, a measure of the free-electron pressure integrated along the line of sight. Cross-correlating the Compton-$y$ field with weak lensing data offers a way to probe the matter-pressure power spectrum in a manner that is complementary to the study of the tSZ auto-power spectrum alone. While the latter is dominated by the most massive haloes at very low redshifts \cite{astro-ph/0205468}, the cross-correlation captures a broader range of redshifts and halo masses, and probes the small-scale correlation in the distribution of matter and gas fluctuations. A number of previous studies have used such WL--tSZ cross-correlations to constrain both cosmology and the impact of baryonic physics on gas pressure \cite{VanWaerbeke2013, Osato2019, Gatti2022, Pandey2022, Troster2022, LaPosta2024insights}.

 A major limitation of these measurements is contamination from other sky signals that correlate with both the CMB frequency maps and the large-scale structure traced by weak lensing, particularly the cosmic infrared background (CIB) sourced by dusty, star-forming galaxies \cite{Partridge1967are, Knox2001probing}. Since the CIB and tSZ effect have distinct but partially overlapping frequency dependence, and because the CIB spectral energy distribution (SED) itself depends on redshift and galaxy population, an imperfect modelling of the CIB can bias the recovered Compton-$y$ cross-correlation. The standard approach to this problem constructs component-separated Compton-$y$ maps, typically using internal linear combination (ILC) methods, from which the cross-correlation with large-scale structure is then measured \cite{Aghanim2015planck, McCarthy2024component, Coulton2024act, Chandran2023improved, Maniyar2026spt}. Deprojecting the CIB from these maps reduces its impact at the cost of an increase in the map-level noise, and requires precise external knowledge of the CIB frequency dependence, which makes residual contamination difficult to quantify \cite{2502.08850,2606.28099}.

 In this work, we take an alternative approach in which the tSZ and CIB contributions are modelled directly at the level of the observed WL cross-power spectra rather than deprojected at the map level. This has two main advantages. First, the uncertainty in the CIB SED enters the error budget of the recovered WL--tSZ cross-power spectra explicitly, through marginalisation over a small set of physical parameters. Second, the CIB cross-correlation is recovered alongside the tSZ signal rather than discarded, so the same measurement also yields the WL--CIB cross-power spectrum and adds some information on the CIB SED itself. Further components, such as radio point sources, can additionally be included by simply extending the mixing matrix of the multi-frequency model.
 
 We use the physically motivated CIB SED model of \cite{Bethermin2013redshift, Bethermin2014evolution, Bethermin2017impact}, calibrated against external measurements of the mean intensity of the dust radiation field in infrared galaxies, and propagate its uncertainty through a maximum-likelihood bandpower-based reconstruction of the WL--tSZ cross-power spectrum, following the strategy of \cite{LaPosta2026joint}. This allows us to marginalise over CIB parameters directly in the cross-spectrum measurement, thereby enabling us to test explicitly the dependence of the recovered WL--tSZ signal on the assumed CIB modelling. We apply this framework to the tomographic cosmic shear data from the first three years of the Dark Energy Survey (DES Y3) \cite{DES2020dark, Abbott2022dark, Garcia-Garcia2024cosmic} cross-correlated with \planck PR4 frequency maps \cite{Akrami2020planck}, and interpret the resulting measurements in terms of constraints on cosmology and baryonic feedback.

 The structure of this paper is as follows. In Section~\ref{sec:methods}, we describe the theoretical background, the CIB SED model, the power-spectrum-based component separation method, and the data used in this analysis. In Section~\ref{sec:results}, we present the resulting WL--tSZ cross-power spectra, the constraints they place on the CIB SED, and a suite of robustness tests against alternative modelling choices. We also compare our measurement of the WL--tSZ cross-spectrum against predictions from the \flamingo hydrodynamical simulations, spanning variations in cosmological and baryonic feedback models. We summarise our conclusions and their implications for future surveys in Section~\ref{sec:conclusions}.

\section{Methods} \label{sec:methods}
 \subsection{Theory} \label{ssec:methods.theory}
   \subsubsection{Angular power spectra}
    Projected cosmological observables can generally be expressed as line-of-sight integrals of three-dimensional fields. A projected field $u(\nv)$, observed along the unit vector $\nv$ on the celestial sphere, is related to its corresponding three-dimensional quantity $\mathcal{U}(\mathbf{x},z)$ through
    \begin{equation} \label{eq:projected_field}
        u(\nv) = \int\mathrm{d}\chi\, W_u(\chi)\, \mathcal{U}(\chi\nv,z(\chi)),
    \end{equation}
    where $\chi$ denotes the comoving radial distance, $z(\chi)$ is the corresponding redshift, and $W_u(\chi)$ is the radial kernel associated with $u$. 
    
    Under the Limber approximation \cite{Limber1954, LoVerde2008}, which is valid for projection kernels that are sufficiently broad in redshift, the angular power spectrum of two projected fields, $u$ and $v$, is related to the three-dimensional power spectrum, $P_{\mathcal{UV}}(k,z)$, through
    \begin{equation} \label{eq:Cl}
        C_{\ell}^{uv} = \int\frac{\mathrm{d}\chi}{\chi^2}\, W_u(\chi)\, W_v(\chi)\, P_{\mathcal{UV}}\left(k=\frac{\ell+1/2}{\chi},z(\chi)\right).
    \end{equation}
    The weak lensing and thermal Sunyaev-Zel'dovich observables considered in this work satisfy the conditions under which the Limber approximation is expected to be valid, so we may use Equation \eqref{eq:Cl} to compute the relevant power spectra.

   \subsubsection{Cosmic shear}
    Weak gravitational lensing arises from the deflection of light by the intervening matter distribution as photons propagate from distant galaxies to the observer \cite{Bartelmann2001weak, Mandelbaum2018weak}. Lensing by the large-scale structure of the Universe imprints a small but spatially coherent distortion on the observed shapes of background galaxies, known as cosmic shear, which is detected statistically by correlating the shapes of many galaxies rather than through the lensing of any individual structure. Because it is sourced by the total matter distribution, cosmic shear probes the projected matter field directly, without requiring a model for the relation between the galaxy and matter distributions.

    The $E$-mode component of the shear field is obtained by projecting the three-dimensional matter overdensity, $\delta_{\rm m}(\mathbf{x},z)$, according to Equation \eqref{eq:projected_field}. Its radial kernel is given by
    \begin{equation}\label{eq:lensing_kernel}
      W_{\gamma}(\chi) = \frac{3}{2} G_\ell H_0^2 \Omega_{\rm m} \frac{\chi}{a(\chi)}\int_{z(\chi)}^\infty \mathrm{d}z'\, p(z')\, \frac{\chi(z')-\chi}{\chi(z')},
    \end{equation}
    where $a=1/(1+z)$ is the scale factor, $H_0$ is the expansion rate today, and $p(z)$ is the redshift distribution of source galaxies. The factor
    \begin{equation}
        G_\ell \equiv \sqrt{\frac{(\ell+2)!}{(\ell-2)!}}\,\frac{1}{(\ell+1/2)^2}
    \end{equation}
    is a purely geometric correction that relates the three-dimensional Laplacian of the gravitational potential to the angular Hessian of the lensing potential \cite{Kilbinger2017precision}. Over the multipole range considered in this work, $G_\ell$ may be safely taken to be unity.

   \subsubsection{\texorpdfstring{Compton-$y$ tSZ maps}{Compton-y tSZ maps}}
    The thermal Sunyaev-Zel'dovich (tSZ) effect is a secondary anisotropy in the CMB resulting from inverse-Compton scattering of CMB photons by thermal free electrons in the intergalactic medium \cite{Sunyaev1972observations}. This process transfers energy from the electrons to the CMB photons, producing a characteristic frequency-dependent distortion of the CMB spectrum that can be isolated using multifrequency observations.
    
    The tSZ signal is described by the Compton-$y$ parameter,
    \begin{equation}\label{eq:comptony}
        y(\nv) = \int \frac{\mathrm{d}\chi}{1+z} \, \frac{\sigma_T}{m_{\rm e} c^2} \, P_{\rm e}^{\rm th}(\chi\nv, z(\chi)),
    \end{equation}
    where $\sigma_T$ is the Thomson scattering cross-section and $P_{\rm e}^{\rm th}$ is the thermal electron pressure. The corresponding radial kernel is therefore
    \begin{equation}\label{eq:sz_kernel}
        W_{y}(\chi) \equiv \frac{\sigma_T}{m_{\rm e} c^2 (1+z)}.
    \end{equation}

   \subsubsection{Cosmic infrared background}
    The cosmic infrared background (CIB) is mainly generated by thermal emission from dust in star-forming galaxies. The ultraviolet light emitted from young, massive stars is partially absorbed by interstellar dust and re-emitted at infrared wavelengths, making the CIB an integrated tracer of the cosmic star formation history.
    
    Following \cite{Jego2022star}, the specific CIB intensity at observing frequency $\nu$ can be written as
    \begin{equation}\label{eq:I_cib}
      I^{\rm CIB}_\nu(\nv) = \int\mathrm{d}\chi\, \frac{\chi^2 S_\nu^{\rm eff}(z)}{K}\, \rho_{\rm SFR}(\chi\nv,z),
    \end{equation}
    where $\rho_{\rm SFR}$ is the three-dimensional star formation rate (SFR) density field, which plays the role of $\mathcal{U}$ in Equation~\eqref{eq:projected_field} and whose average over the sky at fixed redshift is the cosmic SFR density. The quantity $S_\nu^{\rm eff}(z)$ is the effective SED of the infrared galaxy population: the flux density at observed frequency $\nu$ of a galaxy at redshift $z$ per unit infrared luminosity, averaged over the population. Finally, $K$ is the Kennicutt constant relating infrared luminosity to star formation rate, ${\rm SFR} = K \, L_{\rm IR}$ \cite{Kennicutt1998, Kennicutt2012}, for which we adopt $K = 10^{-10}\, M_\odot\, {\rm yr}^{-1}\, L_\odot^{-1}$, appropriate for the Chabrier initial mass function \cite{Chabrier2003} assumed in \cite{Bethermin2017impact}. The CIB can therefore be interpreted as a projected tracer of the SFR density, with radial kernel
    \begin{equation}
        W_{\rm CIB,\nu}(\chi) = \frac{\chi^2 S_\nu^{\rm eff}(z)}{K}.
    \end{equation}

 \subsection{Infrared spectra} \label{ssec:methods.spectra}
  We model the effective CIB SED, $S_\nu^{\rm eff}(z)$, following the physically motivated framework of \cite{Bethermin2013redshift, Bethermin2014evolution, Bethermin2017impact}, in which the infrared galaxy population is separated into main-sequence (MS) and starburst (SB) components. Most star-forming galaxies lie on a tight correlation between star formation rate and stellar mass, the so-called main sequence, along which star formation proceeds in a quasi-steady mode \cite{Speagle2014}. A small minority, typically associated with mergers, form stars more efficiently and lie well above this relation \cite{Sargent2012}. At low redshift, star formation in these starbursts is more compact, exposing their dust to a more intense radiation field than in MS galaxies \cite{Magdis2012evolving}. At high redshift, however, the mean radiation field intensities of the two populations appear to be similar, for reasons that are not yet fully understood. One possibility is that starbursts are more metal-rich and hence dustier, so that the number of UV photons per unit dust mass is comparable in both populations \cite{Bethermin2014evolution}. Another is that starburst dust is warmer but optically thick below $\sim100\,\mu{\rm m}$, which makes the SED appear colder than the true dust temperature \cite{Magdis2020GN20}. Given these differences, the two populations are modelled with separate SED templates.

  \begin{figure}
      \centering
      \includegraphics[width=\linewidth]{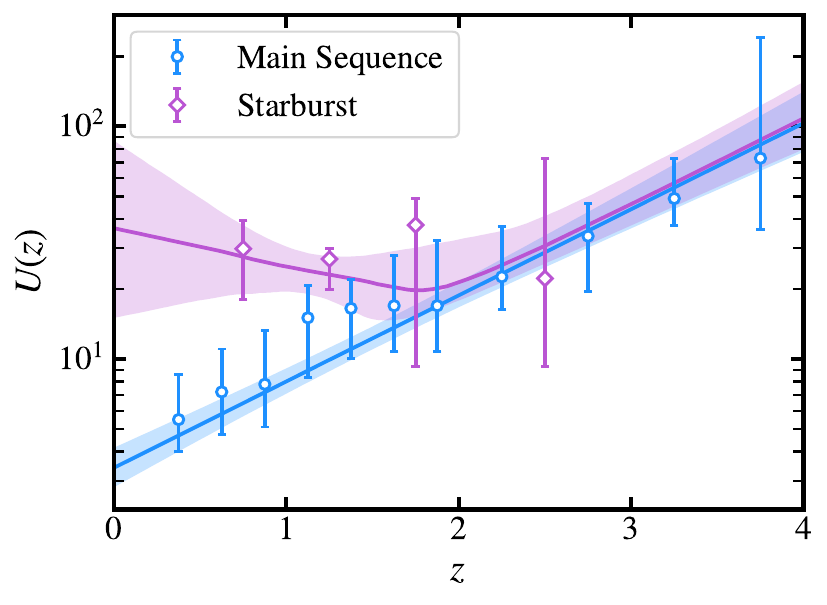}
      \caption{External calibration data on the mean radiation field intensity, $U(z)$, for the MS (blue) and SB (purple) populations \cite{Bethermin2017impact}, together with the $95\%$ credible bands on $U_{\rm MS}(z)$ and $U_{\rm SB}(z)$ obtained from our fiducial $U(z)+C_\ell^{\gamma \nu}$ joint analysis.}
      \label{fig:uz}
  \end{figure}
  
  The SED of each sample is described by the model presented in \cite{2007ApJ...657..810D,2001ApJ...554..778L}, and the resulting SED templates are calibrated by \cite{Magdis2012evolving,Bethermin2014evolution, Bethermin2017impact}. The main parameters governing the SEDs are the intensity of the radiation field $U$ in each sample\footnote{Note that, following \cite{2007ApJ...657..810D,Magdis2012evolving}, $U$ is a dimensionless factor scaling the local Galactic radiative energy density.}, its evolution with redshift, and the relative fraction of SB galaxies. The result is a set of templates for the MS and SB luminosities, each normalised to a total luminosity of $L_\odot$.

  For each population, $U$ evolves with redshift as a log-linear relation,
  \begin{equation}
      \log_{10}U_X(z) = \log_{10}U_{X,0} + \alpha_X\,z,
  \end{equation}
  with $X\in\{{\rm MS},{\rm SB}\}$. The luminosity templates are then determined by four free physical parameters: $\alpha_{\rm MS}$ and $\log_{10} U_{\rm MS,0}$ for the MS population, and $\alpha_{\rm SB}$ and $\log_{10} U_{\rm SB,0}$ for the SB population. Following \cite{Bethermin2017impact}, we further impose two conditions on the redshift evolution: $U_{\rm MS}(z)$ is held fixed at its value at $z=4$ for $z>4$, and $U_{\rm SB}(z)$ is required to be at least as large as $U_{\rm MS}(z)$ at every redshift, $U_{\rm SB}(z) \geq U_{\rm MS}(z)$. We note that these conditions are model-specific choices inherited from \cite{Bethermin2017impact}, rather than requirements with a strong physical motivation. In particular, the second condition was introduced to prevent starbursts from having colder dust temperatures than MS galaxies, which, as discussed above, is not clearly excluded at high redshift. We retain these conditions here for consistency with the calibrated templates.

  The relative contributions of the two populations to the total infrared luminosity density are governed by the SB fraction, $f_{\rm SB}(z)$. We adopt the fiducial redshift evolution of \cite{Bethermin2017impact}, in which $f_{\rm SB}(z)$ increases linearly from $1.5\%$ at $z=0$ to $3.0\%$ at $z=1$ and remains constant at $3.0\%$ for $z>1$. We further rescale this fraction by a fixed calibration factor to account for the higher characteristic infrared luminosity of SB galaxies relative to MS galaxies at a given number density.
  
  The four free parameters ($\{\log_{10}U_{X,0},\alpha_X\}$ with $X\in\{{\rm MS},{\rm SB}\}$) are calibrated against measurements of $U(z)$, the mean intensity of the starlight radiation field heating the dust, obtained by fitting the dust models of \cite{2007ApJ...657..810D} to the observed infrared SEDs of star-forming galaxies. We use the compilation assembled by \cite{Bethermin2017impact}, which comprises $U$ estimates for MS galaxies from the \emph{Herschel} PACS and SPIRE stacking analyses of \cite{Magdis2012evolving} out to $z \approx 2$ and \cite{Bethermin2014evolution} out to $z \approx 4$, together with low-redshift determinations from \cite{Cunha2010exploring, Ciesla2014dust}, and SB estimates from \cite{Magdis2012evolving}.
  
  To constrain the four free parameters we use a split Gaussian likelihood, accounting for the asymmetric uncertainties on the tabulated $U(z)$ measurements. We sample these parameters assuming the broad Gaussian priors listed in Table~\ref{tab:priors}, together with the conditions above, which act as additional hard priors on $U_{\rm SB}(z)$. Note that these Gaussian priors are largely uninformative, as they are much broader than the constraints derived from the data, and are used only to suppress unphysically large values caused by the existing parameter degeneracies. Figure~\ref{fig:uz} shows the calibration data used together with the 95$\%$ credible intervals on our model $U(z)$ obtained from our fiducial analysis, combining the calibration data as well as the multi-frequency cross-correlations with cosmic shear (see Section~\ref{ssec:methods.mlbr}).

  \begin{table}
    \centering
    \begin{tabular}{llc}
      \hline
      \hline
      Parameter & Description & Prior \\
      \hline
      $\alpha_{\rm MS}$ & MS radiation field slope & $\mathcal{N}(0.31, 0.33)$ \\
      $\log_{10} U_{\rm MS,0}$ & MS amplitude at $z=0$ & $\mathcal{N}(0.69, 0.75)$ \\
      $\alpha_{\rm SB}$ & SB radiation field slope & $\mathcal{N}(-0.25, 1.8)$ \\
      $\log_{10} U_{\rm SB,0}$ & SB amplitude at $z=0$ & $\mathcal{N}(1.63, 2.1)$ \\
      \hline
      \hline
    \end{tabular}
    \caption{Broad, largely uninformative priors adopted for the four physical CIB SED parameters, where $\mathcal{N}(\mu,\sigma)$ denotes a Gaussian of mean $\mu$ and standard deviation $\sigma$.}
    \label{tab:priors}
  \end{table}

  Throughout, we refer to the \emph{fiducial} CIB model as the SED templates described above, with the four radiation-field parameters fixed to their best-fit values from the joint $U(z)+C_\ell^{\gamma\nu}$ analysis, and all remaining SED shape parameters fixed to the values of \cite{Bethermin2014evolution}.
  
 \subsection{Power-spectrum-based component separation} \label{ssec:methods.mlbr}
  A standard approach to cross-correlation analyses involving secondary anisotropies of the CMB relies on the construction of component-separated maps, from which the signal of interest is subsequently correlated with large-scale structure tracers \cite{McCarthy2024component, Coulton2024act, Aghanim2015planck, Chandran2023improved, Maniyar2026spt}. Although contaminants with known SEDs can, in principle, be deprojected, this procedure generally increases the variance of the reconstructed maps and, consequently, of the inferred power spectra. Moreover, such deprojection techniques require precise prior knowledge of the frequency dependence of the contaminants, which makes it difficult to interpret any residual contamination if the choice of contaminant SED is not accurate.
  
  In this work, we instead model the contributions from the different astrophysical components directly at the level of the observed power spectra. Assuming that the tSZ effect and the CIB are the only sky components that correlate appreciably with the cosmic shear field, we model the observed map at frequency $\nu$ as
  \begin{equation}\label{eq:map_model}
    m_\nu(\nv)= S_\nu^{\rm tSZ}\,y(\nv) + \bar{S}_\nu^{\rm CIB}(z_\gamma) \, c_{z_\gamma}(\nv) + \tilde{N}_\nu(\nv),
  \end{equation}
  where $y(\nv)$ is the dimensionless Compton-$y$ field, and $c_{z_\gamma}(\nv)$ denotes the contribution to the CIB from the redshift range that overlaps with the cosmic shear kernel. All remaining sky components that do not correlate with the shear field, including instrumental noise, the primary CMB, Galactic foregrounds, and CIB emission outside the relevant redshift range, are combined into the effective noise term, $\tilde{N}_\nu$. $S_\nu^{\rm tSZ}$ is the spectrum of the tSZ effect, given by
  \begin{equation}
      S_\nu^{\rm tSZ} = T_{\rm CMB} \left(x \coth\frac{x}{2} - 4\right), \qquad x \equiv \frac{h\nu}{k_{\rm B} T_{\rm CMB}},
  \end{equation}
  and $\bar{S}_\nu^{\rm CIB}(z_\gamma)$ is the effective spectrum of the $c_{z_\gamma}$ component, which we describe below. Throughout, we work in thermodynamic CMB temperature units, such that $m_\nu$ is measured in $\mu$K, and we integrate both spectral response functions over the corresponding \planck bandpasses. Under this model, the cross-power spectrum of the observed frequency map and the cosmic shear field can be expressed as
  \begin{equation}
      C_\ell^{\gamma \nu} = S_\nu^{\rm tSZ}\,C_\ell^{\gamma y} + \bar{S}_\nu^{\rm CIB}(z_\gamma)\,C_\ell^{\gamma \rm CIB},
  \end{equation}
  where $C_\ell^{\gamma y}$ denotes the cross-power spectrum between the cosmic shear field and the Compton-$y$ map, and $C_\ell^{\gamma \rm CIB}$ is the corresponding cross-power spectrum with $c_{z_\gamma}$.

  We now turn to the interpretation of $c_{z_\gamma}(\nv)$. The CIB contribution to $m_\nu$ is given by Equation~\eqref{eq:I_cib}. Since the CIB radial kernel is proportional to the redshift-dependent SED $S_\nu^{\rm eff}(z)$, the frequency dependence of the CIB enters the Limber integral (Equation~\eqref{eq:Cl}). We will however approximate the CIB contribution to the cross-correlation with the cosmic shear bin denoted by $z_\gamma$ as
  \begin{equation}
    I_\nu^{\rm CIB}(\nv)\simeq \bar{S}^{\rm CIB}_\nu(z_\gamma)\int \mathrm{d}\chi\,\frac{\chi^2}{K}\,\rho_{\rm SFR}(\chi\nv),
  \end{equation}
  where $\bar{S}^{\rm CIB}_\nu(z_\gamma)$ is the CIB SED averaged over the lensing kernel:
  \begin{equation}\label{eq:s_eff_gamma}
    \bar{S}_\nu^{\rm CIB}(z_\gamma) \equiv \frac{\int\mathrm{d}\chi\,\chi^2\, W_\gamma(\chi)\, S_\nu^{\rm eff}(z)}{\int\mathrm{d}\chi\,\chi^2W_\gamma(\chi)},
  \end{equation}
  effectively neglecting the redshift evolution of $S^{\rm eff}_\nu$ over the support of $W_\gamma$. We will assess the validity of this approximation in Section \ref{ssec:results.robustness_cl}. Therefore, in this formalism, the CIB component entering Equation~\eqref{eq:map_model} is $c_{z_\gamma}\equiv \Sigma_{\rm SFR}/K$, where $\Sigma_{\rm SFR}$ is the projected SFR density:
  \begin{equation}\label{eq:sfr_proj}
    \Sigma_{\rm SFR}(\nv)\equiv\int \mathrm{d}\chi\,\chi^2\,\rho_{\rm SFR}(\chi\nv).
  \end{equation}
  Since this quantity has a clearer physical interpretation than $c_{z_\gamma}$, we will present our measurements of the WL--CIB cross-spectrum in terms of $\Sigma_{\rm SFR}$ rather than $c_{z_\gamma}$.
  
  To recover measurements of $C_\ell^{\gamma y}$ and $C_\ell^{\gamma{\rm CIB}}$ we proceed by constructing a data vector ${\bf d}$ composed of all cross-power spectra between the cosmic shear field and the different frequency maps, and building a linear model relating it to the desired component-wise cross-spectra:
  \begin{equation}
      \mathbf{d} = \mathsf{S}\cdot\mathbf{C} + \hat{\bf N}.
  \end{equation}
  Here $\mathbf{C} \equiv (C_\ell^{\gamma y}, C_\ell^{\gamma \rm CIB})^{\rm T}$ is the vector of target cross-power spectra and $\mathsf{S} \equiv (\mathbf{S}^{\rm tSZ}, \mathbf{S}^{\rm CIB})$ is the mixing matrix encoding the SEDs of the contributing components across frequency channels. The vector $\hat{\bf N}$ represents the effective noise contribution, incorporating both instrumental noise and all sky components uncorrelated with the cosmic shear field.

  Assuming that the noise term $\hat{\bf N}$ follows a multivariate Gaussian distribution with zero mean and covariance matrix $\mathsf{\Sigma}$, the likelihood of the data vector can be written in the form
  \begin{equation}
     -2\log p({\bf d}|{\bf C}) = (\mathbf{C} - \overline{\mathbf{C}})^{\rm T} \, \mathsf{\Sigma}_{c}^{-1} \, (\mathbf{C} - \overline{\mathbf{C}}) + {\rm const.},
  \end{equation}
  where $\mathsf{\Sigma}_c$ denotes the covariance matrix of the reconstructed spectra. The maximum-likelihood estimator for the target spectra is then given by the standard least squares solution,
  \begin{equation}\label{eq:leastsquares}
    \overline{\mathbf{C}} = \mathsf{\Sigma}_c \, \mathsf{S}^{\rm T} \, \mathsf{\Sigma}^{-1} \, \mathbf{d}, \quad\quad \mathsf{\Sigma}_c \equiv (\mathsf{S}^{\rm T} \, \mathsf{\Sigma}^{-1} \, \mathsf{S})^{-1}.
  \end{equation}
  We refer to this as the maximum-likelihood bandpower reconstruction (MLBR) method, following a strategy similar to \cite{LaPosta2026joint}.

  To account for the uncertainty in the SED model parameters characterising $U(z)$ (see Section \ref{ssec:methods.spectra}) we combine the likelihoods of the $C_\ell$ and $U(z)$ data. To exploit the quadratic nature of the $C_\ell$ likelihood as a function of ${\bf C}$ (which makes it analytically tractable), we sample only the four $U(z)$-based spectral parameters $\vec{\theta}_U$ using a Metropolis-Hastings Markov chain Monte Carlo (MCMC) approach. For each sample of $\vec{\theta}_U$, we calculate the effective CIB spectra and the associated $\bar{\bf C}$ and $\mathsf{\Sigma}_c$ using Equation~\eqref{eq:leastsquares}. From this, we can reconstruct any relevant statistic of the component-wise power spectra ${\bf C}$ (e.g. their maximum-likelihood value, their mean, and their covariance).
  
 \subsection{Data} \label{ssec:methods.data}
  In this work, we make use of the publicly available weak lensing data from the first three years of observations of the Dark Energy Survey (DES Y3) \cite{DES2020dark, Abbott2022dark}. Specifically, we adopt the same analysis choices presented in \cite{Garcia-Garcia2024cosmic}, which mimic the official DES analysis \citep{2203.07128} (in terms of e.g. tomographic bin definitions, additive and multiplicative bias corrections, etc.). The data are divided into four tomographic redshift bins with approximately equal source number densities, covering the range of source redshifts $z_s\lesssim2$. When interpreting the measured power spectra we use the calibrated redshift distributions available with the data release. Further details may be found in \cite{Garcia-Garcia2024cosmic,2203.07128,DES2020dark}.

  As tracers of the microwave and far-infrared sky, we use temperature maps from the \planck PR4 (NPIPE) data release \cite{Akrami2020planck} at frequencies of 100, 143, 217, 353, 545, and 857 GHz. The 100--545 GHz channels are calibrated using the orbital dipole, while the 857 GHz channel relies on planet-based calibration. In analyses where calibration uncertainties are propagated, we impose Gaussian priors on the calibration coefficients for each frequency channel, with standard deviations given by the quoted uncertainties in \cite{Akrami2020planck}.

  The photometric calibration of the \planck frequency maps assumes a reference source spectrum for which $\nu I_\nu$ is constant across each bandpass. Because the CIB spectrum differs substantially from this reference spectrum, we apply colour corrections to the theoretical predictions in each frequency channel. These corrections account for the difference between the assumed calibration spectrum and the actual CIB SED, integrated over the corresponding bandpass. The correction factors adopted are 1.076, 1.017, 1.119, 1.097, 1.068, and 0.995 for the 100, 143, 217, 353, 545, and 857~GHz maps, respectively \cite{Ade2013planck}. They are computed using the \planck bandpasses \cite{Ade2013planck} and the fiducial CIB SED templates of \cite{Bethermin2013redshift, Bethermin2014evolution, Bethermin2017impact}. Since the corrections depend on the assumed spectral shape, they should in principle be recomputed as the SED parameters are varied. Here, however, since variations between different CIB SED models are small, we fix them to their fiducial values throughout the analysis.

  For comparison and validation purposes, we additionally make use of the \planck 2015 full-sky Compton-$y$ map \cite{Aghanim2015planck}, constructed using the modified internal linear combination algorithm \citep[MILCA;][]{1007.1149}. This method extracts components with known spectral dependence through optimally-weighted linear combinations of the multi-frequency maps.

  \subsubsection{Catalogue-based power spectra}\label{sssec:methods.data.cls}
   We use the DES Y3 weak lensing catalogue directly by implementing catalogue-based fields within the \textsc{Cosmotheka}\footnote{\url{https://github.com/Cosmotheka/Cosmotheka}} framework \cite{Garcia-Garcia2021growth}. Unlike past analyses (e.g. \cite{Garcia-Garcia2024cosmic}), instead of creating pixelised cosmic shear maps, we compute all cosmic shear power spectra using the catalogue-based method described in \cite{2407.21013}, evaluating the harmonic coefficients of the masked shear field directly from the positions and ellipticities of the galaxies. The resulting power spectra are therefore free of the pixel window function and of the potential biases introduced by sub-pixel averaging of the shear field, and are not limited to the multipole range accessible at a given map resolution. 

   Throughout this work, we analyse all unique auto- and cross-correlations between tomographic bins, as well as the cross-correlations with the \planck temperature maps. While the angular power spectra themselves are estimated from catalogue-based fields, the covariance matrix is computed using map-based fields constructed with \hpx \cite{Gorski2004healpix} at $N_{\rm side} = 2048$, corresponding to a maximum reliable multipole of $\ell_{\rm pix} = 2 N_{\rm side} = 4096$, comfortably above the range used in this analysis\footnote{The analysis presented here was well under way when a method for catalogue-based power spectrum covariances was presented in \cite{2607.14843}, and we have thus not adopted it here. Nevertheless, the pixel-based method used here was shown in \cite{Nicola2020cosmic} to be sufficiently accurate on the angular scales used here.}. This is justified because both approaches yield identical power spectra on scales unaffected by pixelisation, where they correspond to the same estimator and therefore have the same statistical uncertainties, so the covariance derived from map-based fields can be consistently applied to the catalogue-based power spectra. Since the catalogue-based estimator is not itself subject to this pixelisation limit, the weak lensing side of each correlation places no restriction on the multipole range used here.

   \begin{figure*}
       \centering
       \includegraphics[width=\textwidth]{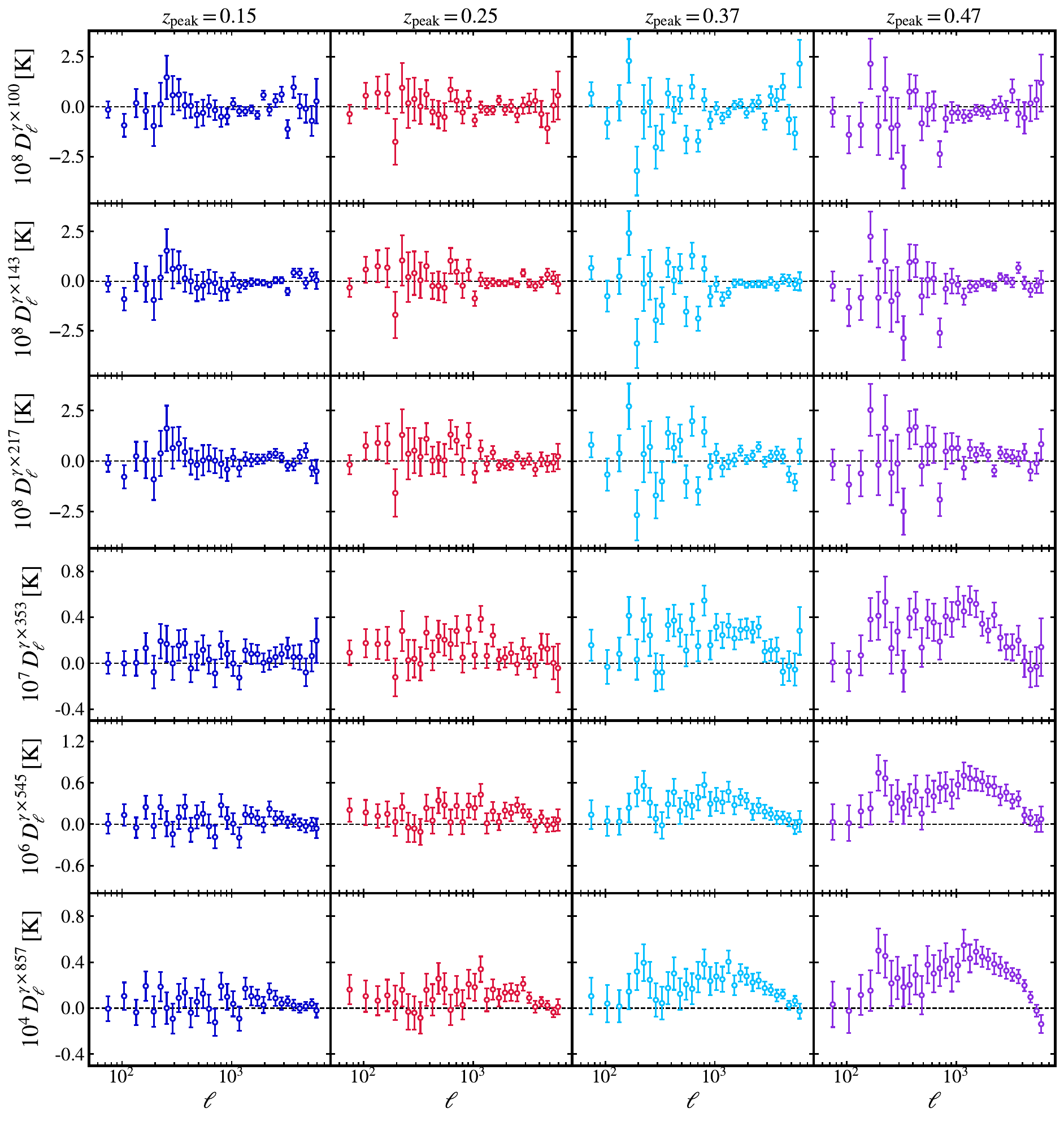}
       \caption{The measured cross-power spectra between cosmic shear and each \planck NPIPE frequency map at 100, 143, 217, 353, 545, and 857~GHz, for each of the four DES~Y3 weak lensing bins. The angular power spectra are expressed in terms of $D_\ell \equiv [\ell(\ell+1)/(2\pi)] C_\ell$.}
       \label{fig:raw_bandpowers}
   \end{figure*}

   On large angular scales, we adopt the scale cut $\ell_{\rm min} = 30$ to exclude the first bandpower, where analytical covariance estimates are typically unreliable. On small angular scales, we restrict all cross-correlations to $\ell \leq \ell_{\rm max} = 2000$. Little information can be recovered on smaller scales, given the \planck beam ($\theta_{\rm FWHM}\approx 5'$) and the relatively large shape noise. The measured power spectra are binned into bandpowers using linearly-spaced bins with $\Delta\ell=30$ over the range $0\leq\ell\leq240$, and logarithmically-spaced bins up to $\ell_{\rm max}$ with $\Delta\log_{10}\ell=0.055$.

   When comparing theoretical predictions with the measured WL--tSZ power spectra, for instance against the \flamingo simulations in Section~\ref{ssec:results.Flamingo}, we account for smoothing effects of the \planck instrumental beam and \hpx pixelisation on the \planck side of each correlation (since the WL power spectra are estimated using catalogue-based fields, no pixel window correction is needed on the WL side). The beam is modelled as a Gaussian with the frequency-dependent FWHM of each \planck channel, while the pixel window function depends only on the map resolution, $N_{\rm side}$, and is common to all frequency channels, capturing the additional damping from discrete map pixels. For each WL--tSZ correlation, the spectrum is therefore multiplied by the product of the relevant channel's beam and the pixel window function, and then projected through the bandpower window functions, which account for residual mode coupling in the pseudo-$C_\ell$ estimator.

  \subsubsection{Covariance}\label{sssec:methods.data.covar}
   The power spectrum covariance matrix receives contributions from three components: the Gaussian covariance, corresponding to the fully disconnected part of the trispectrum; the super-sample covariance (SSC), arising from the coupling between observed modes and modes larger than the survey footprint through the connected trispectrum; and the connected non-Gaussian (cNG) covariance, which accounts for all remaining contributions to the connected trispectrum \cite{Barreira2018accurate}. In the presence of significant shape noise, the Gaussian term typically dominates the covariance \cite{Hamimeche2008likelihood, Sellentin2017insufficiency}. We compute this component using the improved Narrow Kernel Approximation of \cite{Nicola2020cosmic}.
   
   The SSC is generally the next-to-leading contribution. We evaluate it explicitly for the tomographic bin with the highest signal-to-noise, using the halo model framework of \cite{Krause2016cosmolike} as implemented analytically in \textsc{CCL} \cite{Chisari2019core}. We adopt an effective sky fraction of $f_{\rm sky} = 0.12$, appropriate for the DES--\planck overlap. Since the super-sample term scales with the amplitude of the signal, we normalise the fiducial pressure profile so that the halo model reproduces the measured amplitude of $C_\ell^{\gamma y}$. The SSC is largest on the largest scales, where it increases the diagonal errors by $12\%$, and decreases monotonically to below $1\%$ for $\ell \gtrsim 700$. Because the SSC does not average down with bandwidth in the same way as the Gaussian contribution, its relative importance is greatest for the widest, lowest-$\ell$ bandpowers. We therefore neglect the SSC, along with the cNG contribution, which is expected to be smaller still \cite{Barreira2018accurate}.

\section{Results} \label{sec:results}
 \subsection{\texorpdfstring{\boldmath $C_{\ell}^{\gamma y}$}{WL-tSZ} measurements and constraints on CIB SEDs} \label{ssec:results.fiducial}
  Figure~\ref{fig:raw_bandpowers} shows the cross-power spectra between cosmic shear and each \planck frequency map, $C_\ell^{\gamma\nu}$, for the six channels used in this analysis. The cross-spectra at 100, 143, and 217~GHz seem roughly consistent with noise in all four tomographic bins, since at these frequencies neither the tSZ nor the CIB contribution to the cross-correlation is large enough to be individually significant. We can, however, identify similar fluctuations in the power spectra corresponding to different frequency channels, denoting the presence of one or more correlated components across them. A positive signal starts to emerge at 353~GHz in the two highest redshift bins, and becomes more prominent at 545 and 857~GHz, where a clear positive bump around $\ell\sim 1000$ is detected, growing in amplitude with the tomographic bin and frequency channel as expected from the CIB spectrum as well as the redshift dependence of both the CIB and cosmic shear. As in \cite{LaPosta2026joint}, no single frequency map shows visual evidence of the tSZ contribution to the cross-correlation on its own; the Compton-$y$ signal only emerges once the full frequency information is combined via the MLBR reconstruction.

\begin{figure}[t!]
      \centering
      \includegraphics[width=\linewidth]{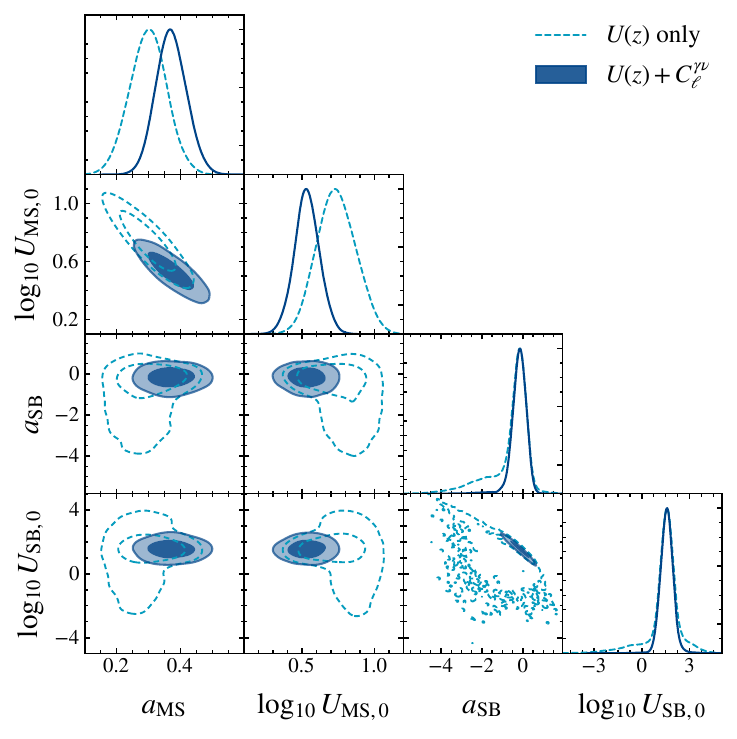}
      \caption{Marginalised posterior constraints on the four physical CIB SED parameters from external $U(z)$ data alone (light blue) and from the combination with the $C_\ell^{\gamma \nu}$ cross-power spectra (dark blue). Contours show the $68\%$ and $95\%$ credible regions.}
      \label{fig:corner_fiducial}
  \end{figure}

\begin{figure}[t!]
      \centering
      \includegraphics[width=\linewidth]{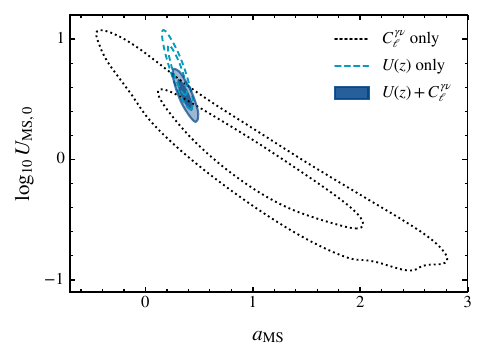}
      \caption{Marginalised posteriors in the $(\alpha_{\rm MS}, \log_{10} U_{\rm MS,0})$ plane from $C_\ell^{\gamma \nu}$ cross-power spectra alone (black), together with the $U(z)$-only (light blue) and joint (dark blue) contours.}
      \label{fig:corner_cl_only}
  \end{figure}

  \begin{table}
      \centering
      \begin{tabular}{lcccc}
          \hline
          \hline
          \noalign{\vskip 2.0pt}
          Dataset & $\alpha_{\rm MS}$ & $\log_{10} U_{\rm MS,0}$ & $\alpha_{\rm SB}$ & $\log_{10} U_{\rm SB,0}$ \\
          \noalign{\vskip 2.0pt}
          \hline
          \noalign{\vskip 3.0pt}
          $U(z)$ only 
          & $0.30^{+0.12}_{-0.11}$ 
          & $0.74^{+0.26}_{-0.25}$ 
          & $-0.6^{+1.3}_{-2.4}$ 
          & $1.4^{+1.9}_{-3.1}$ \\

          $U(z) + C_\ell^{\gamma \nu}$ 
          & $0.371^{+0.098}_{-0.091}$ 
          & $0.53^{+0.18}_{-0.17}$ 
          & $-0.19^{+0.71}_{-0.68}$ 
          & $1.56^{+0.82}_{-0.79}$ \\
          \hline
          \hline
      \end{tabular}
      \caption{Marginalised $68\%$ posterior constraints on the CIB SED from $U(z)$ data alone and the joint fit to $U(z)$ and $C_\ell^{\gamma\nu}$.}
      \label{tab:CIBconstraints}
  \end{table}
  Figure~\ref{fig:corner_fiducial} shows the marginalised posterior constraints on the four physical CIB SED parameters obtained using the external $U(z)$ data alone and in combination with the $C_\ell^{\gamma \nu}$ cross-power spectra. The corresponding marginalised constraints are summarised in Table~\ref{tab:CIBconstraints}. The posterior obtained from the cross-power spectra alone is far broader than either of these, and is shown separately in Figure~\ref{fig:corner_cl_only} for the MS parameters, over a much wider range of parameter values. Taken on their own, the $C_\ell^{\gamma \nu}$ data place only weak constraints on the CIB SED, as expected given that they were not designed as a dedicated CIB probe. For instance, the \planck bands are not able to cover the high-frequency turnover of the approximate modified black body dust spectrum, located at $\nu\sim1000\,{\rm GHz}$. The $C_\ell$ data on their own are essentially uninformative about the SB parameters, which are therefore not shown in this figure. Their posterior distribution for the MS parameters is consistent with the $U(z)$ constraints, but is elongated along a noticeably shallower degeneracy direction in the $(\alpha_{\rm MS}, \log_{10} U_{\rm MS,0})$ plane. The figure shows that the main effect of including the $C_\ell$ data is suppressing the higher values of $\log_{10}U_{\rm MS,0}$ leading to a $0.6\sigma$ positive shift in the value of $\alpha_{\rm MS}$, along the degeneracy direction of the $U(z)$ data. 
  
  The MS parameters are already well constrained by $U(z)$ alone. Including the cross-power spectra tightens these constraints modestly, reducing the marginalised uncertainties by $20$--$30\%$, in addition to the shift towards larger $\alpha_{\rm MS}$ and smaller $\log_{10}U_{\rm MS,0}$ mentioned above. As we show in Section~\ref{ssec:results.robustness_params}, this shift is preferentially driven by the small-scale modes of the two highest-redshift tomographic bins. 
  
  
  The SB parameters, on the other hand, behave quite differently. Their $U(z)$-only posterior is markedly non-Gaussian, with a long tail towards low $\alpha_{\rm SB}$ and extended tails in $\log_{10} U_{\rm SB,0}$ in both directions. This reflects the small number of SB measurements in the external calibration sample: with only two or three constraining data points, with relatively large uncertainties, the amplitude and slope of $U_{\rm SB}$ are strongly degenerate, resulting in a correspondingly non-Gaussian posterior. Adding the $C_\ell^{\gamma\nu}$ data removes most of these tails, and the marginalised $95\%$ credible intervals narrow by factors of $2.7$ and $3.1$ for $\alpha_{\rm SB}$ and $\log_{10} U_{\rm SB,0}$ respectively. These numbers should not be read as the cross-power spectra constraining the SB population directly, since SB galaxies contribute only a few per cent of the total infrared luminosity density and the $C_\ell^{\gamma\nu}$ data on their own are unable to constrain either SB parameter. This effect is instead entirely indirect: as the $\alpha_{\rm SB}$ and $\log_{10} U_{\rm SB,0}$ panels of Figure~\ref{fig:corner_fiducial} show, the tails of the $U(z)$-only posterior are not uniformly distributed in MS parameter space. Instead, they are supported almost exclusively by models with $\alpha_{\rm MS} \lesssim 0.3$ and $\log_{10} U_{\rm MS,0} \gtrsim 0.7$, precisely the region that the cross-power spectra disfavour. In shifting the MS parameters away from this region, the $C_\ell^{\gamma\nu}$ data remove the support for the extended SB tails as a by-product, without adding any direct information on the SB SED itself. We caution, however, that this behaviour is difficult to interpret physically. It may partly be an artefact of the assumed log-linear functional form for $U(z)$, which may not describe the true redshift evolution of either population. Moreover, the SB fraction $f_{\rm SB}(z)$ is held fixed to the fiducial evolution of \cite{Bethermin2017impact}, and since it is itself uncertain, this choice could also bias the inferred SB parameters.

\begin{figure}[t!]
    \centering
    \includegraphics[width=\linewidth]{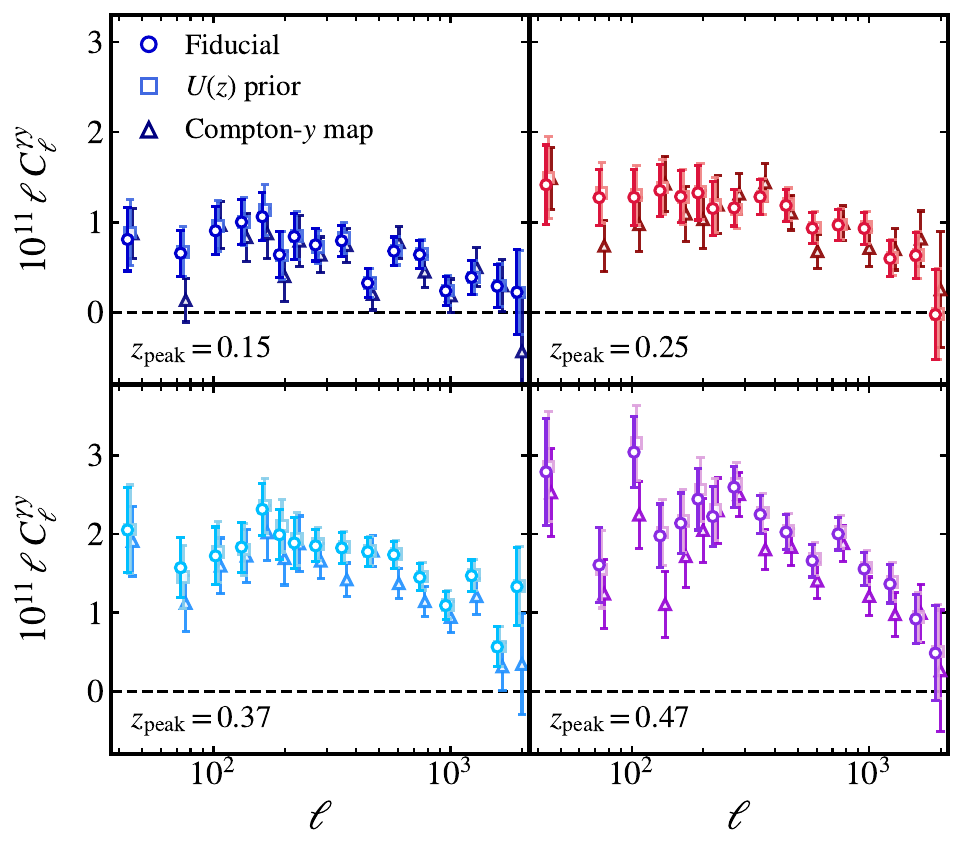}
    \caption{Maximum-likelihood bandpower reconstruction of the WL--tSZ cross-power spectrum, $C_\ell^{\gamma y}$, for each DES~Y3 weak lensing bin. Circles show the fiducial reconstruction, squares the reconstruction obtained with the CIB SED fixed to its $U(z)$-only best fit, and triangles the cross-correlation of cosmic shear with the public \planck MILCA Compton-$y$ map. For clarity of presentation, bandpowers above $\ell=240$ have been merged in pairs using inverse-variance weights, with the uncertainties propagated through the bandpower covariance, and the three sets of points are offset horizontally by $2.5\%$ in $\ell$. Neither affects the analysis, which uses the unmerged bandpowers throughout.}
    \label{fig:mlbr_tsz}
  \end{figure}

  Figure~\ref{fig:mlbr_tsz} shows the resulting maximum-likelihood reconstruction of the WL--tSZ cross-power spectrum, $C_\ell^{\gamma y}$, for each tomographic bin, obtained by applying the MLBR method of Section~\ref{ssec:methods.mlbr} to the bandpowers of Figure~\ref{fig:raw_bandpowers} under the fiducial CIB model defined in Section~\ref{ssec:methods.spectra}. The spectrum is recovered as positive across the full multipole range in all four bins, with the amplitude increasing with the tomographic bin, consistent with the larger integrated Compton-$y$ signal probed by the higher-redshift bins. Also shown are two alternatives to the fiducial reconstruction: fixing the CIB SED to its $U(z)$-only best fit, and cross-correlating cosmic shear with the public \planck MILCA Compton-$y$ map. Both are broadly consistent with the fiducial measurement, although the cross-correlations against the MILCA map seem to be consistently lower than our fiducial measurements in the two higher redshift bins. We quantify these differences as well as the impact of other analysis choices in Section~\ref{ssec:results.robustness_cl}.

\begin{figure}[t!]
    \centering
    \includegraphics[width=\linewidth]{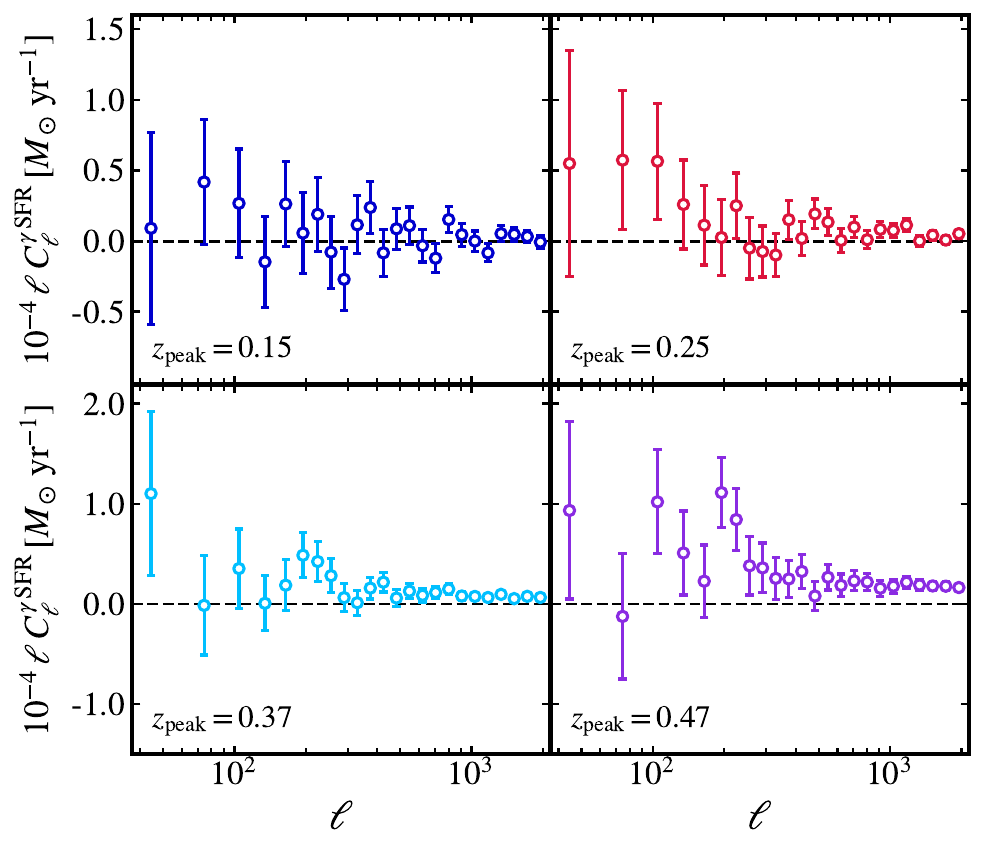}
    \caption{As Figure~\ref{fig:mlbr_tsz}, for the cross-correlation between cosmic shear and the star-formation-rate surface density, $C_\ell^{\gamma{\rm SFR}}$, under the fiducial CIB model. This is obtained from the reconstructed WL--CIB spectrum by multiplying by the Kennicutt constant, $K = 10^{-10}\,M_\odot\,{\rm yr}^{-1}\,L_\odot^{-1}$, which converts infrared luminosity density into a star-formation rate. The result is expressed in $M_\odot\,{\rm yr}^{-1}$ and is independent of the normalisation convention adopted for the CIB SED. Bandpowers above $\ell=240$ are merged in pairs using $\ell$-weighted inverse-variance weights, while lower-$\ell$ bandpowers are left unchanged.}
    \label{fig:mlbr_cib}
  \end{figure}

  Figure~\ref{fig:mlbr_cib} shows the corresponding reconstruction of the WL--CIB cross-power spectrum in terms of the projected SFR density defined in Equation~\eqref{eq:sfr_proj}, $C_\ell^{\gamma{\rm SFR}}$. The signal is clearly recovered in all tomographic bins but the first one, growing in amplitude towards higher redshifts, where both the mean SFR density and the lensing signal rise in amplitude. Although we will not interpret these measurements here, they may be used to study, for example, the abundance and clustering properties of star-forming systems. As an example, it is shown in \cite{Maleubre_inprep} that these measurements are in fact in reasonable agreement with expectations from hydrodynamical simulations that reproduce current measurements of the SFR history and the low-redshift stellar mass function.

 \subsection{Robustness of \texorpdfstring{\boldmath $C_\ell^{\gamma y}$}{WL-tSZ angular power spectra} to CIB modelling} \label{ssec:results.robustness_cl}
\begin{figure*}[t!]
      \centering
      \includegraphics[width=\linewidth]{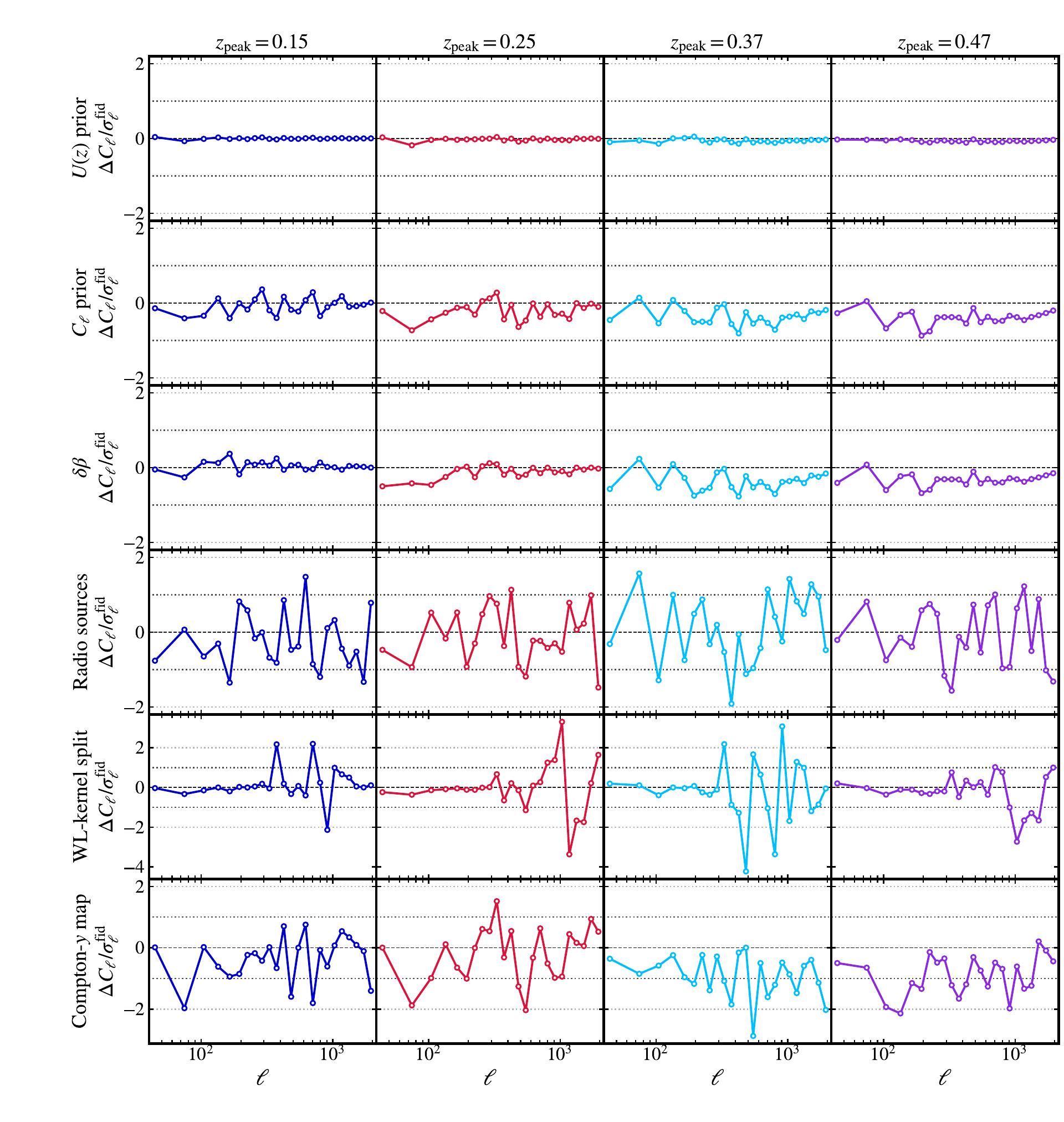}
      \caption{Residuals between the reconstructed WL--tSZ cross-power spectrum, in units of the fiducial statistical uncertainty, under alternative CIB modelling choices and the fiducial result for each of the four WL tomographic bins.}
      \label{fig:robustness_grid}
  \end{figure*}

  We next assess the sensitivity of the recovered WL--tSZ cross-power spectrum, $C_\ell^{\gamma y}$, to the details of the CIB modelling. For each alternative analysis choice, we recompute $C_\ell^{\gamma y}$ and compare it with the fiducial measurement, $C_\ell^{\rm fid}$, in units of the fiducial statistical uncertainty, $\sigma_\ell^{\rm fid}$. Figure~\ref{fig:robustness_grid} shows $\Delta C_\ell/\sigma_\ell^{\rm fid} \equiv (C_\ell - C_\ell^{\rm fid})/\sigma_\ell^{\rm fid}$ as a function of multipole for each of the four tomographic bins. We consider the following variations: fixing the CIB SED to its $U(z)$-only best fit; fixing the CIB SED to its $C_\ell^{\gamma \nu}$-only best fit; introducing an additional free spectral-index offset, $\delta\beta$, in each redshift bin; including a contribution from radio point sources; allowing for the CIB SED to vary within the redshift range covered by the lensing kernel; and replacing our reconstructed $C_\ell^{\gamma y}$ with the cross-correlation of cosmic shear with the public \planck MILCA Compton-$y$ map \cite{Aghanim2015planck}.

  For the spectral-index test, we allow the effective CIB SED in each tomographic bin to be tilted by a free power law,
  \begin{equation}
      \bar{S}_\nu^{\rm CIB} \rightarrow \bar{S}_\nu^{\rm CIB} \left(\frac{\nu}{\nu_0}\right)^{\delta\beta},
  \end{equation}
  where $\nu_0 = 545$~GHz is the pivot frequency and $\delta\beta$ is a free offset to the effective spectral index of the CIB, fitted independently in each of the four WL bins. The choice of pivot frequency affects only the overall normalisation of the recovered $C_\ell^{\gamma \rm CIB}$, and not $C_\ell^{\gamma y}$. Nevertheless, we renormalise the resulting spectrum to ensure the interpretability of the resulting CIB cross-correlation. A non-zero value would therefore indicate a mismatch between the frequency dependence predicted by the CIB model of Section~\ref{ssec:methods.spectra} and that preferred by the data.

  For the radio point-source test, we extend the mixing matrix model to include a third component,
  \begin{equation}
      C_\ell^{\gamma\nu} = S_\nu^{\rm tSZ} C_\ell^{\gamma y} + \bar{S}_\nu^{\rm CIB} C_\ell^{\gamma \rm CIB} + S_\nu^{\rm PS} C_\ell^{\gamma \rm PS},
  \end{equation}
  where the radio source SED is assumed to scale as $S_\nu^{\rm PS} \propto \nu^{\beta_{\rm PS}}$ in the units of the frequency maps (Section~\ref{ssec:methods.data}). Our baseline choice is $\beta_{\rm PS}=-3$, typical of synchrotron-dominated radio sources ($\beta_{\rm synch}\simeq-2.7$). We also consider an extragalactic component with a flatter spectrum, $\beta_{\rm PS}=-2$, closer to free-free emission.

  To quantify the potential impact of the redshift dependence in the CIB SED across the lensing kernel, we divide the kernel into two regions about its maximum and calculate the average SED within each half. We then consider a model with two CIB components (a low-redshift $l$ and a high-redshift one $h$):
  \begin{equation}
    C_\ell^{\gamma\nu} = S_\nu^{\rm tSZ} C_\ell^{\gamma y} + \bar{S}_\nu^l C_\ell^{\gamma l} + \bar{S}_\nu^h C_\ell^{\gamma h}.
  \end{equation}
  Here, $\bar{S}_\nu^l$ and $\bar{S}_\nu^h$ are the CIB SED averaged over the two halves of the lensing kernel.

  \begin{table}
      \centering
      \begin{tabular}{ccc}
         \hline
         \hline
         \noalign{\vskip 2.0pt}
         Bin & $z_{\rm peak}$ & $\delta\beta$ \\
         \noalign{\vskip 2.0pt}
         \hline
         \noalign{\vskip 3.0pt}
         1 & $0.15$ & $0.15 \pm 0.11$ \\
         2 & $0.25$ & $-0.10 \pm 0.09$ \\
         3 & $0.37$ & $-0.15 \pm 0.07$ \\
         4 & $0.47$ & $-0.05 \pm 0.05$ \\
         \hline
         \hline
      \end{tabular}
      \caption{Best-fitting spectral index offsets $\delta\beta$ obtained when allowing a free CIB SED tilt in each tomographic bin. Here, $z_{\rm peak}$ refers to the peak of the WL lensing kernel.}
      \label{tab:deltabeta}
  \end{table}

  As shown in Figure \ref{fig:robustness_grid}, varying the CIB SED within the parameter values allowed by the $U(z)$ data leads to negligible differences with respect to our fiducial measurements (a small fraction of the measurement errors in all cases). When allowing the SED parameters to vary fully (within the range allowed by the $C_\ell$ data themselves), we observe a larger shift in the recovered spectra. In particular the recovered $C_\ell$s are $\sim 0.1$-$0.5\sigma$ lower in the higher redshift bins. This is not entirely surprising: as shown in Figure \ref{fig:corner_cl_only}, the $C_\ell$ data alone are not able to place meaningful constraints on the SED parameters, which allows for extreme models that depart significantly from the observational $U(z)$ data. A similar result is obtained when varying the effective CIB spectral index $\delta\beta$. In fact the resulting deviations are remarkably similar, both in amplitude and scale dependence, to those found when varying the SED model parameters freely, showing that both tests probe the same preference of the $C_\ell$ data for a slightly different CIB SED. The constraints on $\delta\beta$ found in this case are summarised in Table~\ref{tab:deltabeta}. All four measurements are consistent with zero at the $\sim 2\sigma$ level or better, with no clear monotonic trend with redshift. This indicates that there is no statistically significant evidence for a mismatch between the assumed CIB spectral shape and that preferred by the data. Since both tests probe essentially the same trend in the data, and since Figure \ref{fig:corner_cl_only} shows that the $C_\ell$ data are in reasonable agreement with the CIB SED favoured by the $U(z)$ data, we conclude that there is no significant evidence of a systematic deviation with respect to the CIB SED used in our fiducial analysis.

  Introducing a radio point source component, or incorporating the redshift dependence of the CIB SED in the model leads to larger deviations with respect to our fiducial power spectrum measurements, of the order of 1-2$\sigma$. These deviations do not exhibit any particular trend, however, either as a function of scale or redshift, and seem compatible with noise. Their increased amplitude, compared with the deviations found in the previous tests, can be explained by the fact that both of these tests involve including additional components in the model, increasing the number of free parameters in the model (the cross-spectrum amplitudes) by $50\%$, and leading to larger statistical uncertainties in the resulting measurements.

  The comparison with the cross-spectrum obtained from the Compton-$y$ map shows a more systematic pattern, particularly in the two highest-redshift bins, where the recovered spectra are systematically lower by 1-2$\sigma$. This could signal the presence of residual contamination in the MILCA map caused by imperfect component separation. The fact that the CIB amplitude grows with redshift, and that the CIB and tSZ SEDs have opposite signs at low frequency (making the CIB effectively a ``negative'' contribution to the observed $y$ map), would point to it as a likely source of contamination in this case.

  In addition to these variations, we also check the impact of the residual uncertainty in the CIB SED itself. Rather than fixing the four physical SED parameters to a single best-fit value, we marginalise over their full joint $U(z)+C_\ell^{\gamma\nu}$ posterior when constructing the mixing matrix. This changes the recovered bandpowers by no more than $0.19\sigma$ in any tomographic bin or multipole range, confirming that the propagated CIB modelling uncertainty is subdominant to the statistical uncertainty of the measurement.

 \subsection{Robustness of CIB parameters to analysis choices} \label{ssec:results.robustness_params}
  Finally, we test whether the physical CIB parameter constraints themselves depend on the range of angular scales or redshifts included in the fit.
  
  A scale dependence in the effective CIB SED is motivated by the different contributions from haloes of different masses at different scales. The halo model provides a simplified framework to understand this intuitively: on large scales, dominated by the 2-halo term, the contribution from a given halo to the power spectrum on the CIB side is simply proportional to the total SFR in that halo. Large scales are thus sensitive to the average SFR-weighted SED of star-forming sources. However, on small scales, where the 1-halo term dominates, a given halo's contribution to the WL--CIB cross-correlation is proportional to the product of its mass and its SFR. Thus, if the infrared SED depends strongly on halo mass at fixed SFR, CIB cross-correlations may acquire a significant scale-dependent spectral signature. The same effect occurs in other cross-correlations, for example against the galaxy overdensity, where the small-scale signal depends on the effective SED of the galaxy sample targeted, as well as their correlation with star-forming galaxies.
  
\begin{figure}[t!]
      \centering
      \includegraphics[width=\linewidth]{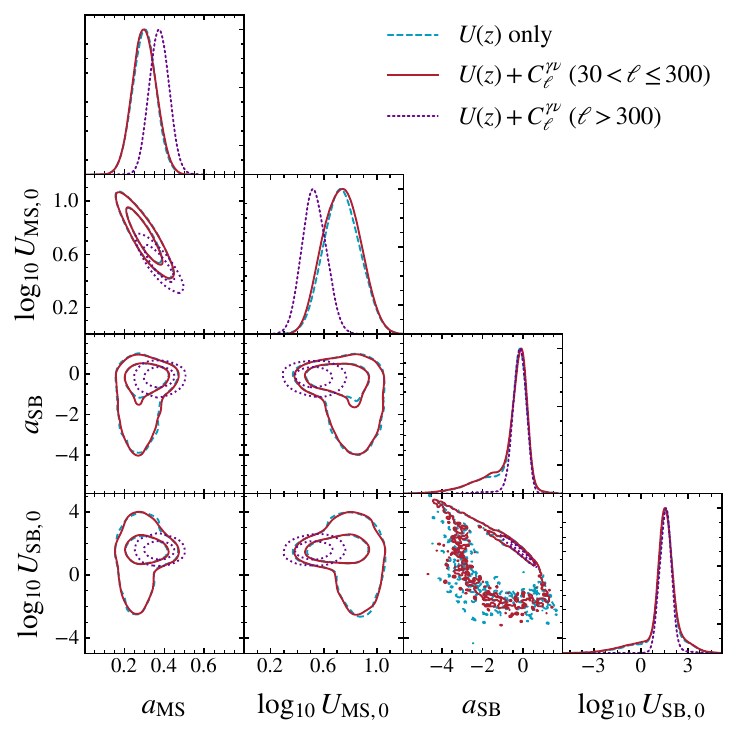}
      \caption{Marginalised posterior constraints on the four physical CIB SED parameters from the $U(z)$ data alone (blue), and together with $C_\ell^{\gamma \nu}$ restricted to large scales, $30<\ell\leq300$ (red), or small scales, $\ell>300$ (purple).}
      \label{fig:corner_ellcut}
  \end{figure}

  For the tomographic bins used here, the angular scale $\ell\simeq300$ roughly separates the 2-halo and 1-halo-dominated regimes (especially in the two highest-redshift bins, where the signal is stronger). Therefore, we explore the constraints on the CIB SED parameters obtained using only large-scale multipoles, in the range $30<\ell\leq300$, and small angular scales, in the range $300<\ell\leq2000$, in both cases in combination with the $U(z)$ data. The result is shown in Figure~\ref{fig:corner_ellcut}. The large-scale-only posterior is essentially indistinguishable from the $U(z)$-only constraint, showing no preference for different MS parameters. This is because the large-scale $C_\ell$ measurements are not sufficiently sensitive to significantly affect the constraints on the $U(z)$ parameters. The small-scale-only posterior, in contrast, reproduces the shift towards higher $\alpha_{\rm MS}$ and lower $\log_{10} U_{\rm MS,0}$ seen in the full joint fit (Figure~\ref{fig:corner_fiducial}). This mild shift is therefore predominantly driven by the small-scale modes of the cross-power spectrum. In summary, we do not find evidence of a significant scale dependence in the effective CIB SED in the case of cosmic shear cross-correlations.

  \begin{figure}[t!]
    \centering
    \includegraphics[width=\linewidth]{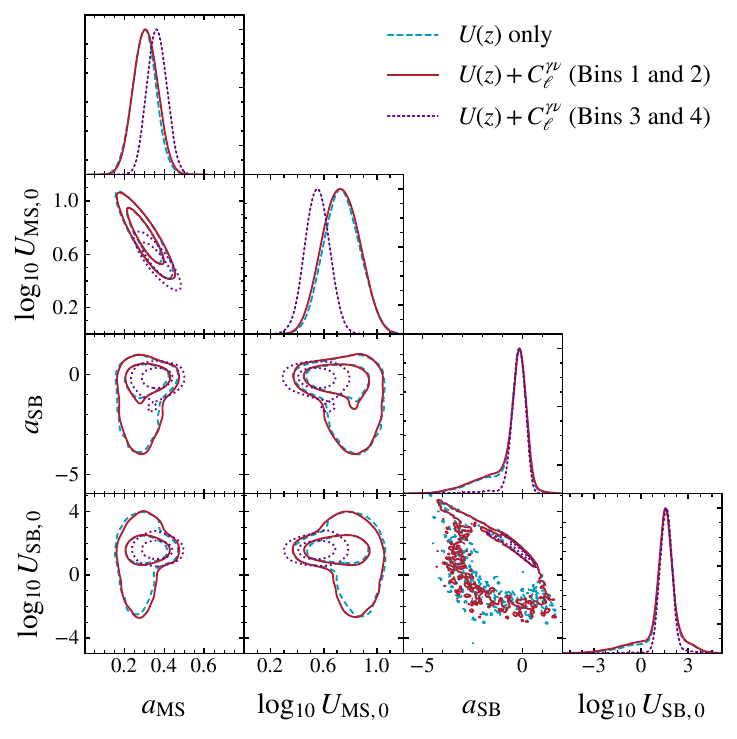}
    \caption{Marginalised posterior constraints on the four physical CIB SED parameters from the $U(z)$ data alone (blue), and together with $C_\ell^{\gamma \nu}$ restricted to the two lowest redshift WL bins (red), or the two highest redshift WL bins (purple).}
    \label{fig:corner_zcut}
  \end{figure}

  We carry out a similar exercise, this time splitting by redshift rather than by scale, to study the potential redshift dependence in the CIB SED parameters beyond the parametrisation described in Section \ref{ssec:methods.spectra}. Since a fit to all four tomographic bins simultaneously cannot itself reveal which bins are responsible for a given shift, while fitting each bin independently does not provide meaningful constraints on the redshift-dependent CIB parameters, we divide our data into the two lowest and two highest redshift bins, and repeat the joint $U(z)+C_\ell^{\gamma \nu}$ fit for each pair. Figure~\ref{fig:corner_zcut} shows that the low-redshift pair yields a posterior essentially indistinguishable from the $U(z)$-only constraint, with no preference for different MS parameters. The high-redshift pair, in contrast, reproduces the shift towards higher $\alpha_{\rm MS}$ and lower $\log_{10} U_{\rm MS,0}$ seen in the full joint fit (Figure~\ref{fig:corner_fiducial}), recovering $88\%$ of the shift in $\alpha_{\rm MS}$ found in the fiducial analysis. This indicates that most of the additional constraining power on the CIB SED beyond $U(z)$ is sourced by the small-scale modes of the two highest-redshift bins (unsurprisingly, given that this is also where $C_\ell^{\gamma \nu}$ has the highest signal-to-noise). We do not find significant evidence of an evolution in the effective CIB SED beyond that allowed by our fiducial $U(z)$ model.

 \subsection{Sensitivity to cosmology and baryonic feedback} \label{ssec:results.Flamingo}
  \begin{figure*}[t!]
    \centering
    \includegraphics[width=\textwidth]{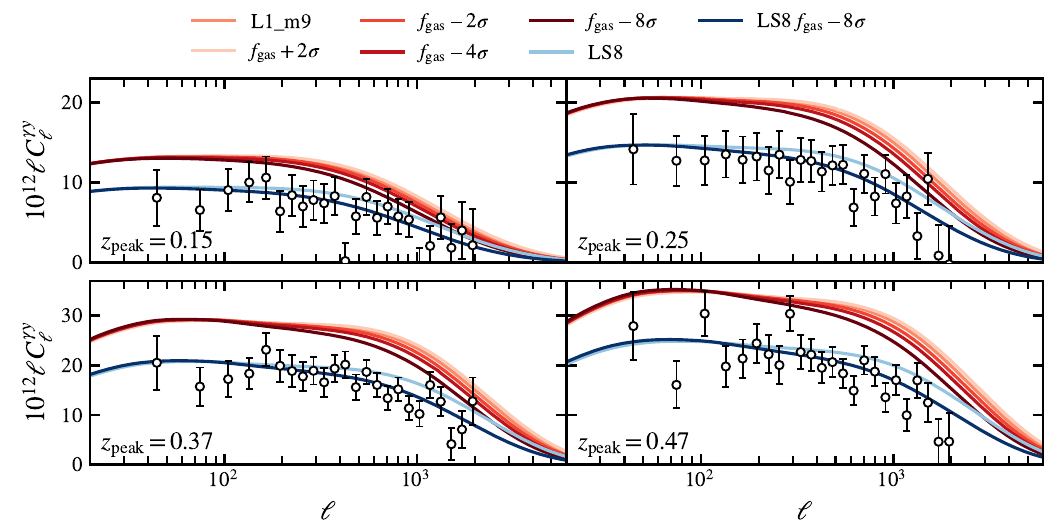}
    \caption{Maximum-likelihood bandpower reconstruction of the WL--tSZ power spectrum presented in Figure~\ref{fig:mlbr_tsz}. We compare the measurements to predictions from the \flamingo simulations~\cite{2306.04024,2306.05492} using the publicly available matter-pressure power spectra~\cite{2604.24324} for the fiducial D3A cosmology (red lines) and the low-$S_8$ cosmology (blue lines), varying the feedback strength.}
    \label{fig:flamingo}
  \end{figure*}
  \begin{table*}[t!]
    \begin{ruledtabular}
      \begin{tabular}{lcccc}
        \noalign{\vskip 2.0pt}
        & {\boldmath$A_{\gamma\mathrm{tSZ}}$} ({\boldmath$30 < \ell \le 2000$}) & {\boldmath$A_{\gamma\mathrm{tSZ}}$} ({\boldmath$30 \le \ell \le 300$}) & {\boldmath $A_{\gamma\mathrm{tSZ}}$} (\boldmath{$300 \le \ell \le 2000$}) & {\boldmath$A_{\gamma\gamma}$}\\
        \noalign{\vskip 2.0pt}
        \hline
        \noalign{\vskip 3.0pt}
        \textbf{Fiducial (L1\_m9)} & $0.598 \pm 0.015$ & $0.656 \pm 0.028$ & $0.573 \pm 0.018$ & $0.872 \pm 0.016$ \\
        {\boldmath$f_{\rm gas}+2\sigma$} & $0.589 \pm 0.015$ & $0.656 \pm 0.028$ & $0.562 \pm 0.018$ & $0.863 \pm 0.016$ \\
        {\boldmath$f_{\rm gas}-2\sigma$} & $0.612 \pm 0.016$ & $0.661 \pm 0.028$ & $0.591 \pm 0.019$ & $0.880 \pm 0.016$ \\
        {\boldmath$f_{\rm gas}-4\sigma$} & $0.629 \pm 0.016$ & $0.667 \pm 0.029$ & $0.612 \pm 0.019$ & $0.887 \pm 0.016$ \\
        {\boldmath$f_{\rm gas}-8\sigma$} & $0.662 \pm 0.017$ & $0.679 \pm 0.029$ & $0.653 \pm 0.021$ & $0.900 \pm 0.017$ \\
        \hline
        \noalign{\vskip 3.0pt}
        \textbf{LS8} & $0.854 \pm 0.022$ & $0.931 \pm 0.040$ & $0.820 \pm 0.026$ & $1.025 \pm 0.019$ \\
        {\textbf{LS8} \boldmath$f_{\rm gas}-8\sigma$} & $0.949 \pm 0.024$ & $0.963 \pm 0.041$ & $0.941 \pm 0.030$ & $1.055 \pm 0.019$ \\
        
      \end{tabular}
    \end{ruledtabular}
    \caption{Measurements of the rescaling amplitude $A_{\gamma\mathrm{tSZ}}$ for different choices of \flamingo simulations and scale cuts. We also report the WL auto power spectra amplitudes $A_{\gamma\gamma}$ for the multipole range $\ell\in(30,2000)$.}\label{tab:amplitudes}
  \end{table*}
  Having established the robustness of our measurements of the WL--tSZ power spectrum, we now proceed to compare them against theoretical predictions as a function of cosmological and baryonic feedback parameters. Unlike most previous analyses, which have typically relied on halo-model-based predictions \citep{Pandey2022,LaPosta2024insights,Troster2022}, here we limit ourselves to comparing our measurements to predictions from synthetic observations extracted from the \flamingo suite of hydrodynamical simulations \cite{2306.04024,2306.05492}. While this limits the range of models we can explore, this comparison avoids most of the inaccuracies and approximations associated with the halo model, and indirectly incorporates information from other astrophysical observations (e.g. the galaxy stellar mass function, X-ray gas fractions) used to calibrate these simulations.

  We focus on predictions from simulations exploring different cosmological models as well as different AGN feedback strengths. Specifically, we consider the fiducial ``D3A'' cosmological model, corresponding to the best-fit from Table~2 of~\cite{Abbott2022dark}, as well as the low-$S_8$ (``LS8'') model, based on \cite{2202.07440}, in which the value of $S_8\equiv\sigma_8\sqrt{\Omega_{\rm m}/0.3}$ is $S_8=0.766$. As a proxy for AGN feedback strength, we also use simulations calibrated to different X-ray gas fractions, shifted with respect to the measurements used to calibrate the fiducial \flamingo model. These are labelled $f_{\rm gas}N\sigma$, with negative values of $N$ corresponding to models with stronger feedback, where a larger fraction of the gas is ejected from haloes.

  To generate theoretical predictions for $C_\ell^{\gamma y}$ we use measurements of the 3D power spectrum between the matter overdensity and the electron thermal pressure\footnote{Note that the released power spectra correspond to the \emph{comoving} thermal pressure, which must be multiplied by $(1+z)^3$ to transform it into the physical pressure entering the tSZ effect (Equation~\eqref{eq:comptony}).}, made available with the \flamingo data release \cite{2604.24324}. We build a two-dimensional interpolator for this power spectrum, as a function of redshift and wavenumber, spanning the redshift range $z\in[0,30]$, and use it to calculate $C^{\gamma y}_\ell$ through the Limber integral (Equation~\eqref{eq:Cl}), with the lensing and tSZ radial kernels defined in Equations~\eqref{eq:lensing_kernel} and \eqref{eq:sz_kernel}.
  
  The result of this comparison is shown in Figure~\ref{fig:flamingo}. Focusing first on the theoretical predictions, we find that, particularly on large scales, the WL--tSZ cross-spectrum is significantly more sensitive to changes in cosmological parameters than baryonic feedback. Specifically, on scales $\ell\lesssim300$, in the 2-halo-dominated regime, the differences between different AGN feedback models are significantly smaller than the measurement errors, while predictions for the D3A and LS8 cosmologies differ by approximately 30\%. Assuming a power-law dependence of the form $C_\ell^{\gamma y}\propto \sigma_8^\alpha$, this corresponds to a very steep scaling, with $\alpha\simeq5.5$. This is in agreement with past analyses (e.g. \cite{Troster2022}), and with physical expectations: due to the strong scaling of thermal energy with mass ($E_{\rm th}\propto M^{5/3}$), the tSZ effect is dominated by the most massive haloes, whose abundance is a steep function of the amplitude of matter fluctuations. On small scales, the power spectrum becomes sensitive to the distribution and temperature of the gas within haloes, and the differences between different AGN feedback strengths become more apparent.

  In spite of these differences, we find that our measurements lie systematically below the theoretical predictions from \flamingo for the D3A cosmological model, regardless of feedback strength, at all redshifts and scales, and display a much better agreement with the predictions from the simulations adopting the LS8 model. To quantify the agreement of a given model with our data, we rescale the theoretical prediction by a free amplitude $A_{\gamma{\rm tSZ}}$ to match our data. The best-fit value of $A_{\gamma{\rm tSZ}}$ and its uncertainty for this linear model can be found as the least-squares solution
  \begin{align}
    &A_{\gamma\mathrm{tSZ}} = {\rm Var}(A_{\gamma{\rm tSZ}})\,{\mathbf{C}^{\gamma y}}^\top\mathsf{\Sigma}^{-1}\hat{\mathbf{C}}^{\gamma y},\\
    &{\rm Var}(A_{\gamma{\rm tSZ}})=({\mathbf{C}^{\gamma y}}^\top\mathsf{\Sigma}^{-1}\mathbf{C}^{\gamma y})^{-1},
  \end{align}
  where ${\bf C}^{\gamma y}$ is the vector of theoretical predictions, $\hat{\bf C}^{\gamma y}$ are the measurements, and $\mathsf{\Sigma}$ is the covariance matrix of the latter. A value of $A_{\gamma {\rm tSZ}}<1$ denotes a model that over-predicts our measurements by a factor $1/A_{\gamma {\rm tSZ}}$, and the departure from $A_{\gamma {\rm tSZ}}=1$ can be used to quantify the level of tension of different models.
  
  We report the measured amplitudes in Table~\ref{tab:amplitudes}. Our measurements disagree strongly with the fiducial \flamingo model, at the $27\sigma$ level. Even the strongest feedback scenario ($f_{\rm gas}-8\sigma$) is in strong disagreement with the data ($20\sigma$) within the D3A cosmology. In turn, the low-$S_8$ models are in better agreement, showing a smaller $6\sigma$ tension for the standard feedback model, and a mere $2\sigma$ deviation in the presence of strong feedback (LS8 $f_{\rm gas}-8\sigma$). This residual level of tension is driven by the small-scale measurements. Using only multipoles $\ell<300$, where the data are largely in the 2-halo regime, both LS8 predictions are in reasonable agreement ($\sim2\sigma$) with the data, whereas the small-scale data ($300<\ell<2000$) drive the preference for the strong-feedback scenario, with the standard-feedback model in tension at the $7\sigma$ level. These constraints must be taken with a pinch of salt: here we have used a rather simplistic single-parameter model, whereas a more complete analysis would marginalise over a broader parameter space, characterising systematic effects (e.g. redshift uncertainties, intrinsic alignments) as well as other cosmological parameters. While this would alleviate the level of disagreement shown in Figure~\ref{fig:flamingo} to some extent, it is unlikely that it would fully resolve it without causing tensions with other cosmological probes. 
  
  In summary, independently of feedback strength, the data favour the low-$S_8$ predictions in \flamingo. This stark preference is greatly aided by the strong scaling of $C_\ell^{\gamma y}$ with $\sigma_8$. To illustrate this, we can repeat this analysis for a standard cosmological probe of the matter fluctuations, the shear power spectrum, which has a comparatively weak dependence on $\sigma_8$ (approximately $C_\ell^{\gamma\gamma}\propto\sigma_8^2$). Specifically, we measure the shear auto- and cross-correlations between the four tomographic bins using the methodology described in \cite{Garcia-Garcia2024cosmic} with the catalogue-based approach outlined in Section \ref{sssec:methods.data.cls}. We then generate predictions for these measurements from \flamingo using the matter power spectra from the simulations, and fit these predictions to our measurements with a free amplitude $A_{\gamma\gamma}$. The results are shown in the rightmost column in Table~\ref{tab:amplitudes}, and the DES Y3 shear-shear power spectra are shown in Appendix \ref{app:extra_figs}. The level of disagreement in the fiducial D3A family of simulations is significantly smaller than observed in the case of $C_\ell^{\gamma y}$, rising from $A_{\gamma\gamma}=0.900\pm0.017$ (6$\sigma$) for $f_{\rm gas}-8\sigma$ to 7.5$\sigma$ for the fiducial simulation\footnote{The caveats described above regarding the limitations of using a single-parameter model also apply when interpreting these tension metrics.}. In turn, the LS8 simulations achieve a better agreement with the $C_\ell^{\gamma\gamma}$ measurements, with $A_{\gamma\gamma}$ compatible with 1 at the 1.3-2.9$\sigma$ level.

  The preference for a lower $S_8$ value from tSZ cross-correlations is qualitatively consistent with past analyses. A similar comparison with \flamingo using KiDS-1000 shear data was made in \cite{2410.19905} (see also \cite{McCarthy2023}), reaching similar conclusions. As a sanity check, we repeated our measurement of $C_\ell^{\gamma y}$ using the KiDS-1000 dataset \cite{1902.11265}, and found the same preference for the low-$S_8$ \flamingo model, albeit with larger uncertainties (e.g. $A_{\gamma\mathrm{tSZ}}=0.630\pm 0.037$ for $f_{\rm gas}-8\sigma$ -- see Appendix \ref{app:extra_figs}). Studies of other tSZ cross-correlations have also found roughly consistent results. Measurements of the large-scale correlation with galaxies in tomographic redshift bins, quantified in terms of the halo-bias-weighted mean electron pressure, $\langle bP_e\rangle$, have shown evidence of a low $S_8$ value, although the picture is less clear at higher redshifts \cite{LaPosta2026joint,ralp_inprep}. Evidence from the tSZ auto-correlation is also inconclusive (see e.g. \cite{1712.00788,2502.10232}), with the impact of foreground contamination being a key source of uncertainty in this case.

\section{Conclusions} \label{sec:conclusions}
 In this work, we developed a physically motivated framework for propagating CIB SED uncertainties through component separation at the power-spectrum level, and used it to test the robustness of WL--tSZ cross-correlations, applying the method to measurements from DES Y3 cosmic shear data and \planck PR4 frequency maps. Rather than deprojecting the CIB at the map level, which increases variance and requires precise external knowledge of the contaminant frequency dependence, we model the tSZ and CIB contributions directly in the cross-power spectrum using the main-sequence and starburst SED model of \cite{Bethermin2017impact}, calibrated jointly against external measurements of the radiation field intensity $U(z)$ and the WL--\planck cross-power spectra themselves.

 We find that the multifrequency cross-power spectra provide marginal but complementary information on the CIB SED beyond the external calibration based on measurements of $U(z)$. Their main effect is a small shift in the preferred main-sequence parameters, which we show is driven almost entirely by the small-scale modes of the two highest redshift tomographic bins. The cross-correlation data also help in breaking degeneracies between the $U(z)$ parameters, suppressing the large non-Gaussian tails in the distribution of the starburst parameters.

 Despite this additional sensitivity to the CIB SED, the recovered WL--tSZ cross-power spectrum itself is remarkably stable. Fixing the SED to its $U(z)$-only or $C_\ell^{\gamma \nu}$-only best fit, adding a radio point source contribution, or allowing for an independent CIB spectral index offset per redshift bin all leave the recovered spectrum consistent with the fiducial result at the $\sim 1$--$2\sigma$ level, with no clear trend in scale or redshift. Explicitly marginalising over the full joint posterior of the four physical SED parameters, rather than fixing them to a single best-fit value, changes the recovered bandpowers by no more than $0.19\sigma$. Splitting the CIB component between the low- and high-redshift halves of the WL kernel, to account for the potential evolution in the CIB SED throughout the kernel, leads to increased scatter in the measurement (due to the presence of an additional component in the model), but does not show evidence of contamination. We also find no evidence of a scale dependence in the effective CIB SED, which would be present if infrared spectra depended strongly on halo mass, nor of a redshift dependence beyond that allowed by the $U(z)$ model. Finally, comparing our reconstruction with the cross-correlation of cosmic shear against the public \planck MILCA Compton-$y$ map reveals a systematically lower amplitude, at the level of 1-2$\sigma$, rising with redshift, potentially due to residual CIB contamination in the Compton-$y$ map.

 Taken together, these results demonstrate that a physically motivated, power-spectrum level treatment of the CIB can be used to recover a WL--tSZ cross-power spectrum that is robust to realistic uncertainties in the CIB SED, its redshift evolution, and residual point source contamination, without resorting to the higher-variance map-level deprojection techniques used in some past analyses. This robustness is particularly encouraging given that the additional information the cross-power spectra provide on the CIB SED is concentrated in exactly the regime where one might otherwise worry that CIB mismodelling could bias the recovered tSZ signal.

 Comparing our measurements of the WL--tSZ cross-correlation with predictions from the \flamingo hydrodynamical simulations, we find that the data clearly favour the \flamingo models with a low value of $S_8$. This is independent of the strength of AGN feedback and is driven by the strong scaling of this cross-correlation with $\sigma_8$. More specifically, the large-scale measurements alone ($\ell<300$), where the cross-correlation is almost completely immune to AGN feedback within the \flamingo models, show a clear preference for low $S_8$. The small-scale data are also consistent with this result, and further seem to favour models with strong AGN feedback. This result is consistent with past literature using tSZ cross-correlations \cite{McCarthy2023,2410.19905,LaPosta2026joint,2605.11083}. The evidence for low $S_8$ is stronger for the WL--tSZ cross-correlation than the shear auto-correlations themselves, driven by the steeper scaling with $\sigma_8$ of the former. This illustrates the potential of the large-scale lensing-SZ cross-correlation as a cosmological probe of the large-scale structure, assuming that accurate predictions for it can be produced.
 
 The framework developed here, in which we jointly marginalise over a physically motivated CIB SED model with external calibration data and the target cross-power spectra, is not specific to DES Y3 and \planck, and can be applied directly to future high-precision weak lensing and CMB surveys. With more sensitive data, some of the assumptions made here will need to be reassessed, however. In particular, the evolution in the global CIB SED within the redshift range covered by the lensing kernel is likely to become an important effect, which must then be forward-modelled within the multi-frequency power spectrum likelihood. Likewise, the dependence of the SEDs on internal halo/galaxy properties (e.g. mass) may lead to an effective scale dependence in the CIB contamination that future datasets may become sensitive to. This is likely a more complex problem for galaxy cross-correlations than it is for cosmic shear. Accounting for this in a model-independent way may be more difficult, although better infrared calibration data could help develop physics-based parametrisations for this effect. Nevertheless, control over astrophysical foregrounds will be essential to fully exploit WL--tSZ cross-correlations as probes of baryonic feedback and cosmology in future observations, and the approach described here represents a promising way forward.

 \begin{acknowledgments}
   AW is supported by a Science and Technology Facilities Council (STFC) studentship. ALP and DA are supported by STFC under grants with reference UKRI1164 and ST/W000903/1, and they further acknowledge support from the Beecroft Trust. BJ is supported by a CDSN doctoral studentship through the ENS Paris-Saclay.
 \end{acknowledgments}

\onecolumngrid
\appendix
\section{Additional comparisons to the \flamingo simulations}\label{app:extra_figs}
  We include here figures showing two other measurements compared with \flamingo and discussed in Section \ref{ssec:results.Flamingo}.
  
\begin{figure}
    \centering
    \includegraphics[width=\textwidth]{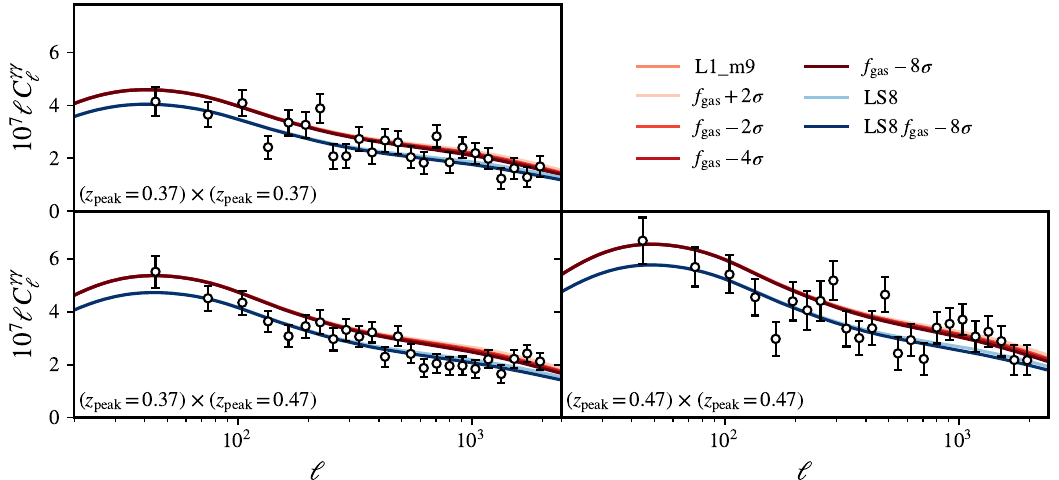}
    \caption{As Figure \ref{fig:flamingo} for the shear-shear power spectra for the two highest redshift bins in DES Y3. We compare the measurements to predictions from the \flamingo simulations~\cite{2306.04024,2306.05492} using the publicly available matter power spectra~\cite{2604.24324} for the fiducial D3A cosmology (red lines) and the low-$S_8$ cosmology.}
    \label{fig:shear-auto-flamingo}
  \end{figure}

  Figure \ref{fig:shear-auto-flamingo} shows the shear-shear power spectra of the two highest-redshift bins in the DES Y3 dataset ($z_{\rm peak}=0.37$ and 0.47), together with the \flamingo predictions for the D3A cosmological model and different thermal feedback scenarios (red lines) as well as the low-$S_8$ cosmological model with its standard and strong feedback models (blue lines).

\begin{figure}
    \centering
    \includegraphics[width=\textwidth]{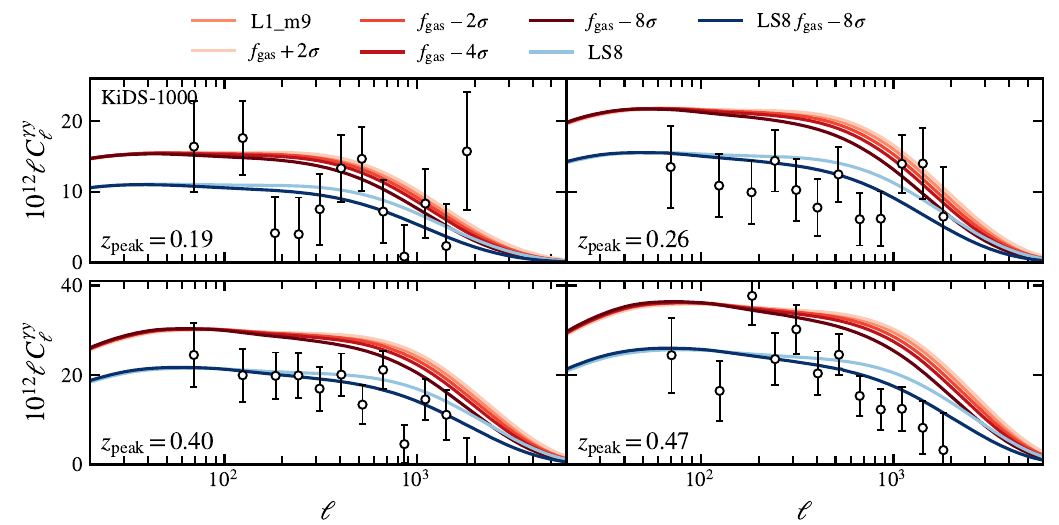}
    \caption{As Figure \ref{fig:flamingo} for the cross-correlation between tSZ and shear from the KiDS-1000 dataset \cite{2007.15633}.}
    \label{fig:flamingo-kids}
  \end{figure}

  Figure~\ref{fig:flamingo-kids} shows the WL--tSZ cross-correlation between \planck and the KiDS-1000 shear sample, extracted from the \planck temperature maps using the methodology presented in this paper. The measurements follow the same trend observed in Figure~\ref{fig:flamingo} in the case of DES: the WL--tSZ cross-correlation favours the \flamingo predictions for cosmologies with a low $S_8$ value, independently of the feedback scenario. A similar comparison was presented in \cite{2410.19905}.

  \twocolumngrid

\bibliography{references}

@string{june = {June}}

@article{1007.1149,
 adsurl = {https://ui.adsabs.harvard.edu/abs/2013A&A...558A.118H},
 archiveprefix = {arXiv},
 author = {{Hurier}, G. and {Mac{\'\i}as-P{\'e}rez}, J.~F. and {Hildebrandt}, S.},
 doi = {10.1051/0004-6361/201321891},
 eid = {A118},
 eprint = {1007.1149},
 journal = {\aap},
 month = {October},
 pages = {A118},
 primaryclass = {astro-ph.IM},
 title = {{MILCA, a modified internal linear combination algorithm to extract astrophysical emissions from multifrequency sky maps}},
 volume = {558},
 year = {2013}
}

@article{1712.00788,
 adsurl = {https://ui.adsabs.harvard.edu/abs/2018MNRAS.477.4957B},
 archiveprefix = {arXiv},
 author = {{Bolliet}, Boris and {Comis}, Barbara and {Komatsu}, Eiichiro and {Mac{\'\i}as-P{\'e}rez}, Juan Francisco},
 doi = {10.1093/mnras/sty823},
 eprint = {1712.00788},
 journal = {\mnras},
 month = {July},
 number = {4},
 pages = {4957-4967},
 primaryclass = {astro-ph.CO},
 title = {{Dark energy constraints from the thermal Sunyaev-Zeldovich power spectrum}},
 volume = {477},
 year = {2018}
}

@article{1902.11265,
 adsurl = {https://ui.adsabs.harvard.edu/abs/2019A&A...625A...2K},
 archiveprefix = {arXiv},
 author = {{Kuijken}, K. and {Heymans}, C. and {Dvornik}, A. and {Hildebrandt}, H. and {de Jong}, J.~T.~A. and {Wright}, A.~H. and {Erben}, T. and {Bilicki}, M. and {Giblin}, B. and {Shan}, H.-Y. and {Getman}, F. and {Grado}, A. and {Hoekstra}, H. and {Miller}, L. and {Napolitano}, N. and {Paolilo}, M. and {Radovich}, M. and {Schneider}, P. and {Sutherland}, W. and {Tewes}, M. and {Tortora}, C. and {Valentijn}, E.~A. and {Verdoes Kleijn}, G.~A.},
 doi = {10.1051/0004-6361/201834918},
 eid = {A2},
 eprint = {1902.11265},
 journal = {\aap},
 month = {May},
 pages = {A2},
 primaryclass = {astro-ph.GA},
 title = {{The fourth data release of the Kilo-Degree Survey: ugri imaging and nine-band optical-IR photometry over 1000 square degrees}},
 volume = {625},
 year = {2019}
}

@article{2001ApJ...554..778L,
 adsurl = {https://ui.adsabs.harvard.edu/abs/2001ApJ...554..778L},
 archiveprefix = {arXiv},
 author = {{Li}, Aigen and {Draine}, B.~T.},
 doi = {10.1086/323147},
 eprint = {astro-ph/0011319},
 journal = {\apj},
 month = {June},
 number = {2},
 pages = {778-802},
 primaryclass = {astro-ph},
 title = {{Infrared Emission from Interstellar Dust. II. The Diffuse Interstellar Medium}},
 volume = {554},
 year = {2001}
}

@article{2007.15633,
 adsurl = {https://ui.adsabs.harvard.edu/abs/2021A&A...645A.104A},
 archiveprefix = {arXiv},
 author = {{Asgari}, Marika and {Lin}, Chieh-An and {Joachimi}, Benjamin and {Giblin}, Benjamin and {Heymans}, Catherine and {Hildebrandt}, Hendrik and {Kannawadi}, Arun and {St{\"o}lzner}, Benjamin and {Tr{\"o}ster}, Tilman and {van den Busch}, Jan Luca and {Wright}, Angus H. and {Bilicki}, Maciej and {Blake}, Chris and {de Jong}, Jelte and {Dvornik}, Andrej and {Erben}, Thomas and {Getman}, Fedor and {Hoekstra}, Henk and {K{\"o}hlinger}, Fabian and {Kuijken}, Konrad and {Miller}, Lance and {Radovich}, Mario and {Schneider}, Peter and {Shan}, HuanYuan and {Valentijn}, Edwin},
 doi = {10.1051/0004-6361/202039070},
 eid = {A104},
 eprint = {2007.15633},
 journal = {\aap},
 month = {January},
 pages = {A104},
 primaryclass = {astro-ph.CO},
 title = {{KiDS-1000 cosmology: Cosmic shear constraints and comparison between two point statistics}},
 volume = {645},
 year = {2021}
}

@article{2007ApJ...657..810D,
 adsurl = {https://ui.adsabs.harvard.edu/abs/2007ApJ...657..810D},
 archiveprefix = {arXiv},
 author = {{Draine}, B.~T. and {Li}, Aigen},
 doi = {10.1086/511055},
 eprint = {astro-ph/0608003},
 journal = {\apj},
 month = {March},
 number = {2},
 pages = {810-837},
 primaryclass = {astro-ph},
 title = {{Infrared Emission from Interstellar Dust. IV. The Silicate-Graphite-PAH Model in the Post-Spitzer Era}},
 volume = {657},
 year = {2007}
}

@article{2202.07440,
 adsurl = {https://ui.adsabs.harvard.edu/abs/2023MNRAS.518..477A},
 archiveprefix = {arXiv},
 author = {{Amon}, A. and {Robertson}, N.~C. and {Miyatake}, H. and {Heymans}, C. and {White}, M. and {DeRose}, J. and {Yuan}, S. and {Wechsler}, R.~H. and {Varga}, T.~N. and {Bocquet}, S. and {Dvornik}, A. and {More}, S. and {Ross}, A.~J. and {Hoekstra}, H. and {Alarcon}, A. and {Asgari}, M. and {Blazek}, J. and {Campos}, A. and {Chen}, R. and {Choi}, A. and {Crocce}, M. and {Diehl}, H.~T. and {Doux}, C. and {Eckert}, K. and {Elvin-Poole}, J. and {Everett}, S. and {Fert{\'e}}, A. and {Gatti}, M. and {Giannini}, G. and {Gruen}, D. and {Gruendl}, R.~A. and {Hartley}, W.~G. and {Herner}, K. and {Hildebrandt}, H. and {Huang}, S. and {Huff}, E.~M. and {Joachimi}, B. and {Lee}, S. and {MacCrann}, N. and {Myles}, J. and {Navarro-Alsina}, A. and {Nishimichi}, T. and {Prat}, J. and {Secco}, L.~F. and {Sevilla-Noarbe}, I. and {Sheldon}, E. and {Shin}, T. and {Tr{\"o}ster}, T. and {Troxel}, M.~A. and {Tutusaus}, I. and {Wright}, A.~H. and {Yin}, B. and {Aguena}, M. and {Allam}, S. and {Annis}, J. and {Bacon}, D. and {Bilicki}, M. and {Brooks}, D. and {Burke}, D.~L. and {Carnero Rosell}, A. and {Carretero}, J. and {Castander}, F.~J. and {Cawthon}, R. and {Costanzi}, M. and {da Costa}, L.~N. and {Pereira}, M.~E.~S. and {de Jong}, J. and {De Vicente}, J. and {Desai}, S. and {Dietrich}, J.~P. and {Doel}, P. and {Ferrero}, I. and {Frieman}, J. and {Garc{\'\i}a-Bellido}, J. and {Gerdes}, D.~W. and {Gschwend}, J. and {Gutierrez}, G. and {Hinton}, S.~R. and {Hollowood}, D.~L. and {Honscheid}, K. and {Huterer}, D. and {Kannawadi}, A. and {Kuehn}, K. and {Kuropatkin}, N. and {Lahav}, O. and {Lima}, M. and {Maia}, M.~A.~G. and {Marshall}, J.~L. and {Menanteau}, F. and {Miquel}, R. and {Mohr}, J.~J. and {Morgan}, R. and {Muir}, J. and {Paz-Chinch{\'o}n}, F. and {Pieres}, A. and {Plazas Malag{\'o}n}, A.~A. and {Porredon}, A. and {Rodriguez-Monroy}, M. and {Roodman}, A. and {Sanchez}, E. and {Serrano}, S. and {Shan}, H. and {Suchyta}, E. and {Swanson}, M.~E.~C. and {Tarle}, G. and {Thomas}, D. and {To}, C. and {Zhang}, Y.},
 doi = {10.1093/mnras/stac2938},
 eprint = {2202.07440},
 journal = {\mnras},
 month = {January},
 number = {1},
 pages = {477-503},
 primaryclass = {astro-ph.CO},
 title = {{Consistent lensing and clustering in a low-S$_{8}$ Universe with BOSS, DES Year 3, HSC Year 1, and KiDS-1000}},
 volume = {518},
 year = {2023}
}

@article{2203.07128,
 adsurl = {https://ui.adsabs.harvard.edu/abs/2022MNRAS.515.1942D},
 archiveprefix = {arXiv},
 author = {{Doux}, C. and {Jain}, B. and {Zeurcher}, D. and {Lee}, J. and {Fang}, X. and {Rosenfeld}, R. and {Amon}, A. and {Camacho}, H. and {Choi}, A. and {Secco}, L.~F. and {Blazek}, J. and {Chang}, C. and {Gatti}, M. and {Gaztanaga}, E. and {Jeffrey}, N. and {Raveri}, M. and {Samuroff}, S. and {Alarcon}, A. and {Alves}, O. and {Andrade-Oliveira}, F. and {Baxter}, E. and {Bechtol}, K. and {Becker}, M.~R. and {Bernstein}, G.~M. and {Campos}, A. and {Carnero Rosell}, A. and {Carrasco Kind}, M. and {Cawthon}, R. and {Chen}, R. and {Cordero}, J. and {Crocce}, M. and {Davis}, C. and {DeRose}, J. and {Dodelson}, S. and {Drlica-Wagner}, A. and {Eckert}, K. and {Eifler}, T.~F. and {Elsner}, F. and {Elvin-Poole}, J. and {Everett}, S. and {Fert{\'e}}, A. and {Fosalba}, P. and {Friedrich}, O. and {Giannini}, G. and {Gruen}, D. and {Gruendl}, R.~A. and {Harrison}, I. and {Hartley}, W.~G. and {Herner}, K. and {Huang}, H. and {Huff}, E.~M. and {Huterer}, D. and {Jarvis}, M. and {Krause}, E. and {Kuropatkin}, N. and {Leget}, P.-F. and {Lemos}, P. and {Liddle}, A.~R. and {MacCrann}, N. and {McCullough}, J. and {Muir}, J. and {Myles}, J. and {Navarro-Alsina}, A. and {Pandey}, S. and {Park}, Y. and {Porredon}, A. and {Prat}, J. and {Rodriguez-Monroy}, M. and {Rollins}, R.~P. and {Roodman}, A. and {Ross}, A.~J. and {Rykoff}, E.~S. and {S{\'a}nchez}, C. and {Sanchez}, J. and {Sevilla-Noarbe}, I. and {Sheldon}, E. and {Shin}, T. and {Troja}, A. and {Troxel}, M.~A. and {Tutusaus}, I. and {Varga}, T.~N. and {Weaverdyck}, N. and {Wechsler}, R.~H. and {Yanny}, B. and {Yin}, B. and {Zhang}, Y. and {Zuntz}, J. and {Abbott}, T.~M.~C. and {Aguena}, M. and {Allam}, S. and {Annis}, J. and {Bacon}, D. and {Bertin}, E. and {Bocquet}, S. and {Brooks}, D. and {Burke}, D.~L. and {Carretero}, J. and {Costanzi}, M. and {da Costa}, L.~N. and {Pereira}, M.~E.~S. and {De Vicente}, J. and {Desai}, S. and {Diehl}, H.~T. and {Doel}, P. and {Ferrero}, I. and {Flaugher}, B. and {Frieman}, J. and {Garc{\'\i}a-Bellido}, J. and {Gerdes}, D.~W. and {Giannantonio}, T. and {Gschwend}, J. and {Gutierrez}, G. and {Hinton}, S.~R. and {Hollowood}, D.~L. and {Honscheid}, K. and {James}, D.~J. and {Kim}, A.~G. and {Kuehn}, K. and {Lahav}, O. and {Marshall}, J.~L. and {Menanteau}, F. and {Miquel}, R. and {Morgan}, R. and {Ogando}, R.~L.~C. and {Palmese}, A. and {Paz-Chinch{\'o}n}, F. and {Pieres}, A. and {Plazas Malag{\'o}n}, A.~A. and {Reil}, K. and {Sanchez}, E. and {Scarpine}, V. and {Serrano}, S. and {Smith}, M. and {Suchyta}, E. and {Swanson}, M.~E.~C. and {Tarle}, G. and {Thomas}, D. and {To}, C. and {Weller}, J. and {DES Collaboration}},
 doi = {10.1093/mnras/stac1826},
 eprint = {2203.07128},
 journal = {\mnras},
 month = {September},
 number = {2},
 pages = {1942-1972},
 primaryclass = {astro-ph.CO},
 title = {{Dark energy survey year 3 results: cosmological constraints from the analysis of cosmic shear in harmonic space}},
 volume = {515},
 year = {2022}
}

@article{2306.04024,
 adsurl = {https://ui.adsabs.harvard.edu/abs/2023MNRAS.526.4978S},
 archiveprefix = {arXiv},
 author = {{Schaye}, Joop and {Kugel}, Roi and {Schaller}, Matthieu and {Helly}, John C. and {Braspenning}, Joey and {Elbers}, Willem and {McCarthy}, Ian G. and {van Daalen}, Marcel P. and {Vandenbroucke}, Bert and {Frenk}, Carlos S. and {Kwan}, Juliana and {Salcido}, Jaime and {Bah{\'e}}, Yannick M. and {Borrow}, Josh and {Chaikin}, Evgenii and {Hahn}, Oliver and {Hu{\v{s}}ko}, Filip and {Jenkins}, Adrian and {Lacey}, Cedric G. and {Nobels}, Folkert S.~J.},
 doi = {10.1093/mnras/stad2419},
 eprint = {2306.04024},
 journal = {\mnras},
 month = {December},
 number = {4},
 pages = {4978-5020},
 primaryclass = {astro-ph.CO},
 title = {{The FLAMINGO project: cosmological hydrodynamical simulations for large-scale structure and galaxy cluster surveys}},
 volume = {526},
 year = {2023}
}

@article{2306.05492,
 adsurl = {https://ui.adsabs.harvard.edu/abs/2023MNRAS.526.6103K},
 archiveprefix = {arXiv},
 author = {{Kugel}, Roi and {Schaye}, Joop and {Schaller}, Matthieu and {Helly}, John C. and {Braspenning}, Joey and {Elbers}, Willem and {Frenk}, Carlos S. and {McCarthy}, Ian G. and {Kwan}, Juliana and {Salcido}, Jaime and {van Daalen}, Marcel P. and {Vandenbroucke}, Bert and {Bah{\'e}}, Yannick M. and {Borrow}, Josh and {Chaikin}, Evgenii and {Hu{\v{s}}ko}, Filip and {Jenkins}, Adrian and {Lacey}, Cedric G. and {Nobels}, Folkert S.~J. and {Vernon}, Ian},
 doi = {10.1093/mnras/stad2540},
 eprint = {2306.05492},
 journal = {\mnras},
 month = {December},
 number = {4},
 pages = {6103-6127},
 primaryclass = {astro-ph.CO},
 title = {{FLAMINGO: calibrating large cosmological hydrodynamical simulations with machine learning}},
 volume = {526},
 year = {2023}
}

@article{2407.21013,
 adsurl = {https://ui.adsabs.harvard.edu/abs/2025JCAP...01..028W},
 archiveprefix = {arXiv},
 author = {{Wolz}, Kevin and {Alonso}, David and {Nicola}, Andrina},
 doi = {10.1088/1475-7516/2025/01/028},
 eid = {028},
 eprint = {2407.21013},
 journal = {\jcap},
 month = {January},
 number = {1},
 pages = {028},
 primaryclass = {astro-ph.CO},
 title = {{Catalog-based pseudo-C$_{{\ensuremath{\ell}}}$ s}},
 volume = {2025},
 year = {2025}
}

@article{2410.19905,
 adsurl = {https://ui.adsabs.harvard.edu/abs/2025MNRAS.540..143M},
 archiveprefix = {arXiv},
 author = {{McCarthy}, Ian G. and {Amon}, Alexandra and {Schaye}, Joop and {Schaan}, Emmanuel and {Angulo}, Raul E. and {Salcido}, Jaime and {Schaller}, Matthieu and {Bigwood}, Leah and {Elbers}, Willem and {Kugel}, Roi and {Helly}, John C. and {Forouhar Moreno}, Victor J. and {Frenk}, Carlos S. and {McGibbon}, Robert J. and {Ondaro-Mallea}, Lurdes and {van Daalen}, Marcel P.},
 doi = {10.1093/mnras/staf731},
 eprint = {2410.19905},
 journal = {\mnras},
 month = {June},
 number = {1},
 pages = {143-163},
 primaryclass = {astro-ph.CO},
 title = {{FLAMINGO: combining kinetic SZ effect and galaxy-galaxy lensing measurements to gauge the impact of feedback on large-scale structure}},
 volume = {540},
 year = {2025}
}

@article{2502.08850,
 adsurl = {https://ui.adsabs.harvard.edu/abs/2025PhRvD.112h3561L},
 archiveprefix = {arXiv},
 author = {{Liu}, R. Henry and {Ferraro}, Simone and {Schaan}, Emmanuel and {Zhou}, Rongpu and {Aguilar}, Jessica Nicole and {Ahlen}, Steven and {Battaglia}, Nicholas and {Bianchi}, Davide and {Brooks}, David and {Claybaugh}, Todd and {Cole}, Shaun and {Coulton}, William R. and {de la Macorra}, Axel and {Dey}, Arjun and {Fanning}, Kevin and {Forero-Romero}, Jaime E. and {Gazta{\~n}aga}, Enrique and {Gong}, Yulin and {Gontcho}, Satya Gontcho A. and {Gruen}, Daniel and {Gutierrez}, Gaston and {Hadzhiyska}, Boryana and {Honscheid}, Klaus and {Howlett}, Cullan and {Kehoe}, Robert and {Kisner}, Theodore and {Kremin}, Anthony and {Kusiak}, Aleksandra and {Lambert}, Andrew and {Landriau}, Martin and {Le Guillou}, Laurent and {Levi}, Michael and {Lokken}, Martine and {Manera}, Marc and {Martini}, Paul and {Meisner}, Aaron and {Miquel}, Ramon and {Moodley}, Kavilan and {Newman}, Jeffrey A. and {Niz}, Gustavo and {Palanque-Delabrouille}, Nathalie and {Percival}, Will and {Prada}, Francisco and {P{\'e}rez-R{\`a}fols}, Ignasi and {Ried Guachalla}, Bernardita and {Rossi}, Graziano and {Sanchez}, Eusebio and {Schlegel}, David and {Schubnell}, Michael and {Seo}, Hee-Jong and {Sif{\'o}n}, Crist{\'o}bal and {Sprayberry}, David and {Tarl{\'e}}, Gregory and {Vavagiakis}, Eve M. and {Weaver}, Benjamin Alan and {Wollack}, Edward J. and {Zou}, Hu},
 doi = {10.1103/jqn8-19gx},
 eid = {083561},
 eprint = {2502.08850},
 journal = {\prd},
 month = {October},
 number = {8},
 pages = {083561},
 primaryclass = {astro-ph.CO},
 title = {{Measurements of the thermal Sunyaev-Zel'dovich effect with ACT and DESI luminous red galaxies}},
 volume = {112},
 year = {2025}
}

@article{2502.10232,
 adsurl = {https://ui.adsabs.harvard.edu/abs/2025MNRAS.540.1055E},
 archiveprefix = {arXiv},
 author = {{Efstathiou}, George and {McCarthy}, Fiona},
 doi = {10.1093/mnras/staf709},
 eprint = {2502.10232},
 journal = {\mnras},
 month = {June},
 number = {1},
 pages = {1055-1068},
 primaryclass = {astro-ph.CO},
 title = {{The power spectrum of the thermal Sunyaev-Zeldovich effect}},
 volume = {540},
 year = {2025}
}

@article{2604.24324,
 adsurl = {https://ui.adsabs.harvard.edu/abs/2026A&C....5701159H},
 archiveprefix = {arXiv},
 author = {{Helly}, John C. and {McGibbon}, Robert J. and {Schaye}, Joop and {Schaller}, Matthieu and {McDonald}, William and {Braspenning}, Joey and {Broxterman}, Jeger C. and {Costello}, Emily E. and {Elbers}, Willem and {Forouhar Moreno}, Victor J. and {Frenk}, Carlos S. and {Jenkins}, Adrian and {Kugel}, Roi and {McCarthy}, Ian G. and {Salcido}, Jaime and {van Daalen}, Marcel P. and {Vandenbroucke}, Bert and {Yang}, Tianyi},
 doi = {10.1016/j.ascom.2026.101159},
 eid = {101159},
 eprint = {2604.24324},
 journal = {Astronomy and Computing},
 month = {October},
 pages = {101159},
 primaryclass = {astro-ph.CO},
 title = {{The FLAMINGO simulations data release}},
 volume = {57},
 year = {2026}
}

@article{2605.11083,
 adsurl = {https://ui.adsabs.harvard.edu/abs/2026arXiv260511083S},
 archiveprefix = {arXiv},
 author = {{Salcido}, Jaime and {Yang}, Tianyi and {McCarthy}, Ian G. and {Costello}, Emily E. and {Conley}, Jonah T. and {Elbers}, Willem and {Frenk}, Carlos S. and {Schaller}, Matthieu and {Schaye}, Joop and {Upadhye}, Amol and {van Daalen}, Marcel P. and {Vandenbroucke}, Bert},
 doi = {10.48550/arXiv.2605.11083},
 eid = {arXiv:2605.11083},
 eprint = {2605.11083},
 journal = {arXiv e-prints},
 month = {May},
 pages = {arXiv:2605.11083},
 primaryclass = {astro-ph.CO},
 title = {{FLAMINGO: The thermal history of the Universe from tSZ effect cross-correlations and its dependencies on cosmology and baryon physics}},
 year = {2026}
}

@article{2606.28099,
 adsurl = {https://ui.adsabs.harvard.edu/abs/2026arXiv260628099Z},
 archiveprefix = {arXiv},
 author = {{Zhao}, Guandi and {Krolewski}, Alex and {Afshordi}, Niayesh},
 doi = {10.48550/arXiv.2606.28099},
 eid = {arXiv:2606.28099},
 eprint = {2606.28099},
 journal = {arXiv e-prints},
 month = {June},
 pages = {arXiv:2606.28099},
 primaryclass = {astro-ph.CO},
 title = {{Thermal Sunyaev-Zel'dovich cross-correlations with unWISE galaxies: disentangling radio contamination, dust properties, and electron pressure}},
 year = {2026}
}

@article{2607.14843,
 adsurl = {https://ui.adsabs.harvard.edu/abs/2026arXiv260714843W},
 archiveprefix = {arXiv},
 author = {{Wolz}, Kevin and {Farah}, Elyas and {Reischke}, Robert and {Alonso}, David and {Nicola}, Andrina},
 doi = {10.48550/arXiv.2607.14843},
 eid = {arXiv:2607.14843},
 eprint = {2607.14843},
 journal = {arXiv e-prints},
 month = {July},
 pages = {arXiv:2607.14843},
 primaryclass = {astro-ph.CO},
 title = {{Analytical covariances for catalogue-based pseudo-$C_\ell$s}},
 year = {2026}
}

@article{Abbott2022dark,
 adsurl = {https://ui.adsabs.harvard.edu/abs/2022PhRvD.105b3520A},
 archiveprefix = {arXiv},
 author = {{Abbott}, T.~M.~C. and {Aguena}, M. and {Alarcon}, A. and {Allam}, S. and {Alves}, O. and {Amon}, A. and {Andrade-Oliveira}, F. and {Annis}, J. and {Avila}, S. and {Bacon}, D. and {Baxter}, E. and {Bechtol}, K. and {Becker}, M.~R. and {Bernstein}, G.~M. and {Bhargava}, S. and {Birrer}, S. and {Blazek}, J. and {Brandao-Souza}, A. and {Bridle}, S.~L. and {Brooks}, D. and {Buckley-Geer}, E. and {Burke}, D.~L. and {Camacho}, H. and {Campos}, A. and {Carnero Rosell}, A. and {Carrasco Kind}, M. and {Carretero}, J. and {Castander}, F.~J. and {Cawthon}, R. and {Chang}, C. and {Chen}, A. and {Chen}, R. and {Choi}, A. and {Conselice}, C. and {Cordero}, J. and {Costanzi}, M. and {Crocce}, M. and {da Costa}, L.~N. and {da Silva Pereira}, M.~E. and {Davis}, C. and {Davis}, T.~M. and {De Vicente}, J. and {DeRose}, J. and {Desai}, S. and {Di Valentino}, E. and {Diehl}, H.~T. and {Dietrich}, J.~P. and {Dodelson}, S. and {Doel}, P. and {Doux}, C. and {Drlica-Wagner}, A. and {Eckert}, K. and {Eifler}, T.~F. and {Elsner}, F. and {Elvin-Poole}, J. and {Everett}, S. and {Evrard}, A.~E. and {Fang}, X. and {Farahi}, A. and {Fernandez}, E. and {Ferrero}, I. and {Fert{\'e}}, A. and {Fosalba}, P. and {Friedrich}, O. and {Frieman}, J. and {Garc{\'\i}a-Bellido}, J. and {Gatti}, M. and {Gaztanaga}, E. and {Gerdes}, D.~W. and {Giannantonio}, T. and {Giannini}, G. and {Gruen}, D. and {Gruendl}, R.~A. and {Gschwend}, J. and {Gutierrez}, G. and {Harrison}, I. and {Hartley}, W.~G. and {Herner}, K. and {Hinton}, S.~R. and {Hollowood}, D.~L. and {Honscheid}, K. and {Hoyle}, B. and {Huff}, E.~M. and {Huterer}, D. and {Jain}, B. and {James}, D.~J. and {Jarvis}, M. and {Jeffrey}, N. and {Jeltema}, T. and {Kovacs}, A. and {Krause}, E. and {Kron}, R. and {Kuehn}, K. and {Kuropatkin}, N. and {Lahav}, O. and {Leget}, P.-F. and {Lemos}, P. and {Liddle}, A.~R. and {Lidman}, C. and {Lima}, M. and {Lin}, H. and {MacCrann}, N. and {Maia}, M.~A.~G. and {Marshall}, J.~L. and {Martini}, P. and {McCullough}, J. and {Melchior}, P. and {Mena-Fern{\'a}ndez}, J. and {Menanteau}, F. and {Miquel}, R. and {Mohr}, J.~J. and {Morgan}, R. and {Muir}, J. and {Myles}, J. and {Nadathur}, S. and {Navarro-Alsina}, A. and {Nichol}, R.~C. and {Ogando}, R.~L.~C. and {Omori}, Y. and {Palmese}, A. and {Pandey}, S. and {Park}, Y. and {Paz-Chinch{\'o}n}, F. and {Petravick}, D. and {Pieres}, A. and {Plazas Malag{\'o}n}, A.~A. and {Porredon}, A. and {Prat}, J. and {Raveri}, M. and {Rodriguez-Monroy}, M. and {Rollins}, R.~P. and {Romer}, A.~K. and {Roodman}, A. and {Rosenfeld}, R. and {Ross}, A.~J. and {Rykoff}, E.~S. and {Samuroff}, S. and {S{\'a}nchez}, C. and {Sanchez}, E. and {Sanchez}, J. and {Sanchez Cid}, D. and {Scarpine}, V. and {Schubnell}, M. and {Scolnic}, D. and {Secco}, L.~F. and {Serrano}, S. and {Sevilla-Noarbe}, I. and {Sheldon}, E. and {Shin}, T. and {Smith}, M. and {Soares-Santos}, M. and {Suchyta}, E. and {Swanson}, M.~E.~C. and {Tabbutt}, M. and {Tarle}, G. and {Thomas}, D. and {To}, C. and {Troja}, A. and {Troxel}, M.~A. and {Tucker}, D.~L. and {Tutusaus}, I. and {Varga}, T.~N. and {Walker}, A.~R. and {Weaverdyck}, N. and {Wechsler}, R. and {Weller}, J. and {Yanny}, B. and {Yin}, B. and {Zhang}, Y. and {Zuntz}, J. and {DES Collaboration}},
 doi = {10.1103/PhysRevD.105.023520},
 eid = {023520},
 eprint = {2105.13549},
 journal = {\prd},
 month = {January},
 number = {2},
 pages = {023520},
 primaryclass = {astro-ph.CO},
 title = {{Dark Energy Survey Year 3 results: Cosmological constraints from galaxy clustering and weak lensing}},
 volume = {105},
 year = {2022}
}

@article{Ade2013planck,
 adsurl = {https://ui.adsabs.harvard.edu/abs/2014A&A...571A..30P},
 archiveprefix = {arXiv},
 author = {{Planck Collaboration} and {Ade}, P.~A.~R. and {Aghanim}, N. and {Armitage-Caplan}, C. and {Arnaud}, M. and {Ashdown}, M. and {Atrio-Barandela}, F. and {Aumont}, J. and {Baccigalupi}, C. and {Banday}, A.~J. and {Barreiro}, R.~B. and {Bartlett}, J.~G. and {Battaner}, E. and {Benabed}, K. and {Beno{\^\i}t}, A. and {Benoit-L{\'e}vy}, A. and {Bernard}, J.-P. and {Bersanelli}, M. and {Bethermin}, M. and {Bielewicz}, P. and {Blagrave}, K. and {Bobin}, J. and {Bock}, J.~J. and {Bonaldi}, A. and {Bond}, J.~R. and {Borrill}, J. and {Bouchet}, F.~R. and {Boulanger}, F. and {Bridges}, M. and {Bucher}, M. and {Burigana}, C. and {Butler}, R.~C. and {Cardoso}, J.-F. and {Catalano}, A. and {Challinor}, A. and {Chamballu}, A. and {Chen}, X. and {Chiang}, H.~C. and {Chiang}, L.-Y. and {Christensen}, P.~R. and {Church}, S. and {Clements}, D.~L. and {Colombi}, S. and {Colombo}, L.~P.~L. and {Couchot}, F. and {Coulais}, A. and {Crill}, B.~P. and {Curto}, A. and {Cuttaia}, F. and {Danese}, L. and {Davies}, R.~D. and {Davis}, R.~J. and {de Bernardis}, P. and {de Rosa}, A. and {de Zotti}, G. and {Delabrouille}, J. and {Delouis}, J.-M. and {D{\'e}sert}, F.-X. and {Dickinson}, C. and {Diego}, J.~M. and {Dole}, H. and {Donzelli}, S. and {Dor{\'e}}, O. and {Douspis}, M. and {Dupac}, X. and {Efstathiou}, G. and {En{\ss}lin}, T.~A. and {Eriksen}, H.~K. and {Finelli}, F. and {Forni}, O. and {Frailis}, M. and {Franceschi}, E. and {Galeotta}, S. and {Ganga}, K. and {Ghosh}, T. and {Giard}, M. and {Giraud-H{\'e}raud}, Y. and {Gonz{\'a}lez-Nuevo}, J. and {G{\'o}rski}, K.~M. and {Gratton}, S. and {Gregorio}, A. and {Gruppuso}, A. and {Hansen}, F.~K. and {Hanson}, D. and {Harrison}, D. and {Helou}, G. and {Henrot-Versill{\'e}}, S. and {Hern{\'a}ndez-Monteagudo}, C. and {Herranz}, D. and {Hildebrandt}, S.~R. and {Hivon}, E. and {Hobson}, M. and {Holmes}, W.~A. and {Hornstrup}, A. and {Hovest}, W. and {Huffenberger}, K.~M. and {Jaffe}, A.~H. and {Jaffe}, T.~R. and {Jones}, W.~C. and {Juvela}, M. and {Kalberla}, P. and {Keih{\"a}nen}, E. and {Kerp}, J. and {Keskitalo}, R. and {Kisner}, T.~S. and {Kneissl}, R. and {Knoche}, J. and {Knox}, L. and {Kunz}, M. and {Kurki-Suonio}, H. and {Lacasa}, F. and {Lagache}, G. and {L{\"a}hteenm{\"a}ki}, A. and {Lamarre}, J.-M. and {Langer}, M. and {Lasenby}, A. and {Laureijs}, R.~J. and {Lawrence}, C.~R. and {Leonardi}, R. and {Le{\'o}n-Tavares}, J. and {Lesgourgues}, J. and {Liguori}, M. and {Lilje}, P.~B. and {Linden-V{\o}rnle}, M. and {L{\'o}pez-Caniego}, M. and {Lubin}, P.~M. and {Mac{\'\i}as-P{\'e}rez}, J.~F. and {Maffei}, B. and {Maino}, D. and {Mandolesi}, N. and {Maris}, M. and {Marshall}, D.~J. and {Martin}, P.~G. and {Mart{\'\i}nez-Gonz{\'a}lez}, E. and {Masi}, S. and {Massardi}, M. and {Matarrese}, S. and {Matthai}, F. and {Mazzotta}, P. and {Melchiorri}, A. and {Mendes}, L. and {Mennella}, A. and {Migliaccio}, M. and {Mitra}, S. and {Miville-Desch{\^e}nes}, M.-A. and {Moneti}, A. and {Montier}, L. and {Morgante}, G. and {Mortlock}, D. and {Munshi}, D. and {Murphy}, J.~A. and {Naselsky}, P. and {Nati}, F. and {Natoli}, P. and {Netterfield}, C.~B. and {N{\o}rgaard-Nielsen}, H.~U. and {Noviello}, F. and {Novikov}, D. and {Novikov}, I. and {Osborne}, S. and {Oxborrow}, C.~A. and {Paci}, F. and {Pagano}, L. and {Pajot}, F. and {Paladini}, R. and {Paoletti}, D. and {Partridge}, B. and {Pasian}, F. and {Patanchon}, G. and {Perdereau}, O. and {Perotto}, L. and {Perrotta}, F. and {Piacentini}, F. and {Piat}, M. and {Pierpaoli}, E. and {Pietrobon}, D. and {Plaszczynski}, S. and {Pointecouteau}, E. and {Polenta}, G. and {Ponthieu}, N. and {Popa}, L. and {Poutanen}, T. and {Pratt}, G.~W. and {Pr{\'e}zeau}, G. and {Prunet}, S. and {Puget}, J.-L. and {Rachen}, J.~P. and {Reach}, W.~T. and {Rebolo}, R. and {Reinecke}, M. and {Remazeilles}, M. and {Renault}, C. and {Ricciardi}, S. and {Riller}, T. and {Ristorcelli}, I. and {Rocha}, G. and {Rosset}, C. and {Roudier}, G. and {Rowan-Robinson}, M. and {Rubi{\~n}o-Mart{\'\i}n}, J.~A.},
 doi = {10.1051/0004-6361/201322093},
 eid = {A30},
 eprint = {1309.0382},
 journal = {\aap},
 month = {November},
 pages = {A30},
 primaryclass = {astro-ph.CO},
 title = {{Planck 2013 results. XXX. Cosmic infrared background measurements and implications for star formation}},
 volume = {571},
 year = {2014}
}

@article{Aghanim2015planck,
 adsurl = {https://ui.adsabs.harvard.edu/abs/2016A&A...594A..22P},
 archiveprefix = {arXiv},
 author = {{Planck Collaboration} and {Aghanim}, N. and {Arnaud}, M. and {Ashdown}, M. and {Aumont}, J. and {Baccigalupi}, C. and {Banday}, A.~J. and {Barreiro}, R.~B. and {Bartlett}, J.~G. and {Bartolo}, N. and {Battaner}, E. and {Battye}, R. and {Benabed}, K. and {Beno{\^\i}t}, A. and {Benoit-L{\'e}vy}, A. and {Bernard}, J.-P. and {Bersanelli}, M. and {Bielewicz}, P. and {Bock}, J.~J. and {Bonaldi}, A. and {Bonavera}, L. and {Bond}, J.~R. and {Borrill}, J. and {Bouchet}, F.~R. and {Burigana}, C. and {Butler}, R.~C. and {Calabrese}, E. and {Cardoso}, J.-F. and {Catalano}, A. and {Challinor}, A. and {Chiang}, H.~C. and {Christensen}, P.~R. and {Churazov}, E. and {Clements}, D.~L. and {Colombo}, L.~P.~L. and {Combet}, C. and {Comis}, B. and {Coulais}, A. and {Crill}, B.~P. and {Curto}, A. and {Cuttaia}, F. and {Danese}, L. and {Davies}, R.~D. and {Davis}, R.~J. and {de Bernardis}, P. and {de Rosa}, A. and {de Zotti}, G. and {Delabrouille}, J. and {D{\'e}sert}, F.-X. and {Dickinson}, C. and {Diego}, J.~M. and {Dolag}, K. and {Dole}, H. and {Donzelli}, S. and {Dor{\'e}}, O. and {Douspis}, M. and {Ducout}, A. and {Dupac}, X. and {Efstathiou}, G. and {Elsner}, F. and {En{\ss}lin}, T.~A. and {Eriksen}, H.~K. and {Fergusson}, J. and {Finelli}, F. and {Forni}, O. and {Frailis}, M. and {Fraisse}, A.~A. and {Franceschi}, E. and {Frejsel}, A. and {Galeotta}, S. and {Galli}, S. and {Ganga}, K. and {G{\'e}nova-Santos}, R.~T. and {Giard}, M. and {Gonz{\'a}lez-Nuevo}, J. and {G{\'o}rski}, K.~M. and {Gregorio}, A. and {Gruppuso}, A. and {Gudmundsson}, J.~E. and {Hansen}, F.~K. and {Harrison}, D.~L. and {Henrot-Versill{\'e}}, S. and {Hern{\'a}ndez-Monteagudo}, C. and {Herranz}, D. and {Hildebrandt}, S.~R. and {Hivon}, E. and {Holmes}, W.~A. and {Hornstrup}, A. and {Huffenberger}, K.~M. and {Hurier}, G. and {Jaffe}, A.~H. and {Jones}, W.~C. and {Juvela}, M. and {Keih{\"a}nen}, E. and {Keskitalo}, R. and {Kneissl}, R. and {Knoche}, J. and {Kunz}, M. and {Kurki-Suonio}, H. and {Lacasa}, F. and {Lagache}, G. and {L{\"a}hteenm{\"a}ki}, A. and {Lamarre}, J.-M. and {Lasenby}, A. and {Lattanzi}, M. and {Leonardi}, R. and {Lesgourgues}, J. and {Levrier}, F. and {Liguori}, M. and {Lilje}, P.~B. and {Linden-V{\o}rnle}, M. and {L{\'o}pez-Caniego}, M. and {Mac{\'\i}as-P{\'e}rez}, J.~F. and {Maffei}, B. and {Maggio}, G. and {Maino}, D. and {Mandolesi}, N. and {Mangilli}, A. and {Maris}, M. and {Martin}, P.~G. and {Mart{\'\i}nez-Gonz{\'a}lez}, E. and {Masi}, S. and {Matarrese}, S. and {Melchiorri}, A. and {Melin}, J.-B. and {Migliaccio}, M. and {Miville-Desch{\^e}nes}, M.-A. and {Moneti}, A. and {Montier}, L. and {Morgante}, G. and {Mortlock}, D. and {Munshi}, D. and {Murphy}, J.~A. and {Naselsky}, P. and {Nati}, F. and {Natoli}, P. and {Noviello}, F. and {Novikov}, D. and {Novikov}, I. and {Paci}, F. and {Pagano}, L. and {Pajot}, F. and {Paoletti}, D. and {Pasian}, F. and {Patanchon}, G. and {Perdereau}, O. and {Perotto}, L. and {Pettorino}, V. and {Piacentini}, F. and {Piat}, M. and {Pierpaoli}, E. and {Pietrobon}, D. and {Plaszczynski}, S. and {Pointecouteau}, E. and {Polenta}, G. and {Ponthieu}, N. and {Pratt}, G.~W. and {Prunet}, S. and {Puget}, J.-L. and {Rachen}, J.~P. and {Reinecke}, M. and {Remazeilles}, M. and {Renault}, C. and {Renzi}, A. and {Ristorcelli}, I. and {Rocha}, G. and {Rossetti}, M. and {Roudier}, G. and {Rubi{\~n}o-Mart{\'\i}n}, J.~A. and {Rusholme}, B. and {Sandri}, M. and {Santos}, D. and {Sauv{\'e}}, A. and {Savelainen}, M. and {Savini}, G. and {Scott}, D. and {Spencer}, L.~D. and {Stolyarov}, V. and {Stompor}, R. and {Sunyaev}, R. and {Sutton}, D. and {Suur-Uski}, A.-S. and {Sygnet}, J.-F. and {Tauber}, J.~A. and {Terenzi}, L. and {Toffolatti}, L. and {Tomasi}, M. and {Tramonte}, D. and {Tristram}, M. and {Tucci}, M. and {Tuovinen}, J. and {Valenziano}, L. and {Valiviita}, J. and {Van Tent}, B. and {Vielva}, P. and {Villa}, F. and {Wade}, L.~A. and {Wandelt}, B.~D. and {Wehus}, I.~K. and {Yvon}, D.},
 doi = {10.1051/0004-6361/201525826},
 eid = {A22},
 eprint = {1502.01596},
 journal = {\aap},
 month = {September},
 pages = {A22},
 primaryclass = {astro-ph.CO},
 title = {{Planck 2015 results. XXII. A map of the thermal Sunyaev-Zeldovich effect}},
 volume = {594},
 year = {2016}
}

@article{Akrami2020planck,
 adsurl = {https://ui.adsabs.harvard.edu/abs/2020A&A...643A..42P},
 archiveprefix = {arXiv},
 author = {{Planck Collaboration} and {Akrami}, Y. and {Andersen}, K.~J. and {Ashdown}, M. and {Baccigalupi}, C. and {Ballardini}, M. and {Banday}, A.~J. and {Barreiro}, R.~B. and {Bartolo}, N. and {Basak}, S. and {Benabed}, K. and {Bernard}, J.-P. and {Bersanelli}, M. and {Bielewicz}, P. and {Bond}, J.~R. and {Borrill}, J. and {Burigana}, C. and {Butler}, R.~C. and {Calabrese}, E. and {Casaponsa}, B. and {Chiang}, H.~C. and {Colombo}, L.~P.~L. and {Combet}, C. and {Crill}, B.~P. and {Cuttaia}, F. and {de Bernardis}, P. and {de Rosa}, A. and {de Zotti}, G. and {Delabrouille}, J. and {Di Valentino}, E. and {Diego}, J.~M. and {Dor{\'e}}, O. and {Douspis}, M. and {Dupac}, X. and {Eriksen}, H.~K. and {Fernandez-Cobos}, R. and {Finelli}, F. and {Frailis}, M. and {Fraisse}, A.~A. and {Franceschi}, E. and {Frolov}, A. and {Galeotta}, S. and {Galli}, S. and {Ganga}, K. and {Gerbino}, M. and {Ghosh}, T. and {Gonz{\'a}lez-Nuevo}, J. and {G{\'o}rski}, K.~M. and {Gruppuso}, A. and {Gudmundsson}, J.~E. and {Handley}, W. and {Helou}, G. and {Herranz}, D. and {Hildebrandt}, S.~R. and {Hivon}, E. and {Huang}, Z. and {Jaffe}, A.~H. and {Jones}, W.~C. and {Keih{\"a}nen}, E. and {Keskitalo}, R. and {Kiiveri}, K. and {Kim}, J. and {Kisner}, T.~S. and {Krachmalnicoff}, N. and {Kunz}, M. and {Kurki-Suonio}, H. and {Lasenby}, A. and {Lattanzi}, M. and {Lawrence}, C.~R. and {Le Jeune}, M. and {Levrier}, F. and {Liguori}, M. and {Lilje}, P.~B. and {Lilley}, M. and {Lindholm}, V. and {L{\'o}pez-Caniego}, M. and {Lubin}, P.~M. and {Mac{\'\i}as-P{\'e}rez}, J.~F. and {Maino}, D. and {Mandolesi}, N. and {Marcos-Caballero}, A. and {Maris}, M. and {Martin}, P.~G. and {Mart{\'\i}nez-Gonz{\'a}lez}, E. and {Matarrese}, S. and {Mauri}, N. and {McEwen}, J.~D. and {Meinhold}, P.~R. and {Mennella}, A. and {Migliaccio}, M. and {Mitra}, S. and {Molinari}, D. and {Montier}, L. and {Morgante}, G. and {Moss}, A. and {Natoli}, P. and {Paoletti}, D. and {Partridge}, B. and {Patanchon}, G. and {Pearson}, D. and {Pearson}, T.~J. and {Perrotta}, F. and {Piacentini}, F. and {Polenta}, G. and {Rachen}, J.~P. and {Reinecke}, M. and {Remazeilles}, M. and {Renzi}, A. and {Rocha}, G. and {Rosset}, C. and {Roudier}, G. and {Rubi{\~n}o-Mart{\'\i}n}, J.~A. and {Ruiz-Granados}, B. and {Salvati}, L. and {Savelainen}, M. and {Scott}, D. and {Sirignano}, C. and {Sirri}, G. and {Spencer}, L.~D. and {Suur-Uski}, A.-S. and {Svalheim}, L.~T. and {Tauber}, J.~A. and {Tavagnacco}, D. and {Tenti}, M. and {Terenzi}, L. and {Thommesen}, H. and {Toffolatti}, L. and {Tomasi}, M. and {Tristram}, M. and {Trombetti}, T. and {Valiviita}, J. and {Van Tent}, B. and {Vielva}, P. and {Villa}, F. and {Vittorio}, N. and {Wandelt}, B.~D. and {Wehus}, I.~K. and {Zacchei}, A. and {Zonca}, A.},
 doi = {10.1051/0004-6361/202038073},
 eid = {A42},
 eprint = {2007.04997},
 journal = {\aap},
 month = {November},
 pages = {A42},
 primaryclass = {astro-ph.CO},
 title = {{Planck intermediate results. LVII. Joint Planck LFI and HFI data processing}},
 volume = {643},
 year = {2020}
}

@article{Amon2022,
 adsurl = {https://ui.adsabs.harvard.edu/abs/2022MNRAS.516.5355A},
 archiveprefix = {arXiv},
 author = {{Amon}, Alexandra and {Efstathiou}, George},
 doi = {10.1093/mnras/stac2429},
 eprint = {2206.11794},
 journal = {\mnras},
 month = {November},
 number = {4},
 pages = {5355-5366},
 primaryclass = {astro-ph.CO},
 title = {{A non-linear solution to the S$_{8}$ tension?}},
 volume = {516},
 year = {2022}
}

@article{Arico2023,
 adsurl = {https://ui.adsabs.harvard.edu/abs/2023A&A...678A.109A},
 archiveprefix = {arXiv},
 author = {{Aric{\`o}}, Giovanni and {Angulo}, Raul E. and {Zennaro}, Matteo and {Contreras}, Sergio and {Chen}, Angela and {Hern{\'a}ndez-Monteagudo}, Carlos},
 doi = {10.1051/0004-6361/202346539},
 eid = {A109},
 eprint = {2303.05537},
 journal = {\aap},
 month = {October},
 pages = {A109},
 primaryclass = {astro-ph.CO},
 title = {{DES Y3 cosmic shear down to small scales: Constraints on cosmology and baryons}},
 volume = {678},
 year = {2023}
}

@article{astro-ph/0205468,
 adsurl = {https://ui.adsabs.harvard.edu/abs/2002MNRAS.336.1256K},
 archiveprefix = {arXiv},
 author = {{Komatsu}, E. and {Seljak}, U.},
 doi = {10.1046/j.1365-8711.2002.05889.x},
 eprint = {astro-ph/0205468},
 journal = {\mnras},
 month = {November},
 number = {4},
 pages = {1256-1270},
 primaryclass = {astro-ph},
 title = {{The Sunyaev-Zel'dovich angular power spectrum as a probe of cosmological parameters}},
 volume = {336},
 year = {2002}
}

@article{astro-ph/9808050,
 adsurl = {https://ui.adsabs.harvard.edu/abs/1999PhR...310...97B},
 archiveprefix = {arXiv},
 author = {{Birkinshaw}, M.},
 doi = {10.1016/S0370-1573(98)00080-5},
 eprint = {astro-ph/9808050},
 journal = {\physrep},
 month = {March},
 number = {2-3},
 pages = {97-195},
 primaryclass = {astro-ph},
 title = {{The Sunyaev-Zel'dovich effect}},
 volume = {310},
 year = {1999}
}

@article{Barreira2018accurate,
       author = {{Barreira}, Alexandre and {Krause}, Elisabeth and {Schmidt}, Fabian},
        title = "{Accurate cosmic shear errors: do we need ensembles of simulations?}",
      journal = {\jcap},
         year = 2018,
        month = oct,
       volume = {2018},
       number = {10},
          eid = {053},
        pages = {053},
          doi = {10.1088/1475-7516/2018/10/053},
archivePrefix = {arXiv},
       eprint = {1807.04266},
 primaryClass = {astro-ph.CO},
       adsurl = {https://ui.adsabs.harvard.edu/abs/2018JCAP...10..053B}
}

@article{Bartelmann2001weak,
 adsurl = {https://ui.adsabs.harvard.edu/abs/2001PhR...340..291B},
 archiveprefix = {arXiv},
 author = {{Bartelmann}, M. and {Schneider}, P.},
 doi = {10.1016/S0370-1573(00)00082-X},
 eprint = {astro-ph/9912508},
 journal = {\physrep},
 month = {January},
 number = {4-5},
 pages = {291-472},
 primaryclass = {astro-ph},
 title = {{Weak gravitational lensing}},
 volume = {340},
 year = {2001}
}

@article{Bethermin2013redshift,
 adsurl = {https://ui.adsabs.harvard.edu/abs/2013A&A...557A..66B},
 archiveprefix = {arXiv},
 author = {{B{\'e}thermin}, Matthieu and {Wang}, Lingyu and {Dor{\'e}}, Olivier and {Lagache}, Guilaine and {Sargent}, Mark and {Daddi}, Emanuele and {Cousin}, Morgane and {Aussel}, Herv{\'e}},
 doi = {10.1051/0004-6361/201321688},
 eid = {A66},
 eprint = {1304.3936},
 journal = {\aap},
 month = {September},
 pages = {A66},
 primaryclass = {astro-ph.CO},
 title = {{The redshift evolution of the distribution of star formation among dark matter halos as seen in the infrared}},
 volume = {557},
 year = {2013}
}

@article{Bethermin2014evolution,
 adsurl = {https://ui.adsabs.harvard.edu/abs/2015A&A...573A.113B},
 archiveprefix = {arXiv},
 author = {{B{\'e}thermin}, Matthieu and {Daddi}, Emanuele and {Magdis}, Georgios and {Lagos}, Claudia and {Sargent}, Mark and {Albrecht}, Marcus and {Aussel}, Herv{\'e} and {Bertoldi}, Frank and {Buat}, V{\'e}ronique and {Galametz}, Maud and {Heinis}, S{\'e}bastien and {Ilbert}, Olivier and {Karim}, Alexander and {Koekemoer}, Anton and {Lacey}, Cedric and {Le Floc'h}, Emeric and {Navarrete}, Felipe and {Pannella}, Maurilio and {Schreiber}, Corentin and {Smol{\v{c}}i{\'c}}, Vernesa and {Symeonidis}, Myrto and {Viero}, Marco},
 doi = {10.1051/0004-6361/201425031},
 eid = {A113},
 eprint = {1409.5796},
 journal = {\aap},
 month = {January},
 pages = {A113},
 primaryclass = {astro-ph.GA},
 title = {{Evolution of the dust emission of massive galaxies up to z = 4 and constraints on their dominant mode of star formation}},
 volume = {573},
 year = {2015}
}

@article{Bethermin2017impact,
 adsurl = {https://ui.adsabs.harvard.edu/abs/2017A&A...607A..89B},
 archiveprefix = {arXiv},
 author = {{B{\'e}thermin}, Matthieu and {Wu}, Hao-Yi and {Lagache}, Guilaine and {Davidzon}, Iary and {Ponthieu}, Nicolas and {Cousin}, Morgane and {Wang}, Lingyu and {Dor{\'e}}, Olivier and {Daddi}, Emanuele and {Lapi}, Andrea},
 doi = {10.1051/0004-6361/201730866},
 eid = {A89},
 eprint = {1703.08795},
 journal = {\aap},
 month = {November},
 pages = {A89},
 primaryclass = {astro-ph.GA},
 title = {{The impact of clustering and angular resolution on far-infrared and millimeter continuum observations}},
 volume = {607},
 year = {2017}
}

@ARTICLE{Magdis2020GN20,
       author = {{Cortzen}, Isabella and {Magdis}, Georgios E. and {Valentino}, Francesco and {Daddi}, Emanuele and {Liu}, Daizhong and {Rigopoulou}, Dimitra and {Sargent}, Mark and {Riechers}, Dominik and {Cormier}, Diane and {Hodge}, Jacqueline A. and {Walter}, Fabian and {Elbaz}, David and {B{\'e}thermin}, Matthieu and {Greve}, Thomas R. and {Kokorev}, Vasily and {Toft}, Sune},
        title = "{Deceptively cold dust in the massive starburst galaxy GN20 at z {\ensuremath{\sim}} 4}",
      journal = {\aap},
         year = 2020,
        month = feb,
       volume = {634},
          eid = {L14},
        pages = {L14},
          doi = {10.1051/0004-6361/201937217},
archivePrefix = {arXiv},
       eprint = {2002.02974},
 primaryClass = {astro-ph.GA},
       adsurl = {https://ui.adsabs.harvard.edu/abs/2020A&A...634L..14C}
}

@article{Chabrier2003,
 adsurl = {https://ui.adsabs.harvard.edu/abs/2003PASP..115..763C},
 archiveprefix = {arXiv},
 author = {{Chabrier}, Gilles},
 doi = {10.1086/376392},
 eprint = {astro-ph/0304382},
 journal = {\pasp},
 month = {July},
 number = {809},
 pages = {763-795},
 primaryclass = {astro-ph},
 title = {{Galactic Stellar and Substellar Initial Mass Function}},
 volume = {115},
 year = {2003}
}

@article{Chandran2023improved,
 adsurl = {https://ui.adsabs.harvard.edu/abs/2023MNRAS.526.5682C},
 archiveprefix = {arXiv},
 author = {{Chandran}, Jyothis and {Remazeilles}, Mathieu and {Barreiro}, R.~B.},
 doi = {10.1093/mnras/stad3156},
 eprint = {2305.10193},
 journal = {\mnras},
 month = {December},
 number = {4},
 pages = {5682-5698},
 primaryclass = {astro-ph.CO},
 title = {{An improved Compton parameter map of thermal Sunyaev-Zeldovich effect from Planck PR4 data}},
 volume = {526},
 year = {2023}
}

@article{Chen2022,
 adsurl = {https://ui.adsabs.harvard.edu/abs/2023MNRAS.518.5340C},
 archiveprefix = {arXiv},
 author = {{Chen}, A. and {Aric{\`o}}, G. and {Huterer}, D. and {Angulo}, R.~E. and {Weaverdyck}, N. and {Friedrich}, O. and {Secco}, L.~F. and {Hern{\'a}ndez-Monteagudo}, C. and {Alarcon}, A. and {Alves}, O. and {Amon}, A. and {Andrade-Oliveira}, F. and {Baxter}, E. and {Bechtol}, K. and {Becker}, M.~R. and {Bernstein}, G.~M. and {Blazek}, J. and {Brandao-Souza}, A. and {Bridle}, S.~L. and {Camacho}, H. and {Campos}, A. and {Carnero Rosell}, A. and {Carrasco Kind}, M. and {Cawthon}, R. and {Chang}, C. and {Chen}, R. and {Chintalapati}, P. and {Choi}, A. and {Cordero}, J. and {Crocce}, M. and {Pereira}, M.~E.~S. and {Davis}, C. and {DeRose}, J. and {Di Valentino}, E. and {Diehl}, H.~T. and {Dodelson}, S. and {Doux}, C. and {Drlica-Wagner}, A. and {Eckert}, K. and {Eifler}, T.~F. and {Elsner}, F. and {Elvin-Poole}, J. and {Everett}, S. and {Fang}, X. and {Fert{\'e}}, A. and {Fosalba}, P. and {Gatti}, M. and {Gaztanaga}, E. and {Giannini}, G. and {Gruen}, D. and {Gruendl}, R.~A. and {Harrison}, I. and {Hartley}, W.~G. and {Herner}, K. and {Hoffmann}, K. and {Huang}, H. and {Huff}, E.~M. and {Jain}, B. and {Jarvis}, M. and {Jeffrey}, N. and {Kacprzak}, T. and {Krause}, E. and {Kuropatkin}, N. and {Leget}, P.-F. and {Lemos}, P. and {Liddle}, A.~R. and {MacCrann}, N. and {McCullough}, J. and {Muir}, J. and {Myles}, J. and {Navarro-Alsina}, A. and {Omori}, Y. and {Pandey}, S. and {Park}, Y. and {Porredon}, A. and {Prat}, J. and {Raveri}, M. and {Refregier}, A. and {Rollins}, R.~P. and {Roodman}, A. and {Rosenfeld}, R. and {Ross}, A.~J. and {Rykoff}, E.~S. and {Samuroff}, S. and {S{\'a}nchez}, C. and {Sanchez}, J. and {Sevilla-Noarbe}, I. and {Sheldon}, E. and {Shin}, T. and {Troja}, A. and {Troxel}, M.~A. and {Tutusaus}, I. and {Varga}, T.~N. and {Wechsler}, R.~H. and {Yanny}, B. and {Yin}, B. and {Zhang}, Y. and {Zuntz}, J. and {Aguena}, M. and {Annis}, J. and {Bacon}, D. and {Bertin}, E. and {Bocquet}, S. and {Brooks}, D. and {Burke}, D.~L. and {Carretero}, J. and {Conselice}, C. and {Costanzi}, M. and {da Costa}, L.~N. and {De Vicente}, J. and {Desai}, S. and {Doel}, P. and {Ferrero}, I. and {Flaugher}, B. and {Frieman}, J. and {Garc{\'\i}a-Bellido}, J. and {Gerdes}, D.~W. and {Giannantonio}, T. and {Gschwend}, J. and {Gutierrez}, G. and {Hinton}, S.~R. and {Hollowood}, D.~L. and {Honscheid}, K. and {James}, D.~J. and {Kuehn}, K. and {Lahav}, O. and {March}, M. and {Marshall}, J.~L. and {Melchior}, P. and {Menanteau}, F. and {Miquel}, R. and {Mohr}, J.~J. and {Morgan}, R. and {Paz-Chinch{\'o}n}, F. and {Pieres}, A. and {Sanchez}, E. and {Smith}, M. and {Suchyta}, E. and {Swanson}, M.~E.~C. and {Tarle}, G. and {Thomas}, D. and {To}, C. and {DES Collaboration}},
 doi = {10.1093/mnras/stac3213},
 eprint = {2206.08591},
 journal = {\mnras},
 month = {February},
 number = {4},
 pages = {5340-5355},
 primaryclass = {astro-ph.CO},
 title = {{Constraining the baryonic feedback with cosmic shear using the DES Year-3 small-scale measurements}},
 volume = {518},
 year = {2023}
}

@article{Chisari2019core,
 adsurl = {https://ui.adsabs.harvard.edu/abs/2019ApJS..242....2C},
 archiveprefix = {arXiv},
 author = {{Chisari}, Nora Elisa and {Alonso}, David and {Krause}, Elisabeth and {Leonard}, C. Danielle and {Bull}, Philip and {Neveu}, J{\'e}r{\'e}my and {Villarreal}, Antonia Sierra and {Singh}, Sukhdeep and {McClintock}, Thomas and {Ellison}, John and {Du}, Zilong and {Zuntz}, Joe and {Mead}, Alexander and {Joudaki}, Shahab and {Lorenz}, Christiane S. and {Tr{\"o}ster}, Tilman and {Sanchez}, Javier and {Lanusse}, Francois and {Ishak}, Mustapha and {Hlozek}, Ren{\'e}e and {Blazek}, Jonathan and {Campagne}, Jean-Eric and {Almoubayyed}, Husni and {Eifler}, Tim and {Kirby}, Matthew and {Kirkby}, David and {Plaszczynski}, St{\'e}phane and {Slosar}, An{\v{z}}e and {Vrastil}, Michal and {Wagoner}, Erika L. and {LSST Dark Energy Science Collaboration}},
 doi = {10.3847/1538-4365/ab1658},
 eid = {2},
 eprint = {1812.05995},
 journal = {\apjs},
 month = {May},
 number = {1},
 pages = {2},
 primaryclass = {astro-ph.CO},
 title = {{Core Cosmology Library: Precision Cosmological Predictions for LSST}},
 volume = {242},
 year = {2019}
}

@article{Ciesla2014dust,
 adsurl = {https://ui.adsabs.harvard.edu/abs/2014A&A...565A.128C},
 archiveprefix = {arXiv},
 author = {{Ciesla}, L. and {Boquien}, M. and {Boselli}, A. and {Buat}, V. and {Cortese}, L. and {Bendo}, G.~J. and {Heinis}, S. and {Galametz}, M. and {Eales}, S. and {Smith}, M.~W.~L. and {Baes}, M. and {Bianchi}, S. and {De Looze}, I. and {di Serego Alighieri}, S. and {Galliano}, F. and {Hughes}, T.~M. and {Madden}, S.~C. and {Pierini}, D. and {R{\'e}my-Ruyer}, A. and {Spinoglio}, L. and {Vaccari}, M. and {Viaene}, S. and {Vlahakis}, C.},
 doi = {10.1051/0004-6361/201323248},
 eid = {A128},
 eprint = {1402.3597},
 journal = {\aap},
 month = {May},
 pages = {A128},
 primaryclass = {astro-ph.GA},
 title = {{Dust spectral energy distributions of nearby galaxies: an insight from the Herschel Reference Survey}},
 volume = {565},
 year = {2014}
}

@article{Coulton2024act,
 adsurl = {https://ui.adsabs.harvard.edu/abs/2024PhRvD.109f3530C},
 archiveprefix = {arXiv},
 author = {{Coulton}, William and {Madhavacheril}, Mathew S. and {Duivenvoorden}, Adriaan J. and {Hill}, J. Colin and {Abril-Cabezas}, Irene and {Ade}, Peter A.~R. and {Aiola}, Simone and {Alford}, Tommy and {Amiri}, Mandana and {Amodeo}, Stefania and {An}, Rui and {Atkins}, Zachary and {Austermann}, Jason E. and {Battaglia}, Nicholas and {Battistelli}, Elia Stefano and {Beall}, James A. and {Bean}, Rachel and {Beringue}, Benjamin and {Bhandarkar}, Tanay and {Biermann}, Emily and {Bolliet}, Boris and {Bond}, J. Richard and {Cai}, Hongbo and {Calabrese}, Erminia and {Calafut}, Victoria and {Capalbo}, Valentina and {Carrero}, Felipe and {Chesmore}, Grace E. and {Cho}, Hsiao-mei and {Choi}, Steve K. and {Clark}, Susan E. and {Rosado}, Rodrigo C{\'o}rdova and {Cothard}, Nicholas F. and {Coughlin}, Kevin and {Crowley}, Kevin T. and {Devlin}, Mark J. and {Dicker}, Simon and {Doze}, Peter and {Duell}, Cody J. and {Duff}, Shannon M. and {Dunkley}, Jo and {D{\"u}nner}, Rolando and {Fanfani}, Valentina and {Fankhanel}, Max and {Farren}, Gerrit and {Ferraro}, Simone and {Freundt}, Rodrigo and {Fuzia}, Brittany and {Gallardo}, Patricio A. and {Garrido}, Xavier and {Givans}, Jahmour and {Gluscevic}, Vera and {Golec}, Joseph E. and {Guan}, Yilun and {Halpern}, Mark and {Han}, Dongwon and {Hasselfield}, Matthew and {Healy}, Erin and {Henderson}, Shawn and {Hensley}, Brandon and {Herv{\'\i}as-Caimapo}, Carlos and {Hilton}, Gene C. and {Hilton}, Matt and {Hincks}, Adam D. and {Hlo{\v{z}}ek}, Ren{\'e}e and {Ho}, Shuay-Pwu Patty and {Huber}, Zachary B. and {Hubmayr}, Johannes and {Huffenberger}, Kevin M. and {Hughes}, John P. and {Irwin}, Kent and {Isopi}, Giovanni and {Jense}, Hidde T. and {Keller}, Ben and {Kim}, Joshua and {Knowles}, Kenda and {Koopman}, Brian J. and {Kosowsky}, Arthur and {Kramer}, Darby and {Kusiak}, Aleksandra and {La Posta}, Adrien and {Lakey}, Victoria and {Lee}, Eunseong and {Li}, Zack and {Li}, Yaqiong and {Limon}, Michele and {Lokken}, Martine and {Louis}, Thibaut and {Lungu}, Marius and {MacCrann}, Niall and {MacInnis}, Amanda and {Maldonado}, Diego and {Maldonado}, Felipe and {Mallaby-Kay}, Maya and {Marques}, Gabriela A. and {van Marrewijk}, Joshiwa and {McCarthy}, Fiona and {McMahon}, Jeff and {Mehta}, Yogesh and {Menanteau}, Felipe and {Moodley}, Kavilan and {Morris}, Thomas W. and {Mroczkowski}, Tony and {Naess}, Sigurd and {Namikawa}, Toshiya and {Nati}, Federico and {Newburgh}, Laura and {Nicola}, Andrina and {Niemack}, Michael D. and {Nolta}, Michael R. and {Orlowski-Scherer}, John and {Page}, Lyman A. and {Pandey}, Shivam and {Partridge}, Bruce and {Prince}, Heather and {Puddu}, Roberto and {Qu}, Frank J. and {Radiconi}, Federico and {Robertson}, Naomi and {Rojas}, Felipe and {Sakuma}, Tai and {Salatino}, Maria and {Schaan}, Emmanuel and {Schmitt}, Benjamin L. and {Sehgal}, Neelima and {Shaikh}, Shabbir and {Sherwin}, Blake D. and {Sierra}, Carlos and {Sievers}, Jon and {Sif{\'o}n}, Crist{\'o}bal and {Simon}, Sara and {Sonka}, Rita and {Spergel}, David N. and {Staggs}, Suzanne T. and {Storer}, Emilie and {Switzer}, Eric R. and {Tampier}, Niklas and {Thornton}, Robert and {Trac}, Hy and {Treu}, Jesse and {Tucker}, Carole and {Ullom}, Joel and {Vale}, Leila R. and {Van Engelen}, Alexander and {Van Lanen}, Jeff and {Vargas}, Cristian and {Vavagiakis}, Eve M. and {Wagoner}, Kasey and {Wang}, Yuhan and {Wenzl}, Lukas and {Wollack}, Edward J. and {Xu}, Zhilei and {Zago}, Fernando and {Zheng}, Kaiwen},
 doi = {10.1103/PhysRevD.109.063530},
 eid = {063530},
 eprint = {2307.01258},
 journal = {\prd},
 month = {March},
 number = {6},
 pages = {063530},
 primaryclass = {astro-ph.CO},
 title = {{Atacama Cosmology Telescope: High-resolution component-separated maps across one third of the sky}},
 volume = {109},
 year = {2024}
}

@article{Cunha2010exploring,
 adsurl = {https://ui.adsabs.harvard.edu/abs/2010A&A...523A..78D},
 archiveprefix = {arXiv},
 author = {{da Cunha}, E. and {Charmandaris}, V. and {D{\'\i}az-Santos}, T. and {Armus}, L. and {Marshall}, J.~A. and {Elbaz}, D.},
 doi = {10.1051/0004-6361/201014498},
 eid = {A78},
 eprint = {1008.2000},
 journal = {\aap},
 month = {November},
 pages = {A78},
 primaryclass = {astro-ph.CO},
 title = {{Exploring the physical properties of local star-forming ULIRGs from the ultraviolet to the infrared}},
 volume = {523},
 year = {2010}
}

@article{DES2020dark,
 adsurl = {https://ui.adsabs.harvard.edu/abs/2021MNRAS.504.4312G},
 archiveprefix = {arXiv},
 author = {{Gatti}, M. and {Sheldon}, E. and {Amon}, A. and {Becker}, M. and {Troxel}, M. and {Choi}, A. and {Doux}, C. and {MacCrann}, N. and {Navarro-Alsina}, A. and {Harrison}, I. and {Gruen}, D. and {Bernstein}, G. and {Jarvis}, M. and {Secco}, L.~F. and {Fert{\'e}}, A. and {Shin}, T. and {McCullough}, J. and {Rollins}, R.~P. and {Chen}, R. and {Chang}, C. and {Pandey}, S. and {Tutusaus}, I. and {Prat}, J. and {Elvin-Poole}, J. and {Sanchez}, C. and {Plazas}, A.~A. and {Roodman}, A. and {Zuntz}, J. and {Abbott}, T.~M.~C. and {Aguena}, M. and {Allam}, S. and {Annis}, J. and {Avila}, S. and {Bacon}, D. and {Bertin}, E. and {Bhargava}, S. and {Brooks}, D. and {Burke}, D.~L. and {Carnero Rosell}, A. and {Carrasco Kind}, M. and {Carretero}, J. and {Castander}, F.~J. and {Conselice}, C. and {Costanzi}, M. and {Crocce}, M. and {da Costa}, L.~N. and {Davis}, T.~M. and {De Vicente}, J. and {Desai}, S. and {Diehl}, H.~T. and {Dietrich}, J.~P. and {Doel}, P. and {Drlica-Wagner}, A. and {Eckert}, K. and {Everett}, S. and {Ferrero}, I. and {Frieman}, J. and {Garc{\'\i}a-Bellido}, J. and {Gerdes}, D.~W. and {Giannantonio}, T. and {Gruendl}, R.~A. and {Gschwend}, J. and {Gutierrez}, G. and {Hartley}, W.~G. and {Hinton}, S.~R. and {Hollowood}, D.~L. and {Honscheid}, K. and {Hoyle}, B. and {Huff}, E.~M. and {Huterer}, D. and {Jain}, B. and {James}, D.~J. and {Jeltema}, T. and {Krause}, E. and {Kron}, R. and {Kuropatkin}, N. and {Lima}, M. and {Maia}, M.~A.~G. and {Marshall}, J.~L. and {Miquel}, R. and {Morgan}, R. and {Myles}, J. and {Palmese}, A. and {Paz-Chinch{\'o}n}, F. and {Rykoff}, E.~S. and {Samuroff}, S. and {Sanchez}, E. and {Scarpine}, V. and {Schubnell}, M. and {Serrano}, S. and {Sevilla-Noarbe}, I. and {Smith}, M. and {Soares-Santos}, M. and {Suchyta}, E. and {Swanson}, M.~E.~C. and {Tarle}, G. and {Thomas}, D. and {To}, C. and {Tucker}, D.~L. and {Varga}, T.~N. and {Wechsler}, R.~H. and {Weller}, J. and {Wester}, W. and {Wilkinson}, R.~D.},
 doi = {10.1093/mnras/stab918},
 eprint = {2011.03408},
 journal = {\mnras},
 month = {July},
 number = {3},
 pages = {4312-4336},
 primaryclass = {astro-ph.CO},
 title = {{Dark energy survey year 3 results: weak lensing shape catalogue}},
 volume = {504},
 year = {2021}
}

@article{Garcia-Garcia2021growth,
       author = {{Garc{\'\i}a-Garc{\'\i}a}, Carlos and {Ruiz-Zapatero}, Jaime and {Alonso}, David and {Bellini}, Emilio and {Ferreira}, Pedro G. and {Mueller}, Eva-Maria and {Nicola}, Andrina and {Ruiz-Lapuente}, Pilar},
        title = "{The growth of density perturbations in the last  10 billion years from tomographic large-scale structure data}",
      journal = {\jcap},
         year = 2021,
        month = oct,
       volume = {2021},
       number = {10},
          eid = {030},
        pages = {030},
          doi = {10.1088/1475-7516/2021/10/030},
archivePrefix = {arXiv},
       eprint = {2105.12108},
 primaryClass = {astro-ph.CO},
       adsurl = {https://ui.adsabs.harvard.edu/abs/2021JCAP...10..030G}
}

@article{Garcia-Garcia2024cosmic,
 adsurl = {https://ui.adsabs.harvard.edu/abs/2024JCAP...08..024G},
 archiveprefix = {arXiv},
 author = {{Garc{\'\i}a-Garc{\'\i}a}, Carlos and {Zennaro}, Matteo and {Aric{\`o}}, Giovanni and {Alonso}, David and {Angulo}, Raul E.},
 doi = {10.1088/1475-7516/2024/08/024},
 eid = {024},
 eprint = {2403.13794},
 journal = {\jcap},
 month = {August},
 number = {8},
 pages = {024},
 primaryclass = {astro-ph.CO},
 title = {{Cosmic shear with small scales: DES-Y3, KiDS-1000 and HSC-DR1}},
 volume = {2024},
 year = {2024}
}

@article{Gatti2022,
 adsurl = {https://ui.adsabs.harvard.edu/abs/2022PhRvD.105l3525G},
 archiveprefix = {arXiv},
 author = {{Gatti}, M. and {Pandey}, S. and {Baxter}, E. and {Hill}, J.~C. and {Moser}, E. and {Raveri}, M. and {Fang}, X. and {DeRose}, J. and {Giannini}, G. and {Doux}, C. and {Huang}, H. and {Battaglia}, N. and {Alarcon}, A. and {Amon}, A. and {Becker}, M. and {Campos}, A. and {Chang}, C. and {Chen}, R. and {Choi}, A. and {Eckert}, K. and {Elvin-Poole}, J. and {Everett}, S. and {Ferte}, A. and {Harrison}, I. and {Maccrann}, N. and {Mccullough}, J. and {Myles}, J. and {Navarro Alsina}, A. and {Prat}, J. and {Rollins}, R.~P. and {Sanchez}, C. and {Shin}, T. and {Troxel}, M. and {Tutusaus}, I. and {Yin}, B. and {Abbott}, T. and {Aguena}, M. and {Allam}, S. and {Andrade-Oliveira}, F. and {Annis}, J. and {Bernstein}, G. and {Bertin}, E. and {Bolliet}, B. and {Bond}, J.~R. and {Brooks}, D. and {Burke}, D.~L. and {Calabrese}, E. and {Carnero Rosell}, A. and {Carrasco Kind}, M. and {Carretero}, J. and {Cawthon}, R. and {Costanzi}, M. and {Crocce}, M. and {da Costa}, L.~N. and {da Silva Pereira}, M.~E. and {De Vicente}, J. and {Desai}, S. and {Diehl}, H.~T. and {Dietrich}, J.~P. and {Doel}, P. and {Dunkley}, J. and {Evrard}, A.~E. and {Ferraro}, S. and {Ferrero}, I. and {Flaugher}, B. and {Fosalba}, P. and {Frieman}, J. and {Garc{\'\i}a-Bellido}, J. and {Gaztanaga}, E. and {Gerdes}, D.~W. and {Giannantonio}, T. and {Gruen}, D. and {Gruendl}, R.~A. and {Gschwend}, J. and {Gutierrez}, G. and {Herner}, K. and {Hincks}, A.~D. and {Hinton}, S.~R. and {Hollowood}, D.~L. and {Honscheid}, K. and {Hughes}, J.~P. and {Huterer}, D. and {Jain}, B. and {James}, D.~J. and {Krause}, E. and {Kuehn}, K. and {Kuropatkin}, N. and {Lahav}, O. and {Lidman}, C. and {Lima}, M. and {Lokken}, M. and {Madhavacheril}, M.~S. and {Maia}, M.~A.~G. and {Marshall}, J.~L. and {Mcmahon}, J.~J. and {Melchior}, P. and {Moodley}, K. and {Mohr}, J.~J. and {Morgan}, R. and {Nati}, F. and {Niemack}, M.~D. and {Page}, L. and {Palmese}, A. and {Paz-Chinch{\'o}n}, F. and {Pieres}, A. and {Plazas Malag{\'o}n}, A.~A. and {Rodriguez-Monroy}, M. and {Romer}, A.~K. and {Sanchez}, E. and {Scarpine}, V. and {Schaan}, E. and {Secco}, L.~F. and {Serrano}, S. and {Sheldon}, E. and {Sherwin}, B.~D. and {Sif{\'o}n}, C. and {Smith}, M. and {Soares-Santos}, M. and {Spergel}, D. and {Suchyta}, E. and {Tarle}, G. and {Thomas}, D. and {To}, C. and {Tucker}, D.~L. and {Varga}, T.~N. and {Weller}, J. and {Wilkinson}, R.~D. and {Wollack}, E.~J. and {Xu}, Z. and {DES} and {ACT Collaboration}},
 doi = {10.1103/PhysRevD.105.123525},
 eid = {123525},
 eprint = {2108.01600},
 journal = {\prd},
 month = {June},
 number = {12},
 pages = {123525},
 primaryclass = {astro-ph.CO},
 title = {{Cross-correlation of Dark Energy Survey Year 3 lensing data with ACT and Planck thermal Sunyaev-Zel'dovich effect observations. I. Measurements, systematics tests, and feedback model constraints}},
 volume = {105},
 year = {2022}
}

@article{Gorski2004healpix,
 adsurl = {https://ui.adsabs.harvard.edu/abs/2005ApJ...622..759G},
 archiveprefix = {arXiv},
 author = {{G{\'o}rski}, K.~M. and {Hivon}, E. and {Banday}, A.~J. and {Wandelt}, B.~D. and {Hansen}, F.~K. and {Reinecke}, M. and {Bartelmann}, M.},
 doi = {10.1086/427976},
 eprint = {astro-ph/0409513},
 journal = {\apj},
 month = {April},
 number = {2},
 pages = {759-771},
 primaryclass = {astro-ph},
 title = {{HEALPix: A Framework for High-Resolution Discretization and Fast Analysis of Data Distributed on the Sphere}},
 volume = {622},
 year = {2005}
}

@article{Hamimeche2008likelihood,
 author = {Hamimeche, Samira and Lewis, Antony},
 journal = {Phys. Rev. D},
 pages = {103013},
 title = {{Likelihood Analysis of CMB power spectra}},
 volume = {77},
 year = {2008}
}

@article{Jego2022star,
 adsurl = {https://ui.adsabs.harvard.edu/abs/2023MNRAS.520.1895J},
 archiveprefix = {arXiv},
 author = {{Jego}, Baptiste and {Ruiz-Zapatero}, Jaime and {Garc{\'\i}a-Garc{\'\i}a}, Carlos and {Koukoufilippas}, Nick and {Alonso}, David},
 doi = {10.1093/mnras/stad213},
 eprint = {2206.15394},
 journal = {\mnras},
 month = {April},
 number = {2},
 pages = {1895-1912},
 primaryclass = {astro-ph.GA},
 title = {{The star-formation history in the last 10 billion years from CIB cross-correlations}},
 volume = {520},
 year = {2023}
}

@article{Kennicutt1998,
 adsurl = {https://ui.adsabs.harvard.edu/abs/1998ARA&A..36..189K},
 archiveprefix = {arXiv},
 author = {{Kennicutt}, Jr., Robert C.},
 doi = {10.1146/annurev.astro.36.1.189},
 eprint = {astro-ph/9807187},
 journal = {\araa},
 month = {January},
 pages = {189-232},
 primaryclass = {astro-ph},
 title = {{Star Formation in Galaxies Along the Hubble Sequence}},
 volume = {36},
 year = {1998}
}

@article{Kennicutt2012,
 adsurl = {https://ui.adsabs.harvard.edu/abs/2012ARA&A..50..531K},
 archiveprefix = {arXiv},
 author = {{Kennicutt}, Robert C. and {Evans}, Neal J.},
 doi = {10.1146/annurev-astro-081811-125610},
 eprint = {1204.3552},
 journal = {\araa},
 month = {September},
 pages = {531-608},
 primaryclass = {astro-ph.GA},
 title = {{Star Formation in the Milky Way and Nearby Galaxies}},
 volume = {50},
 year = {2012}
}

@article{Kilbinger2017precision,
 adsurl = {https://ui.adsabs.harvard.edu/abs/2017MNRAS.472.2126K},
 archiveprefix = {arXiv},
 author = {{Kilbinger}, Martin and {Heymans}, Catherine and {Asgari}, Marika and {Joudaki}, Shahab and {Schneider}, Peter and {Simon}, Patrick and {Van Waerbeke}, Ludovic and {Harnois-D{\'e}raps}, Joachim and {Hildebrandt}, Hendrik and {K{\"o}hlinger}, Fabian and {Kuijken}, Konrad and {Viola}, Massimo},
 doi = {10.1093/mnras/stx2082},
 eprint = {1702.05301},
 journal = {\mnras},
 month = {December},
 number = {2},
 pages = {2126-2141},
 primaryclass = {astro-ph.CO},
 title = {{Precision calculations of the cosmic shear power spectrum projection}},
 volume = {472},
 year = {2017}
}

@article{Knox2001probing,
 adsurl = {https://ui.adsabs.harvard.edu/abs/2001ApJ...550....7K},
 archiveprefix = {arXiv},
 author = {{Knox}, Lloyd and {Cooray}, Asantha and {Eisenstein}, Daniel and {Haiman}, Zoltan},
 doi = {10.1086/319732},
 eprint = {astro-ph/0009151},
 journal = {\apj},
 month = {March},
 number = {1},
 pages = {7-20},
 primaryclass = {astro-ph},
 title = {{Probing Early Structure Formation with Far-Infrared Background Correlations}},
 volume = {550},
 year = {2001}
}

@article{Krause2016cosmolike,
 author = {Krause, Elisabeth and Eifler, Tim},
 journal = {Mon. Not. Roy. Astron. Soc.},
 pages = {2100--2112},
 title = {{cosmolike: cosmological likelihood analyses}},
 volume = {470},
 year = {2017}
}

@article{LaPosta2024insights,
 adsurl = {https://ui.adsabs.harvard.edu/abs/2025PhRvD.112d3525L},
 archiveprefix = {arXiv},
 author = {{La Posta}, Adrien and {Alonso}, David and {Chisari}, Nora Elisa and {Ferreira}, Tassia and {Garc{\'\i}a-Garc{\'\i}a}, Carlos},
 doi = {10.1103/m77z-w7pl},
 eid = {043525},
 eprint = {2412.12081},
 journal = {\prd},
 month = {August},
 number = {4},
 pages = {043525},
 primaryclass = {astro-ph.CO},
 title = {{Insights on gas thermodynamics from the combination of x-ray and thermal Sunyaev-Zel'dovich data cross correlated with cosmic shear}},
 volume = {112},
 year = {2025}
}

@article{LaPosta2026joint,
 adsurl = {https://ui.adsabs.harvard.edu/abs/2026arXiv260304269L},
 archiveprefix = {arXiv},
 author = {{La Posta}, Adrien and {Alonso}, David and {Garc{\'\i}a-Garc{\'\i}a}, Carlos and {Maleubre}, Sara},
 doi = {10.48550/arXiv.2603.04269},
 eid = {arXiv:2603.04269},
 eprint = {2603.04269},
 journal = {arXiv e-prints},
 month = {March},
 pages = {arXiv:2603.04269},
 primaryclass = {astro-ph.CO},
 title = {{Joint tomographic measurement of thermal Sunyaev Zeldovich and the cosmic infrared background}},
 year = {2026}
}

@article{Limber1954,
 adsurl = {https://ui.adsabs.harvard.edu/abs/1954ApJ...119..655L},
 author = {{Limber}, D. Nelson},
 doi = {10.1086/145870},
 journal = {\apj},
 month = {May},
 pages = {655},
 title = {{The Analysis of Counts of the Extragalactic Nebulae in Terms of a Fluctuating Density Field. II.}},
 volume = {119},
 year = {1954}
}

@article{LoVerde2008,
 adsurl = {https://ui.adsabs.harvard.edu/abs/2008PhRvD..78l3506L},
 archiveprefix = {arXiv},
 author = {{LoVerde}, Marilena and {Afshordi}, Niayesh},
 doi = {10.1103/PhysRevD.78.123506},
 eid = {123506},
 eprint = {0809.5112},
 journal = {\prd},
 month = {December},
 number = {12},
 pages = {123506},
 primaryclass = {astro-ph},
 title = {{Extended Limber approximation}},
 volume = {78},
 year = {2008}
}

@article{Magdis2012evolving,
 adsurl = {https://ui.adsabs.harvard.edu/abs/2012ApJ...760....6M},
 archiveprefix = {arXiv},
 author = {{Magdis}, Georgios E. and {Daddi}, E. and {B{\'e}thermin}, M. and {Sargent}, M. and {Elbaz}, D. and {Pannella}, M. and {Dickinson}, M. and {Dannerbauer}, H. and {da Cunha}, E. and {Walter}, F. and {Rigopoulou}, D. and {Charmandaris}, V. and {Hwang}, H.~S. and {Kartaltepe}, J.},
 doi = {10.1088/0004-637X/760/1/6},
 eid = {6},
 eprint = {1210.1035},
 journal = {\apj},
 month = {November},
 number = {1},
 pages = {6},
 primaryclass = {astro-ph.CO},
 title = {{The Evolving Interstellar Medium of Star-forming Galaxies since z = 2 as Probed by Their Infrared Spectral Energy Distributions}},
 volume = {760},
 year = {2012}
}

@article{Maleubre_inprep,
 author = {{Maleubre}, Sara and {Alonso}, David and {others}},
 journal = {},
 title = {{In preparation}},
 year = {2026}
}

@article{Mandelbaum2018weak,
 adsurl = {https://ui.adsabs.harvard.edu/abs/2018ARA&A..56..393M},
 archiveprefix = {arXiv},
 author = {{Mandelbaum}, Rachel},
 doi = {10.1146/annurev-astro-081817-051928},
 eprint = {1710.03235},
 journal = {\araa},
 month = {September},
 pages = {393-433},
 primaryclass = {astro-ph.CO},
 title = {{Weak Lensing for Precision Cosmology}},
 volume = {56},
 year = {2018}
}

@article{Maniyar2026spt,
 adsurl = {https://ui.adsabs.harvard.edu/abs/2026PhRvD.114b3534M},
 archiveprefix = {arXiv},
 author = {{Maniyar}, A.~S. and {Bianchini}, F. and {Wu}, W.~L.~K. and {Raghunathan}, S. and {Anderson}, A.~J. and {Ansarinejad}, B. and {Archipley}, M. and {Balkenhol}, L. and {Barron}, D.~R. and {Barry}, P.~S. and {Benabed}, K. and {Bender}, A.~N. and {Benson}, B.~A. and {Bleem}, L.~E. and {Bocquet}, S. and {Bouchet}, F.~R. and {Bryant}, L. and {Camphuis}, E. and {Campitiello}, M.~G. and {Carlstrom}, J.~E. and {Carron}, J. and {Chang}, C.~L. and {Chaubal}, P. and {Chichura}, P.~M. and {Chokshi}, A. and {Chou}, T.-L. and {Coerver}, A. and {Crawford}, T.~M. and {Daley}, C. and {de Haan}, T. and {Dibert}, K.~R. and {Dobbs}, M.~A. and {Doohan}, M. and {Doussot}, A. and {Dutcher}, D. and {Everett}, W. and {Feng}, C. and {Ferguson}, K.~R. and {Ferree}, N.~C. and {Fichman}, K. and {Foster}, A. and {Galli}, S. and {Gambrel}, A.~E. and {Gao}, A.~K. and {Gardner}, R.~W. and {Ge}, F. and {Goeckner-Wald}, N. and {Gualtieri}, R. and {Guidi}, F. and {Guns}, S. and {Halverson}, N.~W. and {Hivon}, E. and {Ho}, A.~Y.~Q. and {Holder}, G.~P. and {Holzapfel}, W.~L. and {Hood}, J.~C. and {Hryciuk}, A. and {Huang}, N. and {Jhaveri}, T. and {K{\'e}ruzor{\'e}}, F. and {Khalife}, A.~R. and {Knox}, L. and {Korman}, M. and {Kornoelje}, K. and {Kuo}, C.-L. and {Levy}, K. and {Li}, Y. and {Lowitz}, A.~E. and {Lu}, C. and {Lynch}, G.~P. and {Maccarone}, T.~J. and {Martsen}, E.~S. and {Menanteau}, F. and {Millea}, M. and {Montgomery}, J. and {Nakato}, Y. and {Natoli}, T. and {Noble}, G.~I. and {Omori}, Y. and {Ouellette}, A. and {Pan}, Z. and {Paschos}, P. and {Phadke}, K.~A. and {Pollak}, A.~W. and {Prabhu}, K. and {Quan}, W. and {Rahimi}, M. and {Rahlin}, A. and {Reichardt}, C.~L. and {Rouble}, M. and {Ruhl}, J.~E. and {Schiappucci}, E. and {Oliveira}, A.~C. Silva and {Simpson}, A. and {Sobrin}, J.~A. and {Stark}, A.~A. and {Stephen}, J. and {Tandoi}, C. and {Thorne}, B. and {Trendafilova}, C. and {Umilta}, C. and {Vieira}, J.~D. and {Vieregg}, A.~G. and {Vitrier}, A. and {Wan}, Y. and {Whitehorn}, N. and {Young}, M.~R. and {Zebrowski}, J.~A. and {SPT-3G Collaboration}},
 doi = {10.1103/7mdh-jpd3},
 eid = {023534},
 eprint = {2602.11279},
 journal = {\prd},
 month = {July},
 number = {2},
 pages = {023534},
 primaryclass = {astro-ph.CO},
 title = {{SPT-3G D1: Compton-y maps using data from the SPT-3G and Planck surveys}},
 volume = {114},
 year = {2026}
}

@article{McCarthy2023,
 adsurl = {https://ui.adsabs.harvard.edu/abs/2023MNRAS.526.5494M},
 archiveprefix = {arXiv},
 author = {{McCarthy}, Ian G. and {Salcido}, Jaime and {Schaye}, Joop and {Kwan}, Juliana and {Elbers}, Willem and {Kugel}, Roi and {Schaller}, Matthieu and {Helly}, John C. and {Braspenning}, Joey and {Frenk}, Carlos S. and {van Daalen}, Marcel P. and {Vandenbroucke}, Bert and {Conley}, Jonah T. and {Font}, Andreea S. and {Upadhye}, Amol},
 doi = {10.1093/mnras/stad3107},
 eprint = {2309.07959},
 journal = {\mnras},
 month = {December},
 number = {4},
 pages = {5494-5519},
 primaryclass = {astro-ph.CO},
 title = {{The FLAMINGO project: revisiting the S$_{8}$ tension and the role of baryonic physics}},
 volume = {526},
 year = {2023}
}

@article{McCarthy2024component,
 adsurl = {https://ui.adsabs.harvard.edu/abs/2024PhRvD.109b3528M},
 archiveprefix = {arXiv},
 author = {{McCarthy}, Fiona and {Hill}, J. Colin},
 doi = {10.1103/PhysRevD.109.023528},
 eid = {023528},
 eprint = {2307.01043},
 journal = {\prd},
 month = {January},
 number = {2},
 pages = {023528},
 primaryclass = {astro-ph.CO},
 title = {{Component-separated, CIB-cleaned thermal Sunyaev-Zel'dovich maps from Planck PR4 data with a flexible public needlet ILC pipeline}},
 volume = {109},
 year = {2024}
}

@article{Nicola2020cosmic,
       author = {{Nicola}, Andrina and {Garc{\'\i}a-Garc{\'\i}a}, Carlos and {Alonso}, David and {Dunkley}, Jo and {Ferreira}, Pedro G. and {Slosar}, An{\v{z}}e and {Spergel}, David N.},
        title = "{Cosmic shear power spectra in practice}",
      journal = {\jcap},
         year = 2021,
        month = mar,
       volume = {2021},
       number = {3},
          eid = {067},
        pages = {067},
          doi = {10.1088/1475-7516/2021/03/067},
archivePrefix = {arXiv},
       eprint = {2010.09717},
 primaryClass = {astro-ph.CO},
       adsurl = {https://ui.adsabs.harvard.edu/abs/2021JCAP...03..067N}
}

@article{Osato2019,
 adsurl = {https://ui.adsabs.harvard.edu/abs/2020MNRAS.492.4780O},
 archiveprefix = {arXiv},
 author = {{Osato}, Ken and {Shirasaki}, Masato and {Miyatake}, Hironao and {Nagai}, Daisuke and {Yoshida}, Naoki and {Oguri}, Masamune and {Takahashi}, Ryuichi},
 doi = {10.1093/mnras/staa117},
 eprint = {1910.07526},
 journal = {\mnras},
 month = {March},
 number = {4},
 pages = {4780-4804},
 primaryclass = {astro-ph.CO},
 title = {{Cross-correlation of the thermal Sunyaev-Zel'dovich effect and weak gravitational lensing: Planck and Subaru Hyper Suprime-Cam first-year data}},
 volume = {492},
 year = {2020}
}

@article{Pandey2022,
 adsurl = {https://ui.adsabs.harvard.edu/abs/2022PhRvD.105l3526P},
 archiveprefix = {arXiv},
 author = {{Pandey}, S. and {Gatti}, M. and {Baxter}, E. and {Hill}, J.~C. and {Fang}, X. and {Doux}, C. and {Giannini}, G. and {Raveri}, M. and {DeRose}, J. and {Huang}, H. and {Moser}, E. and {Battaglia}, N. and {Alarcon}, A. and {Amon}, A. and {Becker}, M. and {Campos}, A. and {Chang}, C. and {Chen}, R. and {Choi}, A. and {Eckert}, K. and {Elvin-Poole}, J. and {Everett}, S. and {Ferte}, A. and {Harrison}, I. and {Maccrann}, N. and {Mccullough}, J. and {Myles}, J. and {Navarro Alsina}, A. and {Prat}, J. and {Rollins}, R.~P. and {Sanchez}, C. and {Shin}, T. and {Troxel}, M. and {Tutusaus}, I. and {Yin}, B. and {Aguena}, M. and {Allam}, S. and {Andrade-Oliveira}, F. and {Bernstein}, G.~M. and {Bertin}, E. and {Bolliet}, B. and {Bond}, J.~R. and {Brooks}, D. and {Calabrese}, E. and {Carnero Rosell}, A. and {Carrasco Kind}, M. and {Carretero}, J. and {Cawthon}, R. and {Costanzi}, M. and {Crocce}, M. and {da Costa}, L.~N. and {Pereira}, M.~E.~S. and {De Vicente}, J. and {Desai}, S. and {Diehl}, H.~T. and {Dietrich}, J.~P. and {Doel}, P. and {Dunkley}, J. and {Everett}, S. and {Evrard}, A.~E. and {Ferraro}, S. and {Ferrero}, I. and {Flaugher}, B. and {Fosalba}, P. and {Garc{\'\i}a-Bellido}, J. and {Gaztanaga}, E. and {Gerdes}, D.~W. and {Giannantonio}, T. and {Gruen}, D. and {Gruendl}, R.~A. and {Gschwend}, J. and {Gutierrez}, G. and {Herner}, K. and {Hincks}, A.~D. and {Hinton}, S.~R. and {Hollowood}, D.~L. and {Honscheid}, K. and {Hughes}, J.~P. and {Huterer}, D. and {Jain}, B. and {James}, D.~J. and {Jeltema}, T. and {Krause}, E. and {Kuehn}, K. and {Lahav}, O. and {Lima}, M. and {Lokken}, M. and {Madhavacheril}, M.~S. and {Maia}, M.~A.~G. and {Mcmahon}, J.~J. and {Melchior}, P. and {Menanteau}, F. and {Miquel}, R. and {Mohr}, J.~J. and {Moodley}, K. and {Morgan}, R. and {Nati}, F. and {Niemack}, M.~D. and {Page}, L. and {Palmese}, A. and {Paz-Chinch{\'o}n}, F. and {Pieres}, A. and {Plazas Malag{\'o}n}, A.~A. and {Rodriguez-Monroy}, M. and {Romer}, A.~K. and {Sanchez}, E. and {Scarpine}, V. and {Schaan}, E. and {Serrano}, S. and {Sevilla-Noarbe}, I. and {Sheldon}, E. and {Sherwin}, B.~D. and {Sif{\'o}n}, C. and {Smith}, M. and {Soares-Santos}, M. and {Spergel}, D. and {Suchyta}, E. and {Swanson}, M.~E.~C. and {Tarle}, G. and {Thomas}, D. and {To}, C. and {Varga}, T.~N. and {Weller}, J. and {Wollack}, E.~J. and {Xu}, Z. and {DES} and {ACT Collaboration}},
 doi = {10.1103/PhysRevD.105.123526},
 eid = {123526},
 eprint = {2108.01601},
 journal = {\prd},
 month = {June},
 number = {12},
 pages = {123526},
 primaryclass = {astro-ph.CO},
 title = {{Cross-correlation of Dark Energy Survey Year 3 lensing data with ACT and P l a n c k thermal Sunyaev-Zel'dovich effect observations. II. Modeling and constraints on halo pressure profiles}},
 volume = {105},
 year = {2022}
}

@article{Partridge1967are,
 adsurl = {https://ui.adsabs.harvard.edu/abs/1967ApJ...147..868P},
 author = {{Partridge}, R.~B. and {Peebles}, P.~J.~E.},
 doi = {10.1086/149079},
 journal = {\apj},
 month = {March},
 pages = {868},
 title = {{Are Young Galaxies Visible?}},
 volume = {147},
 year = {1967}
}

@article{ralp_inprep,
 author = {{Dunsdon}, Reggie and {La Posta}, Adrien and {Alonso}, David and {others}},
 journal = {},
 title = {{In preparation}},
 year = {2026}
}

@article{Sargent2012,
 adsurl = {https://ui.adsabs.harvard.edu/abs/2012ApJ...747L..31S},
 archiveprefix = {arXiv},
 author = {{Sargent}, M.~T. and {B{\'e}thermin}, M. and {Daddi}, E. and {Elbaz}, D.},
 doi = {10.1088/2041-8205/747/2/L31},
 eid = {L31},
 eprint = {1202.0290},
 journal = {\apjl},
 month = {March},
 number = {2},
 pages = {L31},
 primaryclass = {astro-ph.CO},
 title = {{The Contribution of Starbursts and Normal Galaxies to Infrared Luminosity Functions at z < 2}},
 volume = {747},
 year = {2012}
}

@article{Schneider2021,
 adsurl = {https://ui.adsabs.harvard.edu/abs/2022MNRAS.514.3802S},
 archiveprefix = {arXiv},
 author = {{Schneider}, Aurel and {Giri}, Sambit K. and {Amodeo}, Stefania and {Refregier}, Alexandre},
 doi = {10.1093/mnras/stac1493},
 eprint = {2110.02228},
 journal = {\mnras},
 month = {August},
 number = {3},
 pages = {3802-3814},
 primaryclass = {astro-ph.CO},
 title = {{Constraining baryonic feedback and cosmology with weak-lensing, X-ray, and kinematic Sunyaev-Zeldovich observations}},
 volume = {514},
 year = {2022}
}

@article{Sellentin2017insufficiency,
 author = {Sellentin, Elena and Heavens, Alan F.},
 journal = {Mon. Not. Roy. Astron. Soc.},
 pages = {2355--2363},
 title = {{Non-Gaussian weak lensing likelihoods}},
 volume = {473},
 year = {2018}
}

@article{Speagle2014,
 adsurl = {https://ui.adsabs.harvard.edu/abs/2014ApJS..214...15S},
 archiveprefix = {arXiv},
 author = {{Speagle}, J.~S. and {Steinhardt}, C.~L. and {Capak}, P.~L. and {Silverman}, J.~D.},
 doi = {10.1088/0067-0049/214/2/15},
 eid = {15},
 eprint = {1405.2041},
 journal = {\apjs},
 month = {October},
 number = {2},
 pages = {15},
 primaryclass = {astro-ph.GA},
 title = {{A Highly Consistent Framework for the Evolution of the Star-Forming ``Main Sequence'' from z \raisebox{-0.5ex}\textasciitilde 0-6}},
 volume = {214},
 year = {2014}
}

@article{Sunyaev1972observations,
 adsurl = {https://ui.adsabs.harvard.edu/abs/1972CoASP...4..173S},
 author = {{Sunyaev}, R.~A. and {Zeldovich}, Ya. B.},
 journal = {Comments on Astrophysics and Space Physics},
 month = {November},
 pages = {173},
 title = {{The Observations of Relic Radiation as a Test of the Nature of X-Ray Radiation from the Clusters of Galaxies}},
 volume = {4},
 year = {1972}
}

@article{Troster2022,
 adsurl = {https://ui.adsabs.harvard.edu/abs/2022A&A...660A..27T},
 archiveprefix = {arXiv},
 author = {{Tr{\"o}ster}, Tilman and {Mead}, Alexander J. and {Heymans}, Catherine and {Yan}, Ziang and {Alonso}, David and {Asgari}, Marika and {Bilicki}, Maciej and {Dvornik}, Andrej and {Hildebrandt}, Hendrik and {Joachimi}, Benjamin and {Kannawadi}, Arun and {Kuijken}, Konrad and {Schneider}, Peter and {Shan}, Huan Yuan and {van Waerbeke}, Ludovic and {Wright}, Angus H.},
 doi = {10.1051/0004-6361/202142197},
 eid = {A27},
 eprint = {2109.04458},
 journal = {\aap},
 month = {April},
 pages = {A27},
 primaryclass = {astro-ph.CO},
 title = {{Joint constraints on cosmology and the impact of baryon feedback: Combining KiDS-1000 lensing with the thermal Sunyaev-Zeldovich effect from Planck and ACT}},
 volume = {660},
 year = {2022}
}

@article{VanWaerbeke2013,
 adsurl = {https://ui.adsabs.harvard.edu/abs/2014PhRvD..89b3508V},
 archiveprefix = {arXiv},
 author = {{Van Waerbeke}, Ludovic and {Hinshaw}, Gary and {Murray}, Norman},
 doi = {10.1103/PhysRevD.89.023508},
 eid = {023508},
 eprint = {1310.5721},
 journal = {\prd},
 month = {January},
 number = {2},
 pages = {023508},
 primaryclass = {astro-ph.CO},
 title = {{Detection of warm and diffuse baryons in large scale structure from the cross correlation of gravitational lensing and the thermal Sunyaev-Zeldovich effect}},
 volume = {89},
 year = {2014}
}

\end{document}